# Vacuum Ultraviolet Photoionization Study on the Decomposition of JP-10 (*exo*-Tetrahydrodicyclopentadiene) in a High Temperature Chemical Reactor: Product Identification and Branching Ratios

Harish Kumar Chakravarty

*Department of Chemistry, University of Hawaii at Manoa, Honolulu, Hawaii 96822*

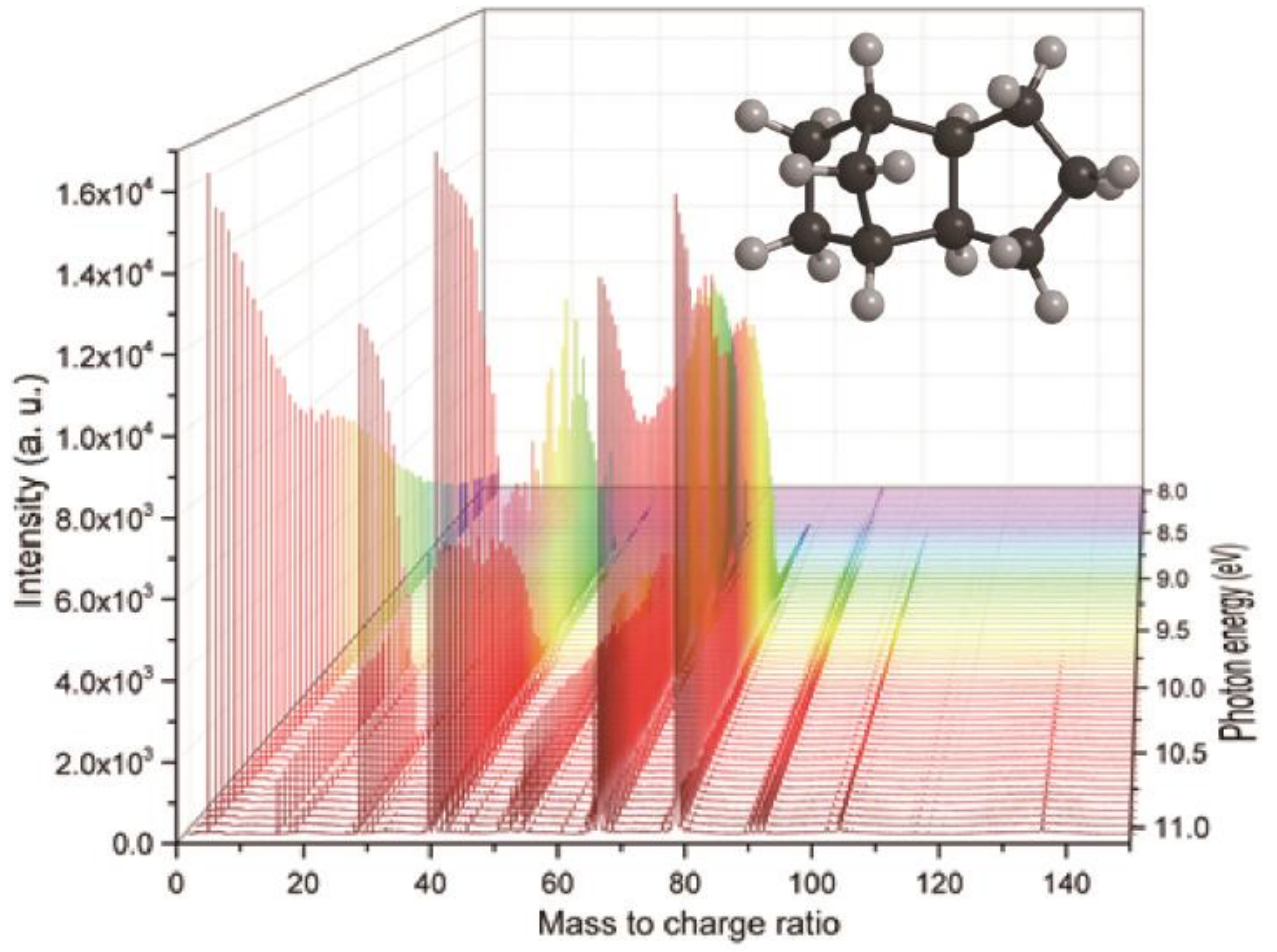

TOC. Mass spectrum of products obtained from the pyrolysis of the tricyclo[5.2.1.0$^{2,6}$]decane (JP-10) molecule recorded at photon energies between 8.0 eV and 11.0 eV at the temperature of 1,600 K.

**ABSTRACT:** Exploiting a high temperature chemical reactor, we explored the pyrolysis of helium-seeded tricyclo[5.2.1.0$^{2,6}$]decane (*exo*-tetrahydrodicyclopentadiene) as the principal constituent of Jet Propellant-10 (JP-10) over a temperature range of 1,100 K to 1,600 K at a pressure of 600 Torr. The nascent products were identified *in situ* in a supersonic molecular beam via single photon vacuum ultraviolet photoionization coupled with a mass spectroscopic analysis of the ions in a reflectron time-of-flight mass spectrometer (ReTOF). Our studies probe for the first time the *initial* reaction products formed in the decomposition of JP-10 – including radicals and thermally labile closed shell species – at reactor residence times of 6 to 7 μs effectively excluding subsequent mass growth processes. The present study identified 44 products in the pyrolysis of JP-10 with 22 molecules probed for the first time: C1 to C7 radicals, which are either acyclic (C1-C3: $CH_3$, $C_2H_3$, $C_2H_5$, $C_3H_3$, $C_3H_5$) (group I) or cyclic (C5-C7: $C_5H_5$, $C_6H_4$, $C_6H_5$, $C_7H_5$) (group II) and (highly) unsaturated (thermally labile) hydrocarbons, which can be subdivided into cyclic structures carrying between two and five double bond equivalents (C4, C6-C8: $C_4H_4$, $C_6H_8$, $C_7H_6$, $C_8H_8$) (group III) and non-cyclic molecules holding two to four double bond equivalents (C4-C5: $C_4H_4$, $C_5H_4$, $C_5H_8$) including cumulenes, dienes, and diynes (group IV). This inventory highlights the potential of our approach to detect novel radical decomposition products and hitherto unobserved closed shell species, which are highly unsaturated (cyclic) and thermally labile. Finally, we also propose potential formation routes of the molecules involving six doublet radicals (R1-R6). This information can be exploited to identify key reaction pathways in the decomposition of JP-10 and tracer molecules/radicals of distinct bond rupture processes.

## 1. INTRODUCTION

During the past decade, the combustion and pyrolysis of aviation fuels has been a fertile research area.[1-3] Tricyclo[5.2.1.0$^{2,6}$]decane or *exo*-tetrahydrodicyclopentadiene (Scheme 1) represents a single component hydrocarbon fuel and is the principal constituent of Jet Propellant-10 (JP-10; $C_{10}H_{16}$) as exploited in detonation engines, missiles, and supersonic combustion ramjets.[1-3] JP-10 holds attractive properties such as a high thermal stability, high-energy density, low freezing point, and high energy storage considering the highly strained tricyclic geometry with a ring strain energy of about 100 kJ mol$^{-1}$.[4] These properties triggered extensive experimental, theoretical, and modeling investigations to examine the features of oxidative and thermal decomposition mechanisms of JP-10[5-14] involving shock tubes, micro-flow tubes coupled with mass spectrometry, jet-stirred reactors, and flow reactors experiments (Tables 1 and 2).[6-13]

The products of a JP-10 pyrolysis were first investigated by Davidson et al. at pressures up to 1.5 bar and temperatures from 1,100 to 1,700 K. These experiments were carried out behind reflected shock waves exploiting a shock tube coupled with a high speed ultra violet (UV) absorption kinetic spectrograph.[6] The authors reported cyclopentene ($C_5H_8$) as the principal product along with acetylene ($C_2H_2$), ethylene ($C_2H_4$), ethane ($C_2H_6$), propene ($C_3H_6$), 1,3-butadiene ($C_4H_6$), cyclopentadiene ($C_5H_6$), and benzene ($C_6H_6$) as the minor products. This study proposed that a carbon-carbon bond rupture represents the first step in the decomposition resulting in the formation of cyclopentene. Hereafter, Nakra et al. explored the thermal decomposition mechanisms of JP-10 employing a micro-flow tube coupled with mass spectrometry; this study reported benzene ($C_6H_6$), cyclopentadiene ($C_5H_6$), propyne ($C_3H_4$), acetylene ($C_2H_2$), and ethylene ($C_2H_4$) as the principal products.[7] However, cyclopentene ($C_5H_8$) was not reported. Striebich and Lawrence investigated the JP-10 decomposition in a high temperature furnace under supercritical conditions at a pressure of 34 atm and within a temperature range of 373 K to 873 K. These studies suggested that even at 873 K, only 10 % of JP-10 undergoes decomposition.[10] Herbinet et al. studied the thermal decomposition of JP-10 in a jet–stirred reactor in the temperature range between 848 K and 933 K at atmospheric pressures; they identified hydrogen ($H_2$), ethylene ($C_2H_4$), propene ($C_3H_6$), cyclopentadiene ($C_5H_6$), benzene ($C_6H_6$), and toluene ($C_6H_5CH_3$) as major primary products.[9] A bi-radical approach involving the cleavage of various carbon-carbon bonds proposed by Tsang[15] was utilized to explain the formation of cyclic and aromatic products in a kinetic model. Subsequently, the thermal cracking of JP-10 was studied at higher pressures by Xing et al.;[11] a mechanism including decomposition and successive bimolecular reactions was proposed to explain the formation of cyclopentene ($C_5H_8$), cyclopentadiene ($C_5H_6$), benzene ($C_6H_6$), cis-bicyclo[3.3.0]oct-2-ene ($C_8H_{12}$), toluene ($C_6H_5CH_3$), and naphthalene ($C_{10}H_8$) as major products. Further, the thermal stability of JP-10 was investigated by Wohlwend et al.[14] in the temperature range from 473 K to 935 K at 34 atm. Based on these experiments, the authors proposed that JP-10 starts to decompose at 473 K with complete decomposition at 723 K. Vandewiele et al.[12] conducted pyrolysis experiments on JP-10 exploiting a continuous flow tubular reactor and gas chromatography coupled with time-of-flight mass spectrometer in the temperature range from 930 K to 1,080 K under atmospheric pressure. This work identified tricyco[5.2.1.0$^{2,6}$]dec-4-ene ($C_{10}H_{16}$) for the first time and also detected hydrogen ($H_2$), ethylene ($C_2H_4$), benzene ($C_6H_6$), and cyclopentadiene ($C_5H_6$) as the major products. Gao et al. performed shock tube pyrolysis experiments in the temperature range from 1,000 to 1,600 K and pressures ranging from 6-8 atm;[13] these authors refined the previous JP-10 combustion kinetic models. Beside the chemical aspects of the JP-10 decomposition, ignition delay times of JP-10 have also been reported at high temperatures and high pressures employing shock tubes with short reaction times of the order of a few milliseconds. First, the ignition delay of JP-10 was investigated by Davidson et al.[16] in the temperature range from 1,200 to 1,700 K, by Mikolaitis et al.[17] between 1,200 K and 2,500 K up to 25 atm, by Wang et al.[18] in the temperature range from 1,000 K to 2,100 K, and by Colket and Spadaccini[19] up to 1,500 K. To explain the delay times, Li et al.[20] first proposed a global JP-10 combustion reaction mechanism, which contains chemical species with up to three carbon atoms. This kinetic model was exploited by Nakra et al.,[7] but failed to explain the experimentally observed species such as cyclopentadiene ($C_5H_6$) and benzene ($C_6H_6$). Additionally, ignition delay experiments do not provide direct information on the reaction mechanisms and product distributions.

Besides these experimental investigations, computational chemistry was also exploited to unravel the decomposition mechanism of JP-10. Herbinet et al. conducted density functional (DFT) calculations to evaluate the theoretical kinetic parameters for initiation and propagation reactions of JP-10 decomposition.[9] Vandewiele et al.[21] developed a detailed kinetic model of JP-10 pyrolysis and refined these data based on rate constant calculations using *ab initio* calculations. Yue et al.[22] exploited DFT calculations to compute barrier heights of plausible decomposition pathways of multiple biradicals formed by carbon-carbon bond scission processes of JP-10. To elucidate the initial decomposition mechanism, Chenoweth et al.[23] carried out molecular dynamic simulations using a reactive force field. This work reported that the decomposition is initiated by carbon-carbon bond scission leading to ethylene ($C_2H_2$) plus $C_8$ hydrocarbons or to two $C_5$ hydrocarbons. Subsequently, Magoon et al.[24] investigated the barrier heights of ring opening processes and intramolecular disproportionation reactions to understand the pyrolysis mechanism of JP-10 reporting barrier heights on the order of a few tens kJ mol$^{-1}$. Thermochemical properties of various radicals and biradicals obtained from abstraction and bond scission of JP-10 were determined by Bozzelli et al.[25] using isodesmic reactions. Zehe et al.[26] studied the thermochemistry of JP-10 employing a variety of quantum chemistry methods suggesting a heat of formation of -126.4 kJ mol$^{-1}$ at 298.15 K.

The compilation of the previous studies suggests that an understanding of the decomposition of JP-10 both from the experimental and theoretical viewpoints is still in its infancy (Tables 1 & 2). *First*, although these investigations yielded valuable information on the formation of closed shell products, the products are mainly analyzed *ex situ* (HPLC, GCMS). This approach excludes detection of transient radicals and thermally labile closed shell molecules formed in the decomposition processes. Therefore, the molecular inventory might have been altered since its formation, crucial reaction intermediates were not sampled, and detailed information on the reaction mechanisms – the key role of radicals and intermediates – cannot always be obtained, but are at best inferred indirectly and qualitatively. *Second*, due to the inherent complexity, a simultaneous *in situ* probing of *all* transient species and closed shell products via spectroscopy such as UVVIS or laser-based techniques has been beyond the scope of any previous combustion experiment; recall that spectroscopic detection schemes like laser induced fluorescence (LIF) and Rydberg tagging (H, D, O) are restricted to species with well-established spectroscopic fingerprints, which are typically smaller, di- and triatomic species. It is therefore not surprising that the present kinetic models of JP-10 pyrolysis are mostly based on the thermochemical data and activation energies *inferred* from rather approximate bond or group additivity schemes or on molecular dynamics simulations with empirical reactive force fields like ReaxFF.[27] *Finally*, the sometimes long residence time of the *initial* reaction products and of radicals in particular can lead to consecutive reactions of the nascent products and even mass growth processes as evident from Table 1. This in turn alters the initial product distribution, and the products detected cannot be classified as 'initial reaction products'. Based on these limitations, a novel methodology to investigate the decomposition of JP-10 is crucial. *This approach requires probing the open and closed shell products in situ without changing the initial 'molecular inventory' and exploiting versatile detection systems so that the complete product spectrum can be sampled simultaneously.* Based on the aforementioned considerations, this endeavor presents a major challenge, as no single study has been conducted to date, in which the *initial* decomposition mechanisms and *overall spectra of newly formed open- and closed-shell product molecules* have been explored *in situ*.

The primary objectives of the present project – the experimental and theoretical exploration of the decomposition ('pyrolysis') of JP-10 (*exo*-tetrahydrodicyclopentadiene) - are therefore threefold. *First*, the decomposition of JP-10 will be studied in a high temperature chemical reactor, in which the decomposition of JP-10 can be probed systematically under combustion-like temperatures up to 1,600 K. The *nascent* product distribution – including radicals and thermally labile closed shell species – are comprehensively identified for the first time *in situ* in a supersonic molecular beam exploiting soft photoionization with single photon vacuum ultraviolet (VUV) light followed by a mass spectroscopic analysis of the ions in a reflectron time-of-flight mass spectrometer (ReTOF).[28-30] *Second*, by limiting the residence time in the reactor, we aim to probe only the *initial reaction products* and attempt to exclude successive (higher order) reactions of the initially formed species. *Third*, by conducting molecular beam experiments and combining these studies with electronic structure calculations, we extract information on the products, their branching ratios, and reaction mechanisms involved in the decomposition of JP-10 over a broad range of combustion relevant temperatures and pressures. Note that besides the basic scientific interest from the combustion community, these studies are also of fundamental interest to the physical (organic) chemistry community to unravel fundamental decomposition mechanisms of complex organic molecules, to probe isomerization processes of organic transient species, and to correlate the fragmentation mechanisms with the molecular structure of JP-10.

The aforementioned approach requires three steps. *First* – as disseminated in the present publication (I) – we report on the experimentally determined product distributions together with their branching ratios over a temperature range from 1,300 K to 1,600 K. *Second*, based on these data, a consecutive manuscript tackles the decomposition mechanism computationally thus rationalizing feasible reaction pathways to the experimentally observed products (II). *Third*, a final publication will combine the results from the experiments and computations and will incorporate these data into models to quantitatively reproduce the products detected along with their branching ratios (III). These data are very much required by the combustion community to unravel the initial processes, which trigger the decomposition of JP-10 and 'switch on' the supply of hydrocarbon radicals along with small closed shell hydrocarbons available for their successive oxidation.

## 2. EXPERIMENTAL METHOD

The experiments were carried out at the Advanced Light Source (ALS) at the Chemical Dynamics Beamline (9.0.2.) utilizing a chemical reactor.[31-42] The schematic diagram of the experimental setup is shown in Figure 1 along with the chemical reactor and the Reflectron Time-of-Flight Mass Spectrometer (ReTOF-MS). Briefly, the high temperature chemical reactor was a resistively heated silicon carbide (SiC) tube of 20 mm in length and 1 mm inner diameter. A gas mixture at a pressure of 600 Torr with 0.05 % JP-10 ($C_{10}H_{16}$) (TCI America; > 94%) in helium carrier gas (He; Airgas; 99.999 %) was prepared by bubbling helium gas through JP-10 kept in a stainless-steel bubbler at 263 ± 3 K. The gas mixture was introduced into the silicon carbide tube at temperatures up to 1,600 ± 10 K as monitored by a Type-C thermocouple. After exiting the reactor, the molecular beam, which contained the pyrolysis products, passed a skimmer and entered a detection chamber, which housed the Wiley–McLaren Reflectron Time-of-Flight Mass Spectrometer. The products were photoionized in the extraction region of the spectrometer by exploiting quasi continuous tunable vacuum ultraviolet (VUV) light from the Chemical Dynamics Beamline 9.0.2 of the Advanced Light Source and detected with a microchannel plate (MCP) detector. Mass spectra were recorded in 0.05 eV intervals from 8.00 eV to 11.00 eV by photoionizing the reactant and products; mass spectra were also taken at 12.00 eV and 15.50 eV to probe acetylene ($C_2H_2$) and molecular hydrogen ($H_2$) formation, respectively. The photoionization efficiency (PIE) curves were then generated by integrating the signal collected at a specific mass-to-charge ratio (m/z) over the range of photon energies and normalized to the incident photon flux. The residence time of the JP-10 ($C_{10}H_{16}$) in the reactor tube (20 mm) under our experimental condition ranged between only 6.8 ± 0.8 μs (1,100 K) and 5.6 ± 0.7 μs (1,600 K).[43] Ellison et al. also modeled the pressure profile in the reactor suggesting that, for example, at axial distances of 10 mm and 15 mm from the inlet, the pressure dropped to about 60 % and 30 % of the inlet pressure. This resulted in typically three to four collisions of a JP-10 molecule with the helium atoms at these distances at 1,600 K.

## 3. RESULTS

Figure 2 depicts characteristic mass spectra recorded at 10.00 eV. The mass-to-charge ratios along with the molecular formulae and assignments of the decomposition products are listed in Table 3. Characteristic photoionization efficiency (PIE) curves together with their fits are depicted in Figure 3 for the highest temperature of 1,600 K, at which the decomposition of JP-10 is quantitative; PIEs for the remaining temperatures are given in the Supplementary Information. It should be stressed that no product can be detected at m/z values higher than 118. Considering that the molecular mass of the JP-10 precursor is 136 amu and no product can be observed beyond m/z = 118, it is evident that our experimental conditions eliminate molecular mass growth processes; this requirement is crucial to exclude subsequent reactions of the nascent decomposition products of JP-10. To extract the nature of the products formed, the individual PIE curves have to be fit with (a linear combination of) known PIE curves of the structural isomers. In each of these graphs, the black line represents the average of three separate PIE scans; the shaded areas define the experimental uncertainties. Literature PIE graphs are taken from the photoionization cross section database[44] or related literatures. The experimentally determined branching ratios as compiled in Table 4 and Figure 4 are corrected for the photoionization cross sections.

**m/z = 15:** Signal at m/z = 15 can be assigned to a species with the molecular formula $CH_3$. The experimentally recorded PIE for m/z = 15 (Fig. 3.1) is nicely reproduced by the reference PIE curve (red line) for the methyl radical.[45] The PIE for the methyl radical exhibits an onset at 9.84 eV. Note that the red shift in the onset of ionization in the PIE curve of the methyl radical might result from rotationally and vibrationally 'hot' methyl radicals.

**m/z = 27:** The signal at m/z = 27 is linked to a species of the molecular formula $C_2H_3$. The PIE for m/z = 27 (Fig. 3.2) is fit with the reference PIE (red line) for the vinyl radical.[46]

**m/z = 28:** The ion counts at m/z = 28 are contributed to by a species with the molecular formula $C_2H_4$. The experimentally recorded PIE for m/z = 28 is fit by the reference PIE curve (red line) for ethylene (Fig. 3.3).[47] The PIE for ethylene exhibits an onset at 10.5 eV.

**m/z = 29:** The signal at m/z = 29 can in principle be connected with a $C_2H_5$ species. However, a second contributor is required to fit the experimental data. Here, the experimentally determined PIE for m/z = 29 is reproducible with a linear combination of the reference PIE curves (red line) for $^{13}C$ substituted ethylene (blue line)[47] and the ethyl radical (green line) (Fig. 3.4).[48] The PIE curves for $^{13}C$-ethylene and the ethyl radical exhibit an onset at 10.5 eV and 8.1 eV, respectively.

**m/z = 39:** The signal at m/z = 39 is associated with a species of the molecular formula $C_3H_3$. The experimentally determined PIE for m/z = 39 is reproducible by the reference PIE curve (red line) of the propargyl radical[46] (Fig. 3.5) – a novel detection of this species in JP-10 decomposition studies. This PIE for the propargyl radical depicts an onset at 8.67 eV.

**m/z = 40:** The signal at m/z = 40 is assigned to the molecular formula $C_3H_4$. The experimentally recorded PIE for m/z = 40 is fit with a linear combination of reference PIE curves (red line) for propyne (green line)[47] and for allene (blue line)[47] (Fig. 3.6). It should be noted here that the cyclic isomer cyclopropene ($C_3H_4$) has no contribution to these PIE.

**m/z = 41:** The signal at m/z = 41 is linked to the molecular formula $C_3H_5$. The experimentally recorded PIE for m/z = 41 is reproducible via the reference PIE curve (red line) of the allyl radical with the PIE for the allyl radical exhibiting an onset at 8.18 eV (Fig. 3.7).[49] The allyl radical has never been observed before in JP-10 related experiments.

**m/z = 42:** The signal at m/z = 42 is attributable to the molecular formula $C_3H_6$. From 9.7 eV on, the experimentally recorded PIEs for m/z = 42 can be matched by the reference PIE curve (red line) for propene (Fig. 3.8).[47] The reference PIE for propene exhibits an onset at 9.73 eV. It should be noted that the red shift in the ionization energy to 8.2 eV might be linked to internally excited ('hot') propene molecules.

**m/z = 52:** The signal at m/z = 52 can be attributed to the formula $C_4H_4$. The PIE for m/z = 52 cannot be fit by a single contribution, but require a linear combination of reference PIE curves (red line) of cyclobutadiene (green line),[44] vinylacetylene (blue line),[50] and 1,2,3-butatriene (orange line)[44] (Fig. 3.9). The experimental PIEs for cyclobutadiene, vinylacetylene, and 1,2,3-butatriene show onsets at 8.16 eV, 9.58 eV, and 9.25 eV, respectively.

**m/z = 54:** The signal at m/z = 54 is connected with various isomers of $C_4H_6$. The PIE for m/z = 54 can be matched by the composite of experimental reference PIE curves (red line) for 1,3-butadiene (blue line)[51] and of 1,2-butadiene (green line)[52] (Fig. 3.10). The reference PIE for 1,3-butadiene and 1,2-butadiene exhibit onsets at 9.07 eV and 9.25 eV, respectively.

**m/z = 64:** The signal at m/z = 64 is assigned to four isomers of $C_5H_4$ – none of them observed in prior JP-10 related decomposition studies. The PIE for m/z = 64 is well-matched by the composite of the experimental reference PIE curves (red line) for 1,3-pentadiyne (blue line),[53] ethynylallene (green line),[44] 1,2,3,4-pentatetraene (orange line),[44] and 1,4-pentadiyne (pink line)[44] (Fig. 3.11). The experimental PIEs for 1,3-pentadiyne, ethynylallene, 1,2,3,4-pentatetra-ene, and 1,4-pentadiyne exhibit onsets at 9.5 eV, 9.22 eV, 8.67 eV, and 10.27 eV, respectively.

**m/z = 65:** The PIE at m/z = 65 ($C_5H_5$) is linked to the experimental reference PIE curve (red line) of the cyclopentadienyl radical ($C_5H_5$)[53] – a novel detection in JP-10 pyrolysis studies (Fig. 3.12).

**m/z = 66:** The PIE curves of m/z = 66 ($C_5H_6$) can be fit with the reference PIE of the 1,3-cyclopentadiene (red line) isomer (Fig. 3.13).[53] The acyclic molecules 3-pent-1-yne, pent-3-yne, and 1,2,4-pentatriene have no contributions to the signal at m/z = 66.

**m/z = 67:** The PIE at m/z = 67 (Fig. 3.14) was found to be $^{13}$C-substituted 1,3-cyclopentadiene (red line) isomer, but not from $C_5H_7$.

**m/z = 68:** The signal at m/z = 68 is assigned to $C_5H_8$. The PIE for m/z = 68 is reproduced by the composite of the experimental reference PIE curves (red line) for 1,3-pentadiene (blue line)[51] and cyclopentene (green line)[50] at temperatures from 1,300 to 1,500 K (Supplementary Information), but can be produced solely by 1,3-pentadiene (red line) at the temperature of 1,600K (Fig. 3.15). The experimental PIEs for 1,3-pentadiene and of cyclopentene exhibit onsets at 8.59 eV and 9.01 eV, respectively. 1,3-Pentadiene represents a first detection in the JP-10 pyrolysis.

**m/z = 76:** Signal at m/z = 76 is assigned to $C_6H_4$. The PIE for m/z = 76 is matched by the reference PIE curve (red line) of o-benzyne (Fig. 3.16).[44, 54] The experimental PIE for benzyne exhibits an onset at 9.03 eV. The red shift of the PIE to 8.25 eV cannot be fit with a supersonically cooled benzyne species, but might suggest internally excited benzyne and/or other $C_6H_4$ isomers. Benzyne also represents a first detection in the pyrolysis of JP-10.

**m/z = 77:** The PIE recorded at m/z = 77 ($C_6H_5$) can be reproduced nicely by the literature data of the PIE of the phenyl radical[55] up to a photon energy of 10.5 eV – a novel detection of an aromatic radical in the pyrolysis of JP-10 (Fig. 3.17).

**m/z =78:** The signal at m/z = 78 is linked to $C_6H_6$. The PIE for m/z = 78 are well-matched by the composite of reference PIE curves (red line) for benzene (blue line)[50] and fulvene (green line)[56] (Fig. 3.18). The PIE for benzene exhibits an onset at 9.2 eV and that of fulvene at 8.4 eV. Note that the experimentally determined branching fractions of benzene and fulvene indicate fractions between 11.92 ± 2.24 % and 32.06 ± 1.53 % for benzene and 3.85 ± 0.18 % and 22.57 ± 4.25 % for the fulvene isomer. From 1,400 K to 1,600 K, the fraction of benzene increases which fulvene decreases. This might suggest that the formation pathways of benzene and fulvene are coupled.

**m/z =79:** The PIE at m/z = 79 can be attributed to the $^{13}$C substituted benzene and fulvene species (Fig. 3.19).

**m/z = 80:** The signal at m/z = 80 is assigned to $C_6H_8$. The experimental PIE is well-fitted with a linear combination of PIEs (red) of 1,3-cyclohexadiene (blue) [57] and 1,4-cyclohexadiene (green)[44] depicting onsets at 8.25 eV and 8.8 eV, respectively (Fig. 3.20). Ion counts at photon energies higher than 10.2 eV originate from dissociative photoionization of the non-pyrolyzed precursor molecules. Whereas 1,3-cyclohexadiene is a common product in the pyrolysis of JP-10, its 1,4-cyclohexadiene isomer has not been observed prior to the present study.

**m/z = 89:** The signal at m/z = 89 is linked to the $C_7H_5$ radical formed at 1,600 K. Similar to its fulvenallene precursor, this species represents a novel observation in the decomposition of JP-10. The onset of the experimental PIE matches well with the PIE of the fulvenallenyl radical exhibiting an onset at 8.2 eV (Fig. 3.21).[44] Since no photoionization cross sections exist in the literature, the branching ratio of this radical cannot be quantified.

**m/z = 90:** The signal at m/z = 90 can be attributed to $C_7H_6$. The experimental PIE curve recorded for m/z = 90 is well-fitted by the reference PIE of fulvenallene (red line) (Fig. 3.22).[44] This represents the first detection of fulvenallene in pyrolysis experiments of JP-10. The PIE for fulvenallene depicts an onset at 8.29 eV.

**m/z = 92:** The signal at m/z = 92 can be assigned to $C_7H_8$. The experimental PIE can be fit by the composite of PIEs (red line) of toluene (blue line)[58] and of 1,3,5-cycloheptatriene (green line) (Fig. 3.23). The PIEs for toluene and 1,3,5-cycloheptatriene exhibit onsets at 8.83 eV and 8.29 eV, respectively. Note that the branching ratios of 1,3,5-cycloheptatriene could not be extracted since its photoionization cross section is unknown.

**m/z = 93:** The PIEs at m/z = 93 represent the $^{13}C$ signal of toluene and 1,3,5-cycloheptatriene with no contribution from the $C_7H_9$ radical (Fig. 3.24).

**m/z = 102:** The signal at m/z = 102 is correlated with $C_8H_6$. The experimental PIE recorded for m/z = 102 can be reproduced from 8.82 eV to 11.00 eV is matched by the PIE of phenylacetylene (red line) (Fig. 3.25).[58] The PIE for phenylacetylene exhibits an onset at 8.82 eV. At the present stage, we are unable to account for the ion signals from 8.00 eV to 8.82 eV. This signal possible originates from internally excited phenylacetylene.

**m/z = 104:** The signal at m/z 104 is allocated to $C_8H_8$. The experimental PIE recorded for m/z = 104 is reproduced by a composite PIEs (red line) of styrene (blue line)[58] and of cyclooctatetraene (green line) (Fig. 3.26). The PIEs for styrene and cyclooctatetraene exhibit onsets at 8.46 eV and 8.03 eV, respectively. Note that branching ratios of cyclooctatetraene could not be calculated since its photoionization cross section is not known. However, this presents the very first time that cyclooctatetraene – a formally anti-aromatic 8π molecule which is unstable in the bulk at room temperature – has been identified in the decomposition of JP-10.

**m/z = 116:** The signal at m/z = 116 is well matched with a molecule of the formula $C_9H_8$. The experimental PIE is nicely reproduced by the PIE of indene (red line) depicting an onset at 8.14 eV (Fig. 3.27).[41, 58]

**m/z = 118:** The ion counts at m/z = 118 can be linked to a molecule holding the molecular formula $C_9H_{10}$. The experimental PIE is well reproduced by the PIE of indane (red line) (Fig. 3.28).[58] This PIE exhibits an onset at 8.54 eV. However, the formation of indane was not observed from 1,300 K to 1,500 K, but only at 1,600 K.

It is also important to address the detection of molecular hydrogen ($H_2$) and of acetylene ($C_2H_2$) which has ionization energies of 15.43 eV and 11.4 eV, respectively. Here, mass spectra were recorded at photon energies of 12.00 eV and 15.50 eV to probe acetylene ($C_2H_2$) and molecular hydrogen ($H_2$), respectively. Corrected for background counts, these data suggest that signal at m/z = 2 can be attributed to the molecular hydrogen observed over the complete temperature range from 1,400 K to 1,600 K. Signal at m/z = 26 can be assigned to a molecular species with the molecular formula $C_2H_2$ and/or $^{13}C_2$. Since the ionization onset for both acetylene and dicarbon are 11.4 eV,[44] but no ion counts were observed at m/z = 24 ($C_2^+$), signal at m/z = 26 originates from acetylene with overall small branching ratios of less than 0.02 %.

We would like to address briefly the sections of some PIE curves, which could not be fit yet: the red shifted tails at lower photon energies and excess ion counts beyond 10.5 eV. Low molecular weight molecules such as m/z = 15 and 42 might be internally excited thus shifting the onset of the ionization to lower photon energies. Alternatively, we attempted to fit these sections of the PIEs with non-traditional isomers such as carbenes of m/z = 42 (Supplementary Information). However, even these isomers could not reproduce the red shifts satisfactorily. Likewise, the 'red shifted' PIEs of m/z = 76 and 102 might contain contributions from hitherto unknown benzyne or styrene isomers. Finally, the excess ion counts at photon energies higher than typically 10.5 eV likely originate from dissociative photoionization of non-pyrolyzed JP-10 precursor molecules. As evident from the temperature dependent PIE curves (Figure 3,

Supplementary Information), this signal decreases with rising temperature as the pyrolysis of the JP-10 becomes more complete, and less of the precursor 'survives'.

## 4. DISCUSSION

*First*, we demonstrated that the pyrolysis of the JP-10 depicts a strong temperature dependence starting at about 1,300 K and shifts from pyrolysis fractions of 0 % (1,100 K), 0 % (1,200 K), 4.4 ± 0.8 % (1,300 K), 53.2 ± 15.6 % (1,400 K), and 94.5 ± 3.1 % (1,500 K) to eventually 100 % (1,600 K).

*Second*, the dilute JP-10 concentration in our beam along with the low residence time in the chemical reactor effectively eliminates mass growth processes as evident from the absence of signal at masses higher than the JP-10 reactant. This is quite distinct from previous studies on the decomposition of JP-10 (Tables 1 & 2) depicting even polycyclic aromatic hydrocarbons such as naphthalene ($C_{10}H_8$; 128 amu), fluorene ($C_{13}H_{10}$; 166 amu), and phenanthrene ($C_{14}H_{10}$; 178 amu), which are clearly absent in our studies. Therefore, our apparatus is uniquely designed to probe the initial decomposition products of JP-10, their branching fractions, and how the branching ratios of the pyrolysis products depend on the temperature (Fig. 4).

*Third*, we have identified 44 products in the pyrolysis of JP-10, among them five $^{13}C$-substituted species originating from naturally occurring $^{13}C$, and characterized their temperature dependence (Tables 3-4; Figures 3-4). Among these 44 species, 22 product molecules – including $^{13}C$-isotopically substituted molecules – were characterized for the very first time as decomposition products of JP-10. These can be arranged into four main groups: C1 to C7 radicals, which are either acyclic (C1-C3: $CH_3$, $C_2H_3$, $C_2H_5$, $C_3H_3$, $C_3H_5$) (group I) or cyclic (C5-C7: $C_5H_5$, $C_6H_4$, $C_6H_5$, $C_7H_5$) (group II) and (highly) unsaturated (thermally labile) hydrocarbons, which can be subdivided into cyclic structures carrying between two and five double bond equivalents (C4-C6, C8: $C_4H_4$, $C_6H_8$, $C_7H_6$, $C_8H_8$) (group III) and acyclic molecules carrying between two and four double bond equivalents (C4-C5: $C_4H_4$, $C_5H_4$, $C_5H_8$) including cumulenes, dienes, and diynes (group IV).

*Fourth*, the overall decomposition products can be categorized in ten classes (Fig. 5) ranging from radicals (class I), cumulenes and their ethylene stem compound (class II), substituted ethylenes (class III), substituted methylacetylenes along with their parent compound (class IV), substituted allenes together with their stem compound (class V), unsaturated molecules with C5-rings (class VI), unsaturated molecules with C6-rings (class VII), monocyclic molecules carrying the benzene core (class VIII), highly unsaturated C4 to C8 cyclic molecules (class IX), and bicyclic molecules (class X).

*Fifth*, a comparison of these classes with the pyrolysis products detected in previous experiments (Tables 1 & 2) leads to interesting findings. All products detected in previous studies, which have masses higher that 118 amu (about one third of all molecules), must be higher order products formed via successive reactions of the nascent pyrolysis species (Tables 3 & 4). Further, those methyl- and ethyl-substituted products (Table 2), which were not observed in our experiments, can be rationalized by radical – radical recombination products involving methyl and ethyl radicals as a result of higher order reactions in previous experiments. Finally, successive hydrogenation reactions could account for the detection of, for example, methane, ethane, propane, n-butane, cyclopentene, 1-butene, 2-butene, butyne, and 1,3-cycloheptadiene.

*Sixth*, we would like to address briefly possible formation routes of the molecules detected in our studies. Recall that a comprehensive computational study of the decomposition pathways of JP-10 via doublet radicals R1 to R6 and through carbon-carbon bond rupture processes leading to biradicals BR1 to BR7 (Figure 6) is beyond the scope of the present work and presented in a forthcoming publication in this journal (II). In an attempt to characterize key radicals and molecular tracers of decomposition pathways, we established a matrix reporting the detected molecules versus the radical precursor(s) (Table 5). Let us discuss potential mechanisms for the formation of the observed products. Here we focus on the decomposition pathways of R1 to R6 occurring predominantly via β-scission processes leading to ring opening and/or dissociation (Figure 6, Supporting Information). Biradical pathways via BR1 – BR6 are not expected to play a major role as a recent theoretical analysis of the reaction pathways by Vandewiele et al. has provided compelling evidence that the contribution of these bond rupture processes does not exceed 19 % overall.[12]

**C1–C2:** Small unsaturated hydrocarbon molecules and radicals including acetylene ($C_2H_2$), vinyl ($C_2H_3$), ethylene ($C_2H_4$), and allyl ($C_3H_5$) can be produced through multiple dissociation steps involving β carbon-carbon bond scissions. Here, acetylene, vinyl, and ethylene can be formed from all six initial radicals R1-R6, whereas allyl can be formed from R1, R4, R5, and R6. The formation of the methyl radical ($CH_3$) cannot be explained within the frame of the β-scission mechanisms. The initial JP-10 molecule and its radicals (Figure 5) do not carry any methyl ($CH_3$) groups and hence hydrogen atom shifts are required to produce methyl. Our analysis (Supplementary Information) suggests that methyl ($CH_3$) can be formed in conjunction with dihydroindane (after two hydrogen shifts in R1-3, R2-2, R4-4), or together with 1,3,5-cycloheptatriene via a multistep pathway. The later starts from R2-2 and involves expansion to a nine-member ring, ring opening, elimination of ethylene ($C_2H_4$) leading to a branched octatrienyl radical ($C_8H_{11}$), which then undergoes a seven-member ring closure, three consecutive hydrogen migrations and methyl ($CH_3$) elimination. The methyl radical can be also formed together with toluene ($C_6H_5CH_3$) following ethylene ($C_2H_4$) losses in the reaction pathways initiating from R2-2, R4-1, and R4-4. Here, the parent species of toluene and the methyl radical are six-member ring $C_8H_{11}$ radicals (substituted cyclohexenes with two external $CH_2$ groups). These radicals have to undergo numerous hydrogen migrations to convert both out-of-ring $CH_2$ groups to $CH_3$ groups, and one of them is then split to form toluene plus the methyl radical. These complicated pathways are expected to be entropically unfavorable, but they may be replaced by a series of hydrogen additions/eliminations. No pathways leading to ethyl radicals ($C_2H_5$) can be seen in β-scission pathways from R1-R6; it is plausible that ethyl radical is formed by recombination of two methyl ($CH_3$) radicals followed by a hydrogen atom loss.

**C3:** The propargyl radical ($C_3H_3$) can originate from R4-1 through the initial ethylene ($C_2H_4$) loss by β-scission, five-member ring openings and dissociation to 1,4-pentadiene ($C_5H_6$) plus propargyl ($C_3H_3$) (Supplementary Information). Allene ($C_3H_4$) can be produced via several pathways beginning from R2-1, R2-2, and R4-1 after up to two β-scission dissociation steps together with $C_2H_3$, $C_7H_{11}$, or $C_3H_5$ radicals. Alternatively, decomposition of allyl radical ($C_3H_5$) by hydrogen loss can also form allene. No β-scission pathways exist to propyne ($CH_3CCH$), and it is more likely to be produced by isomerization of allyl ($C_3H_5$) followed by hydrogen atom elimination; more complex mechanisms involving hydrogen shifts to create the methyl ($CH_3$) group might be present as well. Recombination of the allyl radical ($C_3H_5$) with atomic hydrogen can lead to the production of propene ($C_3H_6$); no one step pathways to propylene ($C_3H_6$) can be found in β-scission mechanisms initiated from R1-R6.

**C4:** The dissociation channels leading to the $C_4H_4$ isomers vinylacetylene and 1,2,3-butatriene are numerous and originate from R1-R5 and R2 and R4, respectively. In the majority of these channels, the precursor of vinylacetylene and 1,2,3-butatriene is the $CH_2C^{\bullet}CHCH_2$ radical. Before losing atomic hydrogen, this $C_4H_5$ radical may undergo a four-member ring closure and then the hydrogen loss would result in the production of cyclobutadiene ($C_4H_4$) as detected experimentally. No one step pathways to cyclobutadiene can be seen in the β-scission schemes. The most stable isomer of $C_4H_6$, 1,3-butadiene, can be formed through a variety of pathways starting from R1, R2, R4, and R5. Alternatively, the less stable 1,2-butadiene isomer can be produced only from R2 and R4, by hydrogen atom additions to the $H_2C{=}C{=}CH{-}CH_2^{\bullet} \leftrightarrow H_2C{=}C^{\bullet}{-}CH{=}CH_2$ β-scission product.

**C5:** Considering the $C_5H_x$ species, let us begin with the most abundant cyclopentene ($C_5H_8$), cyclopentadiene ($C_5H_6$), and the cyclopentadienyl radical ($C_5H_5$). The first two molecules are formed through multiple β-scission pathways initiating from R1-R3 and R6 (cyclopentene) and all six initial radicals (cyclopentadiene); in particular, cyclopentene ($C_5H_8$) can be formed from R1-2, R1-3, R2-1, R2-2, R3-2, and R6-1, whereas cyclopentadiene ($C_5H_6$) can be produced from R1-2, R3-2, R4-5, and R5-1 by atomic hydrogen loss from the cyclopentyl radical ($C_5H_7$) or otherwise by two consecutive hydrogen atom losses from cyclopentene ($C_5H_8$) (Supplementary Information). Next, the cyclopentadienyl radical can originate from atomic hydrogen elimination or abstraction from cyclopentadiene ($C_5H_6$). The acyclic $C_5H_8$ molecule, 1,3-pentadiene, originates from R2, via R2-1 and R2-2. Dissociation of these radicals by β-scission pathways can lead to the production of $CH_2{=}CH{-}CH{=}CH{-}CH_2^{\bullet}$, which can then form 1,3-pentadiene by hydrogen atom addition. As illustrated in Figures S8-S26 in Supplementary Information, the acyclic $C_5H_4$ isomers, 1,2,3,4-pentatetraene, ethynylallene, 1,3-pentadiyne, 1,4-pentadiyne are likely to originate from the $C_5H_7$ radicals. For instance, the $CH_2{=}CH{-}CH{=}CH{-}CH_2^{\bullet} \leftrightarrow CH_2{=}CH{-}CH^{\bullet}{-}CH{=}CH_2$ radical, which can be formed by β-scission dissociation pathways from R2, R4, and R5 or by five-member ring opening in cyclopentyl can lose atomic hydrogen to form 1,2,4-pentatriene, which in turn can eliminate two hydrogens producing 1,2,3,4-pentatetraene and/or ethynylallene. A hydrogen shift in $CH_2{=}CH{-}CH{=}CH{-}CH_2^{\bullet}$ leads to the $CH_2{=}C^{\bullet}{-}CH_2{-}CH{=}CH_2$ radical, which can also be formed from R2 (via R2-1 and R2-2). The latter can dissociate to penta-1-en-4-yne, two consecutive hydrogen losses from which may result in 1,4-pentadiyne. Alternatively, another hydrogen shift in $CH_2$=CH-CH=CH-$CH_2^•$ leads to $CH_2$=CH-CH=$C^•$-$CH_3$, which dissociates to penta-1-en-3-yne, and then two hydrogen atom ejections can result in 1,3-pentadiyne. Apparently, the acyclic $C_5H_6$ isomers, 1,2,4-pentatriene, penta-1-en-4-yne, and penta-1-en-3-yne do not survive long enough under the experimental conditions and give rise to the observed $C_5H_4$ species by dehydrogenation.

**C6:** The pathways to benzene ($C_6H_6$) and its fulvene isomer originate from R2, R3, and R4 and are coupled. Two consecutive ethylene ($C_2H_4$) eliminations by β-scissions starting from R2-3 lead to a five-member ring $C_6H_7$ isomer with an out-of-ring CH group or to the acyclic CH≡C-$CH_2$-$CH^•$-CH=$CH_2$ structure, both of which then can isomerize and lose a hydrogen atom forming benzene ($C_6H_6$) or fulvene ($C_6H_6$). The fact that rearrangements on the $C_6H_7$ potential energy surface followed by hydrogen elimination can lead both to fulvene and benzene via hydrogen atom assisted isomerization is well documented in the literature.[59] Similarly, the $C_6H_7$ isomers – precursors of both benzene ($C_6H_6$) and fulvene ($C_6H_6$) – can be formed from R2-4, R3-1, and R4-3 (Supplementary Information). Additionally, fulvene ($C_6H_6$) can be formed from dehydrogenation of various $C_6H_8$ precursor via R4-1 decomposition pathways. Considering the structure of the JP-10 precursor, no one step pathways to the phenyl radical ($C_6H_5$) and to benzyne ($C_6H_4$) are found; we expect these species to be produced by consecutive dehydrogenation of benzene ($C_6H_6$). The other cyclic $C_6$ species, cyclohexadienes ($C_6H_8$) and cyclohexenes ($C_6H_{10}$), can be produced via R1 and R5 from the same or different $C_6H_9$ cyclic radical precursors. For instance, R1-1 can ring-open in two steps to a chain $C_{10}H_{15}$ radical, which then loses 1,3-butadiene ($C_4H_6$) by β-scission forming a chain $C_6H_9$ isomer and the latter ring-closes to cyclohex-1-en-3-yl. The cyclic $C_6H_9$ radical can lose a hydrogen atom to produce 1,3-cyclohexadiene ($C_6H_{10}$) or add a hydrogen to form cyclohexene ($C_6H_{10}$). An alternative pathway from R1-1 or a channel from R1-2 lead to vinyl substituted cyclohex-1-en-4-yl, which a precursor of both styrene ($C_6H_5C_2H_3$) and 1,4-cyclohexadiene ($C_6H_8$); the latter is produced by splitting the external vinyl ($C_2H_3$) group. Pathways from R5-1 and R5-2 can lead to cyclohex-1-en-4-yl following eliminations of ethylene ($C_2H_4$) and acetylene ($C_2H_2$) and the cyclic $C_6H_9$ can form either 1,3- or 1,4-cyclohexadiene ($C_6H_{10}$) by hydrogen atom losses or add an hydrogen atom to produce cyclohexene ($C_6H_{10}$) (Supplementary Information). The acyclic $C_6$ molecule, 2,4-hexadiene ($C_6H_8$), can be formed only from R1 via R1-3 through a complicated pathway involving several hydrogen shifts. R1-1 opens up to a chain $C_{10}H_{15}$ structure, which after a series of hydrogen migrations dissociates to 2,4-hexadiene ($C_6H_{10}$) plus the $C_4H_5$ radical.

**C7:** Fulveneallene ($C_7H_6$) can be formed in R4 and R5 pathways. Opening of one of the five-member rings and elimination of allyl in R4-5 or R5-1 result in a vinyl-substituted cyclopentene, $C_7H_{11}$, which can produce fulveneallene ($C_7H_6$) by consecutive splitting of four hydrogen atoms (Supplementary Information). Further, fulveneallene can lose one more hydrogen atoms producing the fulveneallenyl radical ($C_7H_5$). The $C_7H_8$ isomers toluene and cycloheptatriene can be formed in conjunction with the methyl radical and the pathways leading to them are described above.

**C8:** The reaction channels producing phenylacetylene ($C_8H_6$) originate from R2 and R4. A pathway from R2-2 leads to an open-chain $C_8H_{11}$ radical after ethylene ($C_2H_4$) elimination and the radical undergoes a six-member ring closure to a $H_2C\text{-}C^{\bullet}$ substituted cyclohexene (Supplementary Information). The latter splits a hydrogen atom forming cyclic $C_8H_{10}$ isomers, which can eventually produce phenylacetylene ($C_6H_5CCH$) after four hydrogen atom eliminations. Again, this is plausible only if the cyclic $C_8H_{10}$ species are thermally unstable. The pathway from R4-3 similarly leads to the cyclic $C_8H_{10}$ precursors of phenylacetylene, but in this case ethylene ($C_2H_4$) elimination leads to an eight-member ring $C_8H_{11}$ radical, which first ring-opens and then undergoes a six-member ring closure. Cyclooctatetraene can form from R1 and R3. R1-1 can produce an eight-member ring radical by β-scission, which eliminates the external vinyl group forming cyclooctadiene ($C_8H_{12}$) (Supplementary Information). If the latter is thermally unstable, it might consecutively eject four hydrogens forming cyclooctatetraene ($C_8H_8$). Alternatively, R3-1 first loses ethylene ($C_2H_4$), then ring-opens to the cyclooctadienyl radical and loses a hydrogen atom to produce cyclooctatriene ($C_8H_{10}$). The latter is a more plausible precursor of cyclooctatetraene because only two additional hydrogen eliminations are required in this case. The second, much more stable $C_8H_8$ isomer, styrene, can be produced from R1 and R5. The pathways from R1-1 and R1-2 are shared with those leading 1,4-cyclohexadiene as described above, but the last step is a hydrogen loss on the contrary to vinyl ($C_2H_3$) elimination. 1-Vinyl-1,4-cyclohexadiene (dihydrostyrene) is formed in this channel and two hydrogen atom eliminations are additionally required. The alternative pathway from R5-3 seems less plausible because it leads to vinylcyclohexene after splitting two vinyl ($C_2H_3$) groups and hence elimination of four hydrogen atoms is required to form styrene.

**C9:** The $C_9$ molecules, indane and indene, are formed by hydrogen atom elimination from dihydroindane ($C_9H_{12}$), which can be produced from R1, R2, and R4 together with methyl radical and the corresponding pathways are described above. It should be emphasized that these mechanisms discussed are only proposed reaction pathways at this stage and are based on the favorability of β-scissions in hydrocarbon radicals. In the future work, we intend to actually compute accurate potential energy surfaces for the proposed mechanism and evaluate temperature- and pressure-dependent rate constants for the formation of various fragments. This would allow us to model the experimental data observed here and to predict the JP-10 pyrolysis rate constants and relative product yields under different conditions relevant to combustion.

## 5. CONCLUSIONS

To conclude, we conducted a comprehensive investigation of the thermal decomposition of JP-10 in a high-temperature chemical reactor in the temperature range from 1,100 K to 1,600 K with residence times of the JP-10 molecules from about 6 to 7 μs at a pressure of 600 Torr. Most important, 44 nascent products have been detected exploiting by interrogating the supersonically cooled pyrolysis products by tunable vacuum ultraviolet (VUV) light. Among them, 22 products were characterized for the very first time. These can be arranged into four groups: C1 to C7 radicals, which are either acyclic (C1-C3: $CH_3$, $C_2H_3$, $C_2H_5$, $C_3H_3$, $C_3H_5$) (groups I) or cyclic (C5-C7: $C_5H_5$, $C_6H_4$, $C_6H_5$, $C_7H_5$) (groups II) and (highly) unsaturated (thermally labile) hydrocarbons, which can be subdivided into cyclic structures carrying between two and five double bond equivalents (C4, C6-C8: $C_4H_4$, $C_6H_8$, $C_7H_6$, $C_8H_8$) (groups III) and non-cyclic molecules with two to four double bond equivalents (C4-C5: $C_4H_4$, $C_5H_4$, $C_5H_8$) including cumulenes, dienes, and diynes (class IV). Finally, we also proposed formation routes of those molecules involving six doublet radicals (R1-R6). Those information can be utilized to assign key reaction pathways in the decomposition of JP-10 and define tracers molecules/radicals of distinct bond rupture processes. These data are crucial to the combustion community to untangle the initial bond rupture processes, which trigger the decomposition of JP-10 and ‘switch on’ the supply of radicals in the combustion of JP-10 jet fuel.

**ASSOCIATED CONTENT**

The Supplementary Information is available free of charge

**AUTHOR INFORMATION**

**Corresponding Authors**

**Email: hkchakravarty@gmail.com**

**The authors declare no competing financial interest.**

**Table 1.** Compilation of previous experimental studies on the pyrolysis of JP-10.

| Group | Method | Temperature (K) | Pressure (bar) | Residence time | Ref. |
|---|---|---|---|---|---|
| Davidson et al. | Shock tube | 1100-1700 | 1.2-1.5 | 50-200 μs | 6, 16 |
| Vandewiele et al. | Flow tubular reactor | 930-1080 | 1.7 | - | 12 |
| Xing et al. | Batch reactor | 823-903 | 1-38 | 0.48-26.4 s | 11 |
| Herbinet et al. | Jet-stirred reactor | 848-933 | 1 | 0.5-6 s | 9 |
| Rao et al. | Annular tubular reactor | 903-968 | 1 | 0.68-6.4 s | 8 |
| Nakra et al. | Flow tube reactor | 298-1498 | 0.002-0.004 | 2.10-9.35 ms | 7 |
| Gao et al. | Shock tube | 1000-1600 | 6-8 | 500 μs | 13 |
| Bruno et al. | Thermal block | 623-698 | 34.5 | 4min-20h | 5 |
| Wohlwend et al. | System for thermal diagnostic studies | 473-935 | 34 | 1.8 s at 473 K | 14 |
| Striebich et al. | System for thermal diagnostic studies | 373-873 | 34 | 1-5 s | 10 |
| Devener et al. | Alumina flow-tube reactor | 300-1400 | 0.03 | 3.3-9.1 ms | 4 |

**Table 2.** The 79 products reported in previous experimental studies on the pyrolysis of JP-10.

| Molecule | Formula | Mass | Structure |
|---|---|---|---|
| Hydrogen | $H_2$ | 2 | H—H |
| Methane | $CH_4$ | 16 | $CH_4$ |
| Acetylene | $C_2H_2$ | 26 | 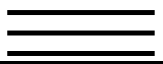 |
| Ethylene | $C_2H_4$ | 28 |  |
| Ethane | $C_2H_6$ | 30 |  |
| Propyne | $C_3H_4$ | 40 | 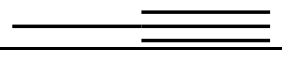 |
| Allene | $C_3H_4$ | 40 | 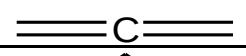 |
| Propene | $C_3H_6$ | 42 | 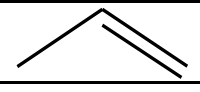 |
| Propane | $C_3H_8$ | 44 | 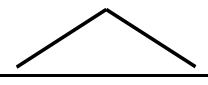 |
| 1,3-Butadiyne | $C_4H_2$ | 50 |  |
| Vinylacetylene | $C_4H_4$ | 52 | 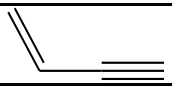 |
| 1-Butyne | $C_4H_6$ | 54 | 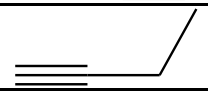 |
| 1,3-Butadiene | $C_4H_6$ | 54 | 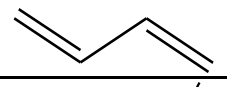 |
| 1,2-Butadiene | $C_4H_6$ | 54 | 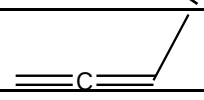 |
| 1-Butene | $C_4H_8$ | 56 | 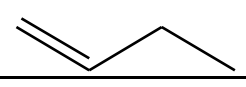 |
| 2-Butene | $C_4H_8$ | 56 | 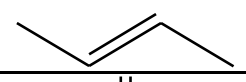 |
| Isobutylene | $C_4H_8$ | 56 | 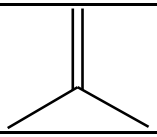 |
| n-Butane | $C_4H_{10}$ | 56 | 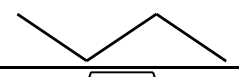 |
| 1,3-Cyclopentadiene | $C_5H_6$ | 66 | 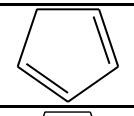 |
| Cyclopentene | $C_5H_8$ | 68 | 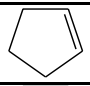 |
| Cyclopentane | $C_5H_{10}$ | 70 | 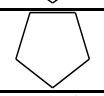 |
| Fulvene | $C_6H_6$ | 78 | 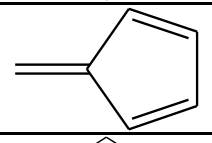 |
| Benzene | $C_6H_6$ | 78 | 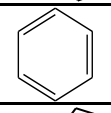 |
| 1-Methylcyclopentadiene | $C_6H_8$ | 80 | 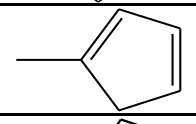 |
| 2-Methylcyclopentadiene | $C_6H_8$ | 80 | 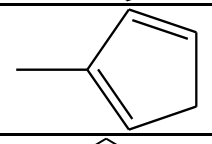 |
| 1,3-Cyclohexadiene | $C_6H_8$ | 80 | 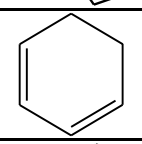 |
| Methylcyclopentadiene | $C_6H_8$ | 80 | 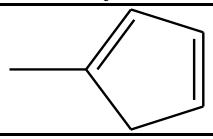 |
| 1,5-Hexadiene | $C_6H_{10}$ | 82 | 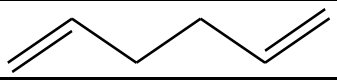 |
| 1-Methylcyclopentene | $C_6H_{10}$ | 82 | 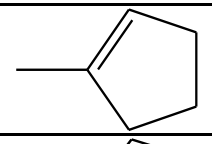 |
| Methylcyclopentane | $C_6H_{10}$ | 84 | 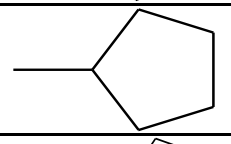 |
| 3-Ethynyl-cyclopentene | $C_7H_8$ | 92 | 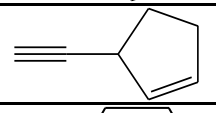 |
| Toluene | $C_7H_8$ | 92 | 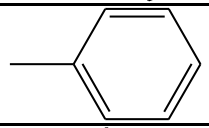 |
| 1,3,5-Cycloheptatriene | $C_7H_8$ | 92 | 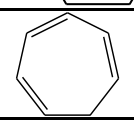 |
| 2-Propenylidene cyclobutene | $C_7H_8$ | 92 | 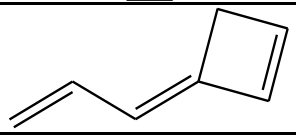 |
| Bicyclo(4.1.0)hept-2-ene | $C_7H_{10}$ | 94 | 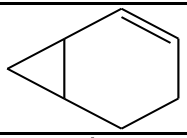 |
| 1,3-Cycloheptadiene | $C_7H_{10}$ | 94 | 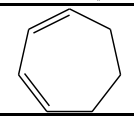 |
| 3-Ethenylcyclopentene | $C_7H_{10}$ | 94 | 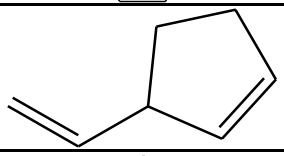 |
| 2-Norbornene | $C_7H_{10}$ | 94 | 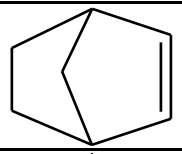 |
| 1,2-Dimethylcyclopentadiene | $C_7H_{10}$ | 94 | 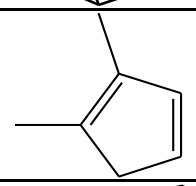 |
| 1-Ethylcyclopentene | $C_7H_{12}$ | 96 | 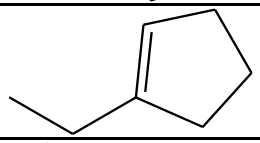 |
| Ethylidenecyclopentane | $C_7H_{12}$ | 96 | 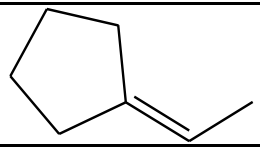 |
| Ethylcyclopentane | $C_7H_{14}$ | 98 | 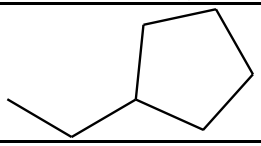 |
| Phenylacetylene | $C_8H_6$ | 102 | 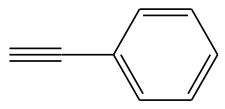 |
| Styrene | $C_8H_8$ | 104 | 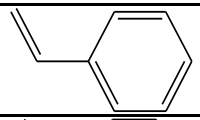 |
| Ethylbenzene | $C_8H_{10}$ | 106 | 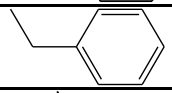 |
| o-Xylene | $C_8H_{10}$ | 106 | 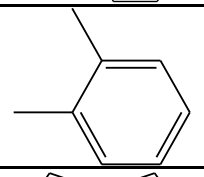 |
| 1,2,3,3a,4,6a-Hexahydropentalene | $C_8H_{12}$ | 108 | 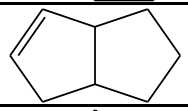 |
| cis-Octahydropentalene | $C_8H_{14}$ | 110 | 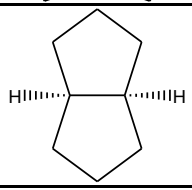 |
| Propylcyclopentane | $C_8H_{16}$ | 112 | 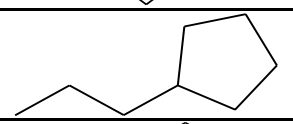 |
| Indene | $C_9H_8$ | 116 | 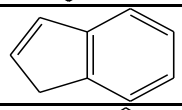 |
| Indane | $C_9H_{10}$ | 118 | 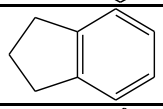 |
| 2,3,4,5,6,7-Hexahydro-1H-indene | $C_9H_{14}$ | 122 | 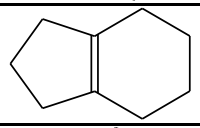 |
| 3-Methylcyclooctene | $C_9H_{16}$ | 124 | 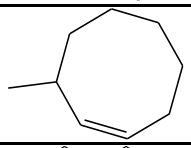 |
| Naphthalene | $C_{10}H_8$ | 128 | 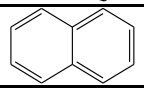 |
| 2-Methylindene | $C_{10}H_{10}$ | 130 | 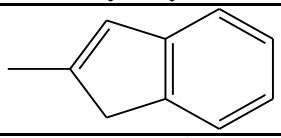 |
| Dicyclopentadiene | $C_{10}H_{12}$ | 132 | 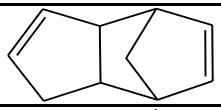 |
| 2,3-Dihydro-4-methyl-1H-indene | $C_{10}H_{12}$ | 132 | 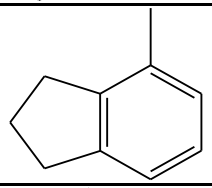 |
| 2,3-Dihydro-1-methyl-1H-indene | $C_{10}H_{12}$ | 132 | 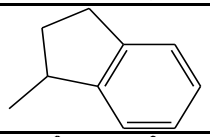 |
| 1,2,3,4-Tetrahydronaphthalene | $C_{10}H_{12}$ | 132 | 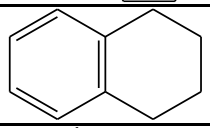 |
| Tricyclodec-4-ene | $C_{10}H_{14}$ | 134 | 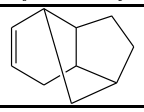 |
| 3a,4,5,6,7,7a-Hexahydro-4,7-methanoindene | $C_{10}H_{14}$ | 134 | 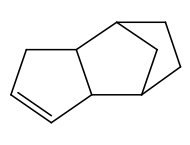 |
| Bicyclopentyl-1,1'-diene | $C_{10}H_{14}$ | 134 | 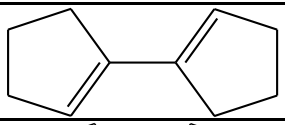 |
| 3-Cyclopentylcyclopentene | $C_{10}H_{16}$ | 136 | 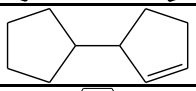 |
| 1,2-Divinylcyclohexane | $C_{10}H_{16}$ | 136 | 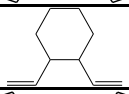 |
| 1-Cyclopentylcyclopentene | $C_{10}H_{16}$ | 136 | 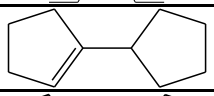 |
| Bicyclopentylidene | $C_{10}H_{16}$ | 136 | 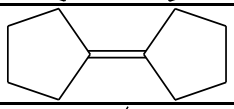 |
| 4-Ethyl-3-ethylidene cyclohexene | $C_{10}H_{16}$ | 136 | 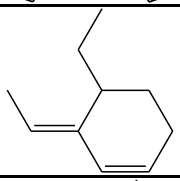 |
| 4-Methyl-2,3,4,5,6,7-hexahydro-1H-indene | $C_{10}H_{16}$ | 136 | 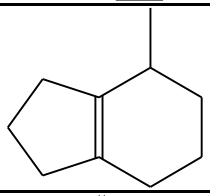 |
| 2,2-Dimethyl-5-methylenebicyclo[2.2.1]heptane | $C_{10}H_{16}$ | 136 | 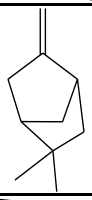 |
| Tricyclo[5.2.1.0(4,8)]decane | $C_{10}H_{16}$ | 136 | 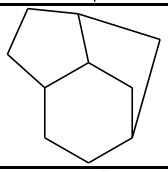 |
| Tricyclo[4.2.1.1(2,5)]decane | $C_{10}H_{16}$ | 136 | 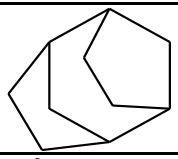 |
| Tricyclo[4.4.0.0(2,8)]decane | $C_{10}H_{16}$ | 136 | 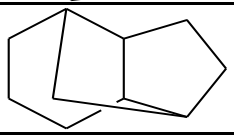 |
| Cyclopentylcyclopentane | $C_{10}H_{18}$ | 138 | 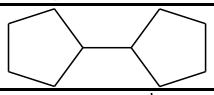 |
| Methylnaphthalene | $C_{11}H_{10}$ | 142 | 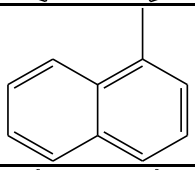 |
| 1,2,3,6,7,8-Hexahydro-as-indacene | $C_{12}H_{14}$ | 158 | 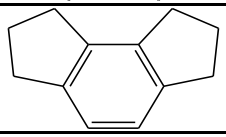 |
| trans-Syn-tricyclo[7.3.0.0(2,6)]-8-dodecene | $C_{12}H_{18}$ | 162 | 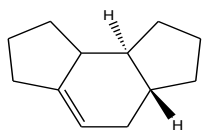 |
| cis-8-Ethyl-exo-tricyclo[5.2.1.0(2.6)]decane | $C_{12}H_{20}$ | 164 | 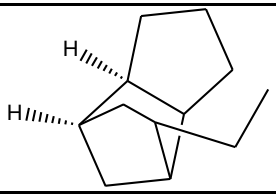 |
| Fluorene | $C_{13}H_{10}$ | 166 | 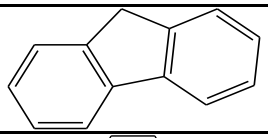 |
| Phenanthrene | $C_{14}H_{10}$ | 178 | 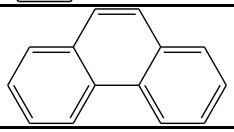 |

**Table 3.** Compilation of products observed in the present experiments on the decomposition of JP-10. 44 species were observed, among them 22 'first time' detections highlighted in bold.

| Molecule | Formula | Mass | Structure |
|---|---|---|---|
| Hydrogen | $H_2$ | 2 | H—H |
| **Methyl radical** | **$CH_3$** | **15** | $CH_3$• |
| Acetylene | $C_2H_2$ | 26 | 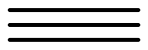 |
| **Vinyl radical** | **$C_2H_3$** | **27** | 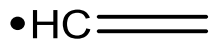 |
| Ethylene | $C_2H_4$ | 28 |  |
| **Ethyl radical** | **$C_2H_5$** | **29** | 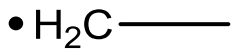 |
| **Propargyl radical** | **$C_3H_3$** | **39** | 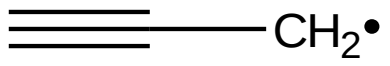 |
| Allene | $C_3H_4$ | 40 | 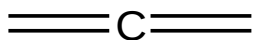 |
| Propyne | | | 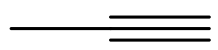 |
| **Allyl radical** | **$C_3H_5$** | **41** | 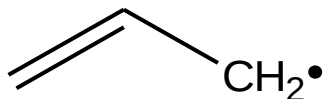 |
| Propene | $C_3H_6$ | 42 | 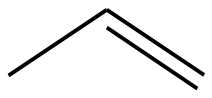 |
| **Cyclobutadiene** | **$C_4H_4$** | **52** | 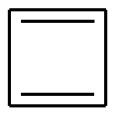 |
| Vinylacetylene | $C_4H_4$ | 52 | 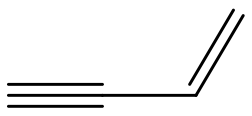 |
| **1,2,3-Butatriene** | **$C_4H_4$** | **52** | 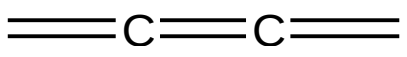 |
| 1,3-Butadiene | $C_4H_6$ | 54 | 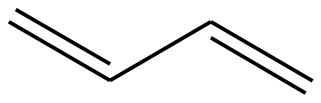 |
| 1,2-Butadiene | | | 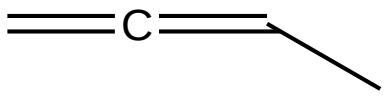 |
| **1,3-Pentadiyne** | **$C_5H_4$** | **64** | 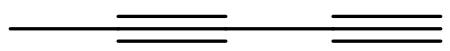 |
| **Ethynylallene** | | | 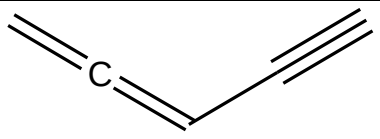 |
| **1,2,3,4-Pentatetraene** | | |  |
| **1,4-Pentadiyne** | | | 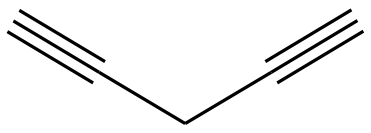 |
| **Cyclopentadienyl radical** | **$C_5H_5$** | **65** | 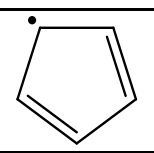 |
| 1,3-Cyclopentadiene | $C_5H_6$ | 66 | 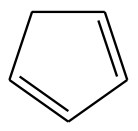 |
| **1,3-Cyclopentadiene ($^{13}$C)** | **$^{13}CC_4H_6$** | **67** | 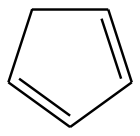 |
| **1,3-Pentadiene** | **$C_5H_8$** | **68** | 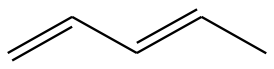 |
| Cyclopentene | $C_5H_8$ | 68 | 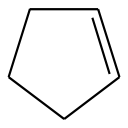 |
| **Benzyne** | **$C_6H_4$** | **76** | 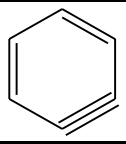 |
| **Phenyl radical** | **$C_6H_5$** | **77** | 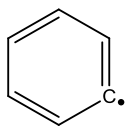 |
| Benzene | $C_6H_6$ | 78 | 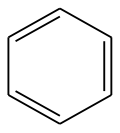 |
| Fulvene | | | 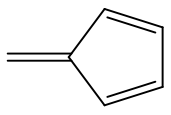 |
| Benzene ($^{13}$C) | $^{13}CC_5H_6$ | 79 | 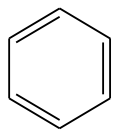 |
| Fulvene ($^{13}$C) | | | 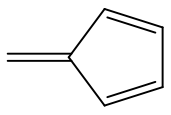 |
| 1,3-Cyclohexadiene | $C_6H_8$ | 80 | 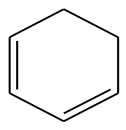 |
| **1,4-Cyclohexadiene** | **$C_6H_8$** | **80** | 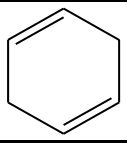 |
| **Fulveneallenyl radical** | **$C_7H_5$** | **89** | 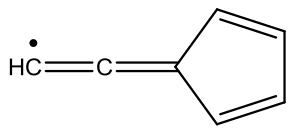 |
| **Fulveneallene** | **$C_7H_6$** | **90** | 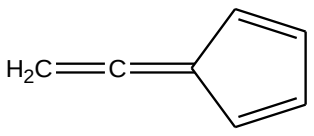 |
| 1,3,5-Cycloheptatriene | $C_7H_8$ | 92 | 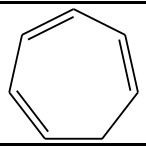 |
| Toluene | | | 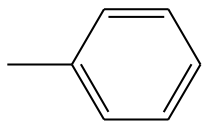 |
| **1,3,5-Cycloheptatriene ($^{13}$C)** | **$^{13}CC_6H_8$** | **93** | 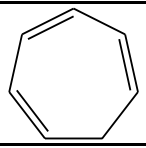 |
| **Toluene ($^{13}$C)** | | | 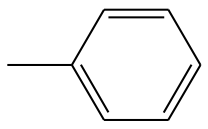 |
| Phenylacetylene | $C_8H_6$ | 102 | 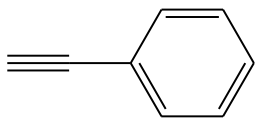 |
| **Cyclooctatetraene** | **$C_8H_8$** | **104** | 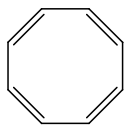 |
| Styrene | $C_8H_8$ | 104 | 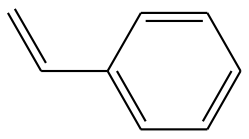 |
| Indene | $C_9H_8$ | 116 | 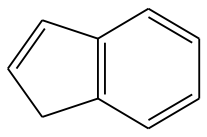 |
| Indane | $C_9H_{10}$ | 118 | 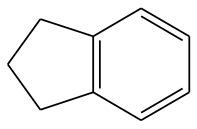 |

**Table 4.** Branching fractions of the products formed in the decomposition of JP-10 at 600 Torr in the chemical reactor at 1,300, 1,400, 1,500, and 1,600 K.

| Molecule* | Mass | Temperature, K | | | |
|---|---|---|---|---|---|
| | | 1,300 | 1,400 | 1,500 | 1,600 |
| Hydrogen | 2 | | 0.004±0.001 | 0.007±0.001 | 0.001±0.001 |
| Methyl radical | 15 | 0.06±0.01 | 0.08±0.02 | 0.11±0.01 | 0.17±0.01 |
| Acetylene | 26 | | 0.001±0.001 | 0.014±0.001 | 0.016±0.001 |
| Vinyl radical | 27 | 22.59±0.23 | 0.42±0.03 | 0.33±0.01 | 0.18±0.03 |
| Ethylene | 28 | | 0.05±0.03 | 0.15±0.03 | 0.05±0.01 |
| Ethyl radical | 29 | | 0.30±0.01 | 0.01±0.00 | 0.04±0.01 |
| Propargyl radical | 39 | | 1.54±0.03 | 2.65±0.73 | 8.82±0.30 |
| Allene | 40 | 1.41±0.03 | 3.93±0.01 | 10.37±1.14 | 12.34±0.32 |
| Propyne | 40 | 0.24±0.01 | 0.70±0.07 | 4.61±0.51 | 7.54±0.20 |
| Allyl radical | 41 | 1.28±0.01 | 0.86±0.13 | 0.37±0.13 | 0.17±0.11 |
| Propene | 42 | 1.00±0.01 | 1.11±0.21 | 1.06±0.14 | 0.80±0.17 |
| 1,2,3-Butatriene | 52 | | 0.19±0.10 | 0.42±0.11 | 1.02±0.12 |
| Cyclobutadiene | 52 | | 0.14±0.08 | 0.21±0.05 | 0.17±0.01 |
| Vinylacetylene | 52 | | 0.05±0.02 | 0.29±0.07 | 0.53±0.06 |
| 1,3-Butadiene | 54 | 0.07±0.01 | 0.08±0.01 | 0.10±0.01 | 0.09±0.01 |
| 1,2-Butadiene | 54 | 0.02±0.01 | 0.02±0.01 | 0.03±0.01 | 0.01±0.01 |
| 1,3-Pentadiene | 64 | | | | 0.01±0.01 |
| Ethynylallene | 64 | | | | 0.12±0.03 |
| 1,2,3,4-Pentatetraene | 64 | | | | 0.43±0.11 |
| 1,4-Pentadiyne | 64 | | | | 0.08±0.02 |
| Cyclopentadienyl radical | 65 | 1.44±0.01 | 1.67±0.32 | 3.71±0.35 | 6.15±0.24 |
| 1,3-Cyclopentadiene | 66 | 35.81±0.36 | 14.06±2.35 | 17.59±0.69 | 12.03±1.75 |
| 1,3-Pentadiene | 68 | 1.51±0.01 | 2.96±0.67 | 1.25±0.32 | 1.08±0.51 |
| Cyclopentene | 68 | 2.93±0.01 | 3.16±0.71 | 1.20±0.31 | |
| Benzyne | 76 | | | | 0.51±0.22 |
| Phenyl radical | 77 | | | 1.44±0.01 | 0.35±0.35 |
| Benzene | 78 | 15.13±0.07 | 11.92±2.24 | 23.22±0.30 | 32.06±1.53 |
| Fulvene | 78 | 16.51±0.08 | 22.57±4.25 | 13.27±0.15 | 3.85±0.18 |
| 1,3-Cyclohexadiene | 80 | | 6.46±1.34 | 1.05±0.25 | 0.17±0.09 |
| 1,4-Cyclohexadiene | 80 | | 3.53±0.72 | 1.34±0.31 | 0.43±0.25 |
| Fulveneallene | 90 | | | 0.49±0.14 | 0.94±0.06 |
| Toluene | 92 | | 5.85±1.20 | 8.41±1.09 | 2.41±0.08 |
| Phenylacetylene | 102 | | | | 0.25±0.16 |
| Styrene | 104 | | 18.35±2.25 | 6.32±0.58 | 7.08±0.02 |
| Indene | 116 | | | | 0.07±0.01 |
| Indane | 118 | | | | 0.05±0.01 |

* Photoionization cross sections of the fulveneallenyl radical (89 amu), 1,3,5-cycloheptatriene (92 amu), and cyclooctatetraene (104 amu) are not available; branching ratios of these species could not be calculated.

**Table 5.** Matrix linking the decomposition products to R1-R6.

| Molecule | Formula | Mass | R1 | R2 | R3 | R4 | R5 | R6 |
|---|---|---|---|---|---|---|---|---|
| Hydrogen | $H_2$ | 2 | X | X | X | X | X | X |
| Methyl radical | $CH_3$ | 15 | X | X | | X | | |
| Acetylene | $C_2H_2$ | 26 | X | X | X | X | X | X |
| Vinyl radical | $C_2H_3$ | 27 | X | X | X | X | X | X |
| Ethylene | $C_2H_4$ | 28 | X | X | X | X | X | X |
| Ethyl radical | $C_2H_5$ | 29 | | | | | | |
| Propargyl radical | $C_3H_3$ | 39 | | | | X | | |
| Allene | $C_3H_4$ | 40 | | X | | X | | |
| Propyne | $C_3H_4$ | 40 | | | | | | |
| Allyl radical | $C_3H_5$ | 41 | X | | | X | X | X |
| Propene | $C_3H_6$ | 42 | | | | | | |
| Cyclobutadiene | $C_4H_4$ | 52 | | | | | | |
| Vinylacetylene | $C_4H_4$ | 52 | X | X | X | X | X | |
| 1,2,3-Butatriene | $C_4H_4$ | 52 | | X | | X | | |
| 1,3-Butadiene | $C_4H_6$ | 54 | X | X | | X | X | |
| 1,2-Butadiene | $C_4H_6$ | 54 | | X | | X | | |
| 1,3-Pentadiyne | $C_5H_4$ | 64 | X | | | | | |
| Ethynylallene | $C_5H_4$ | 64 | | X | | X | X | |
| 1,2,3,4-Pentatetraene | $C_5H_4$ | 64 | | X | | X | X | |
| 1,4-Pentadiyne | $C_5H_4$ | 64 | | X | | X | | |
| Cyclopentadienyl radical | $C_5H_5$ | 65 | X | X | X | X | X | X |
| 1,3-Cyclopentadiene | $C_5H_6$ | 66 | X | X | X | X | X | X |
| 1,3-Pentadiene | $C_5H_8$ | 68 | | X | | | | |
| Cyclopentene | $C_5H_8$ | 68 | X | X | X | | | X |
| Benzyne | $C_6H_4$ | 76 | | | | | | |
| Phenyl radical | $C_6H_5$ | 77 | | | | | | |
| Benzene | $C_6H_6$ | 78 | | X | X | X | | |
| Fulvene | $C_6H_6$ | 78 | | X | X | X | | |
| 1,3-Cyclohexadiene | $C_6H_8$ | 80 | X | | | | X | |
| 1,4-Cyclohexadiene | $C_6H_8$ | 80 | X | | | | X | |
| Fulveneallenyl radical | $C_7H_5$ | 89 | | | | X | X | |
| Fulveneallene | $C_7H_6$ | 90 | | | | X | X | |
| 1,3,5-Cycloheptatriene | $C_7H_8$ | 92 | | X | | | | |
| Toluene | $C_7H_8$ | 92 | | X | | X | | |
| Phenylacetylene | $C_8H_6$ | 102 | | X | | X | | |
| Cyclooctatetraene | $C_8H_8$ | 104 | X | | X | | | |
| Styrene | $C_8H_8$ | 104 | X | | | | X | |
| Indene | $C_9H_8$ | 116 | X | X | | X | | |
| Indane | $C_9H_{10}$ | 118 | X | X | | X | | |

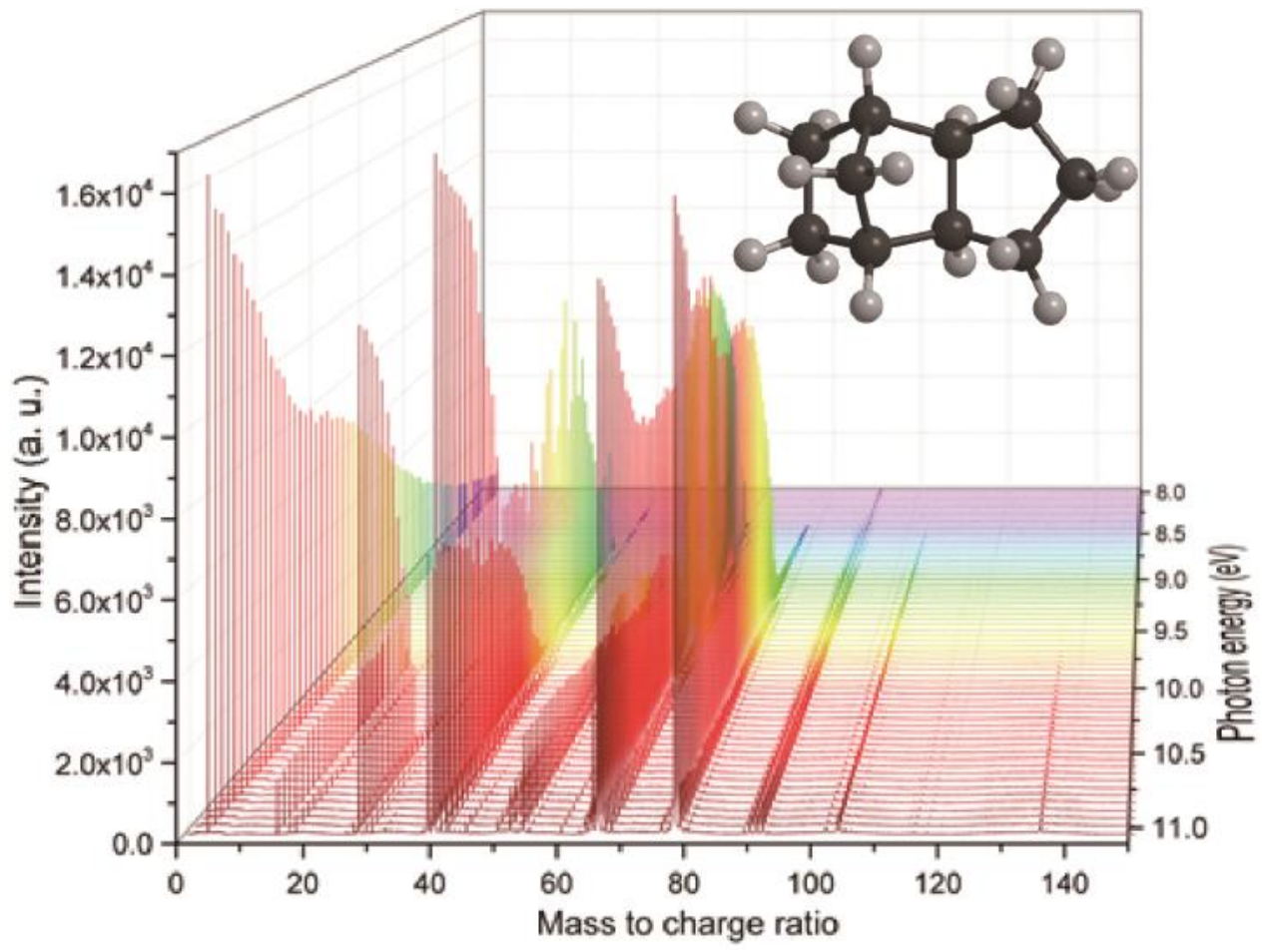

TOC. Mass spectrum of products obtained from the pyrolysis of the tricyclo[5.2.1.0$^{2,6}$]decane (JP-10) molecule recorded at photon energies between 8.0 eV and 11.0 eV at the temperature of 1,600 K.

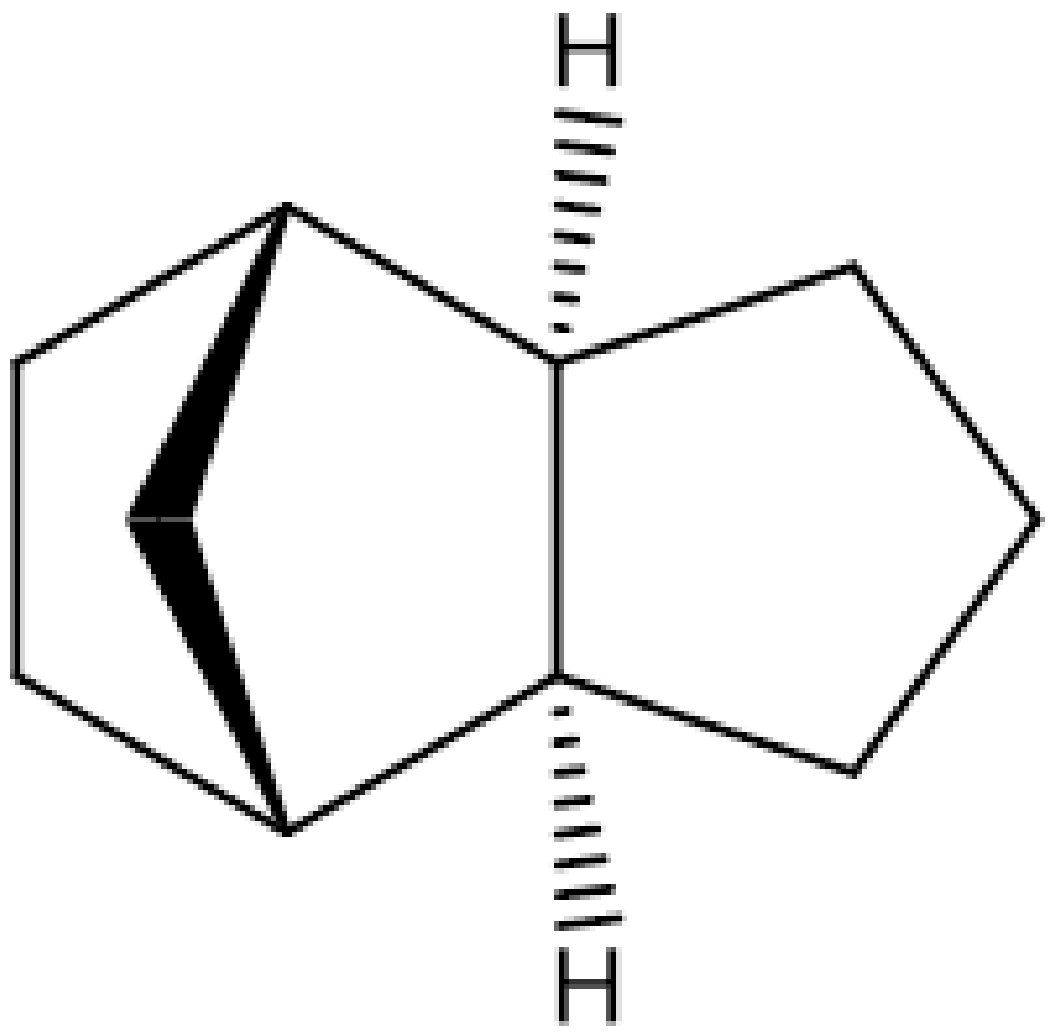

**Scheme 1.** Structure of the tricyclo[5.2.1.0$^{2,6}$]decane (JP-10) molecule.

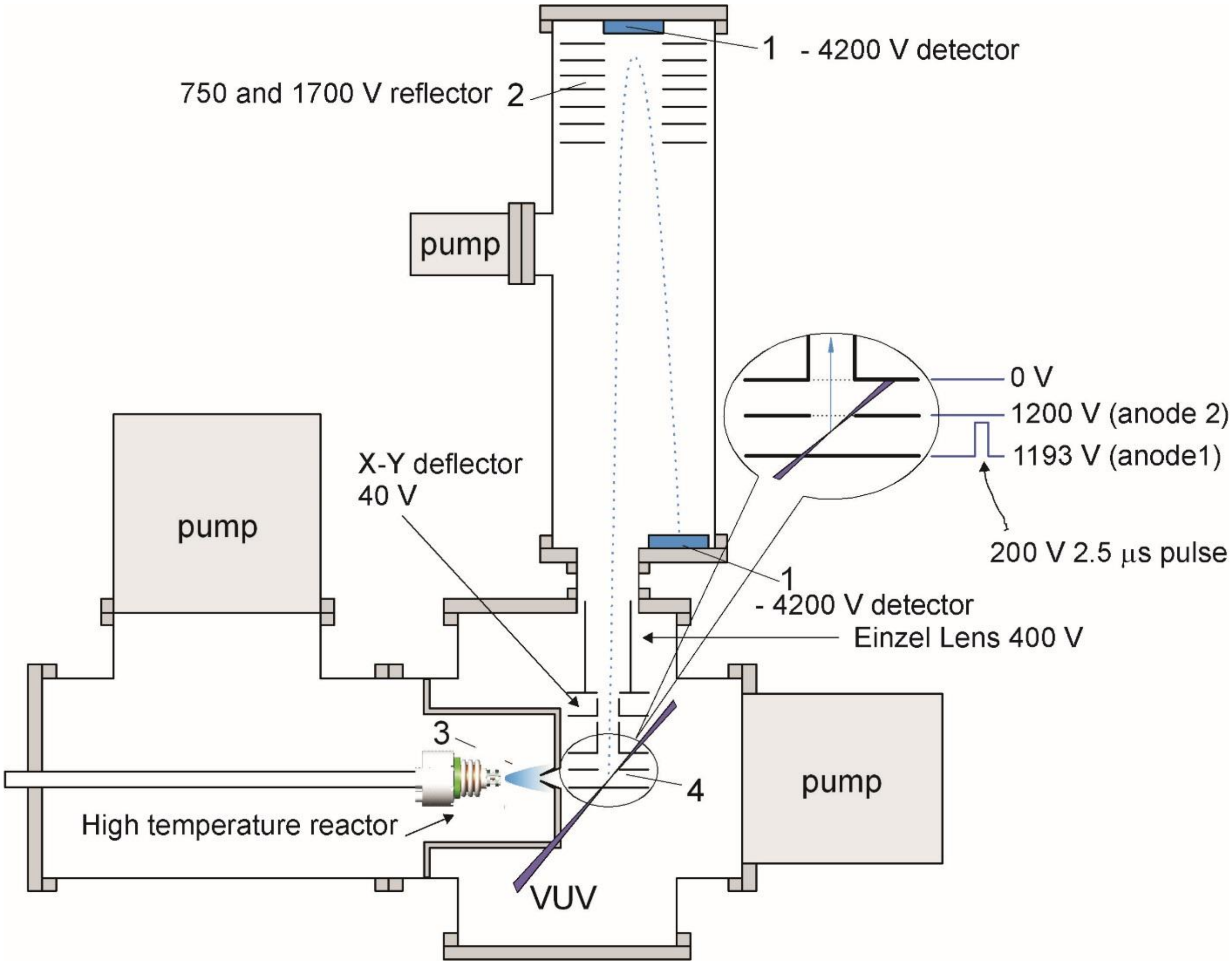

**Figure 1.** Schematic experimental setup depicting the chemical reactor, tunable vacuum ultraviolet light beam, molecular beam, turbo molecular pumps, and reflectron time of flight mass spectrometer.

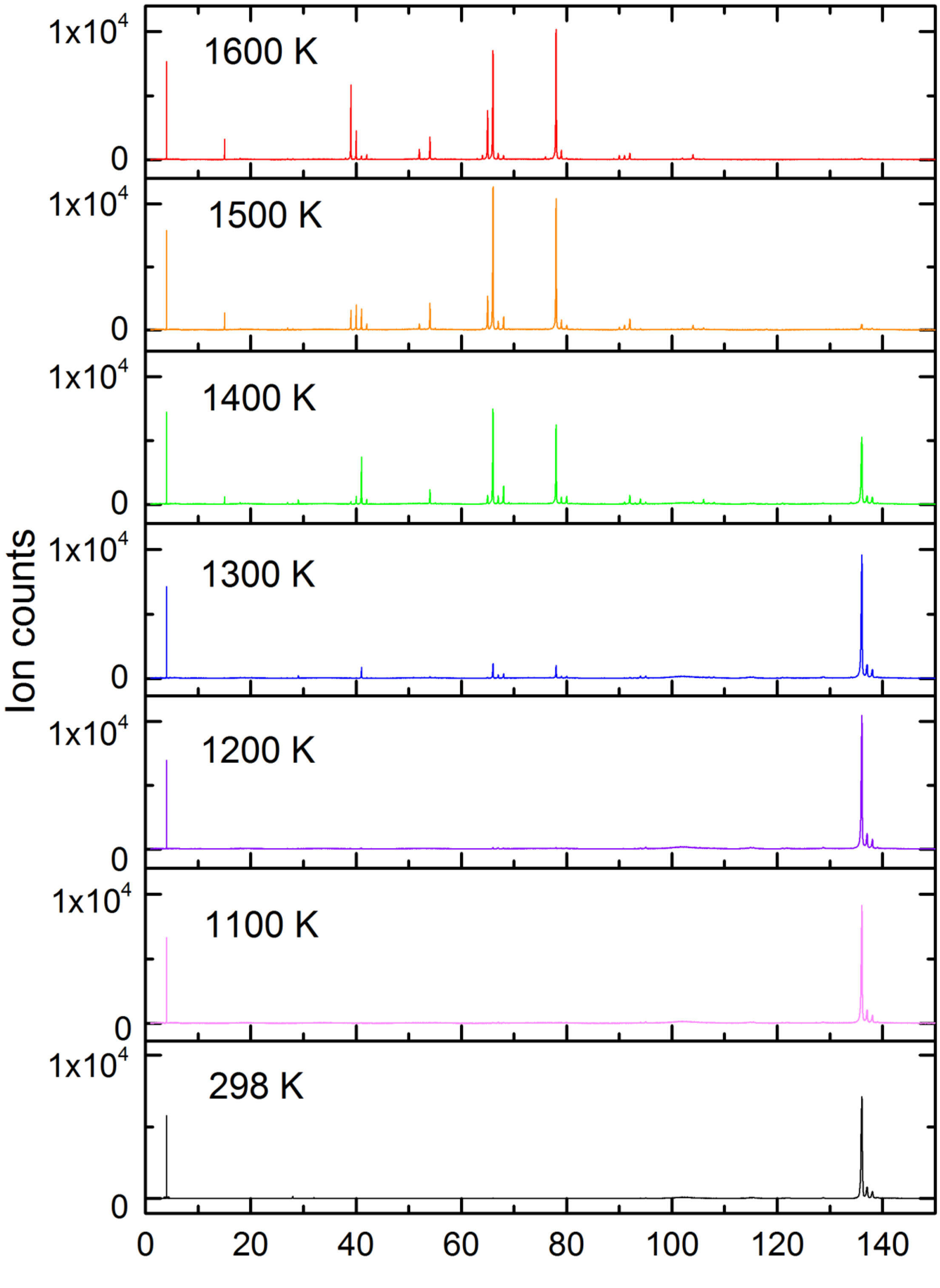

**Figure 2.** Mass spectra of the products obtained from the decomposition of JP-10 recorded at a photon energy of 10.0 eV at different temperatures from 298 K to 1,600 K.

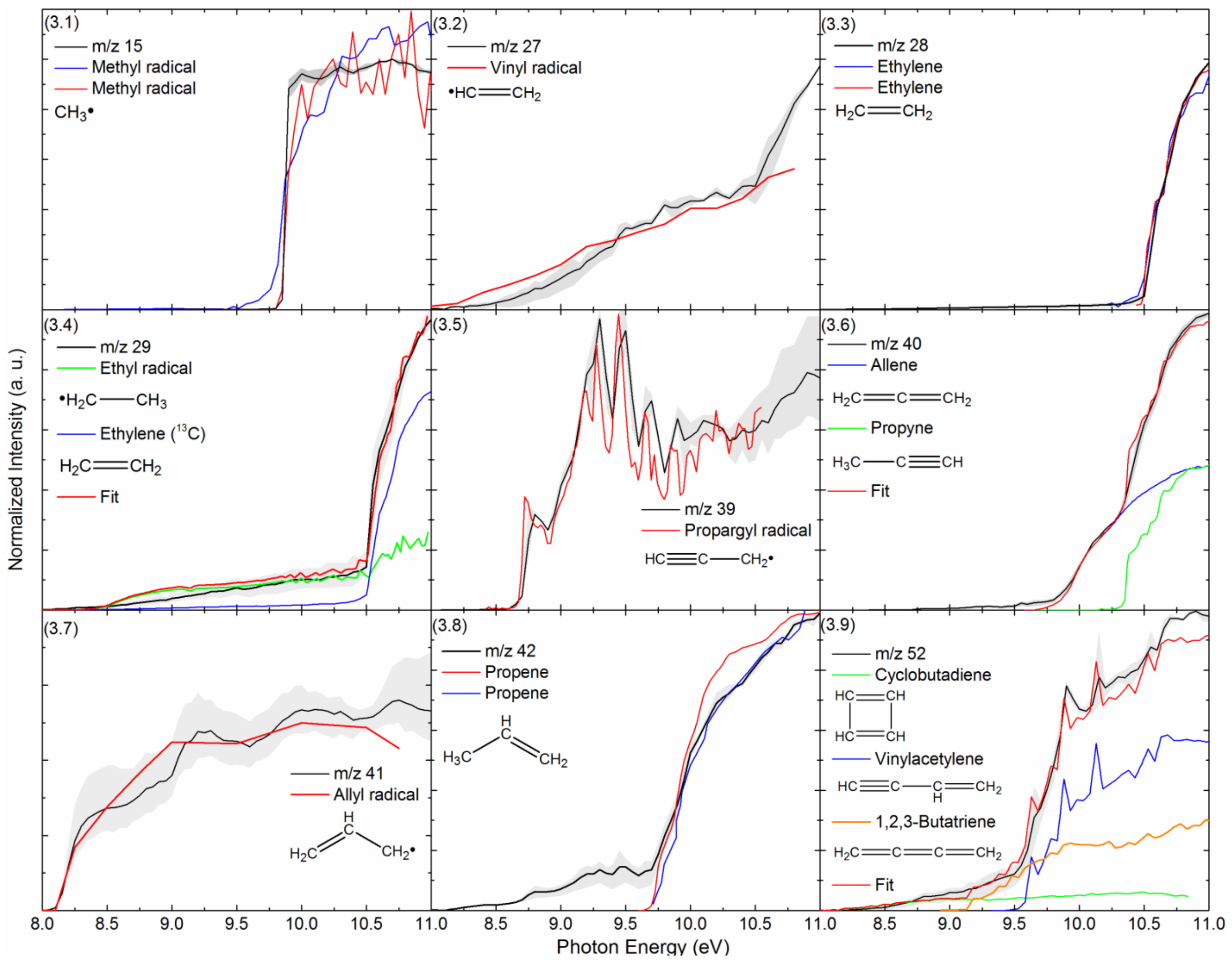

**Figure 3.** Experimental photoionization efficiency curves (PIE) recorded from the decomposition of JP-10 at 1,600 K at mass-to-charges of 15, 27, 28, 29, 39, 40, 41, 42, and 52 (black line) along with the experimental errors (grey area), and reference PIE curves (blue, orange, green, or red). In case of multiple contributions to one PIE curve, the red line resembles the overall fit.

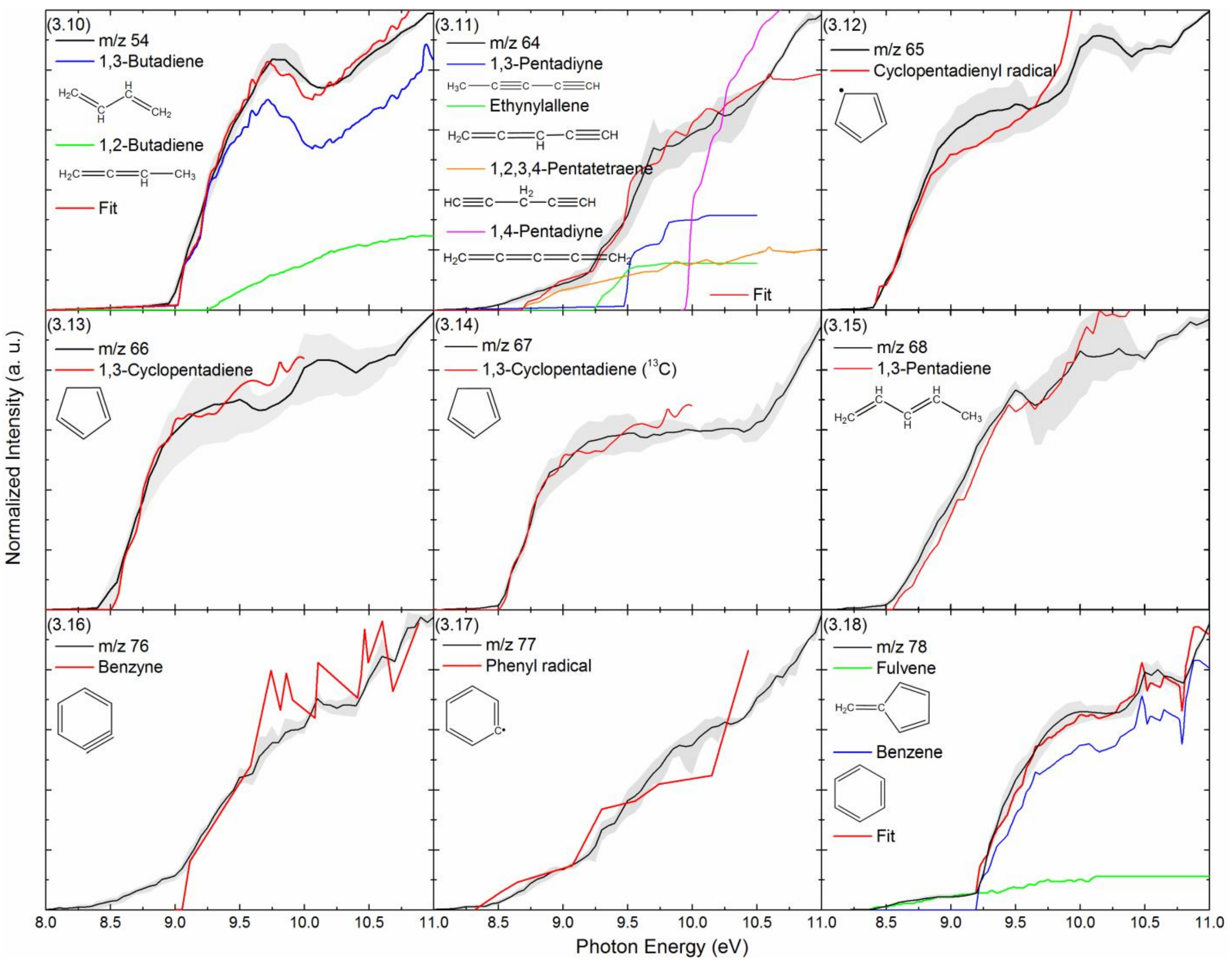

**Figure 3 (cont.).** Experimental photoionization efficiency curves (PIE) recorded from the decomposition of JP-10 at 1,600 K at mass-to-charges of 54, 64, 65, 66, 67, 68, 76, 77, and 78 (black line) along with the experimental errors (grey area), and the reference PIE curves (blue, orange, green, pink or red). In case of multiple contributions to one PIE curve, the red line resembles the overall fit.

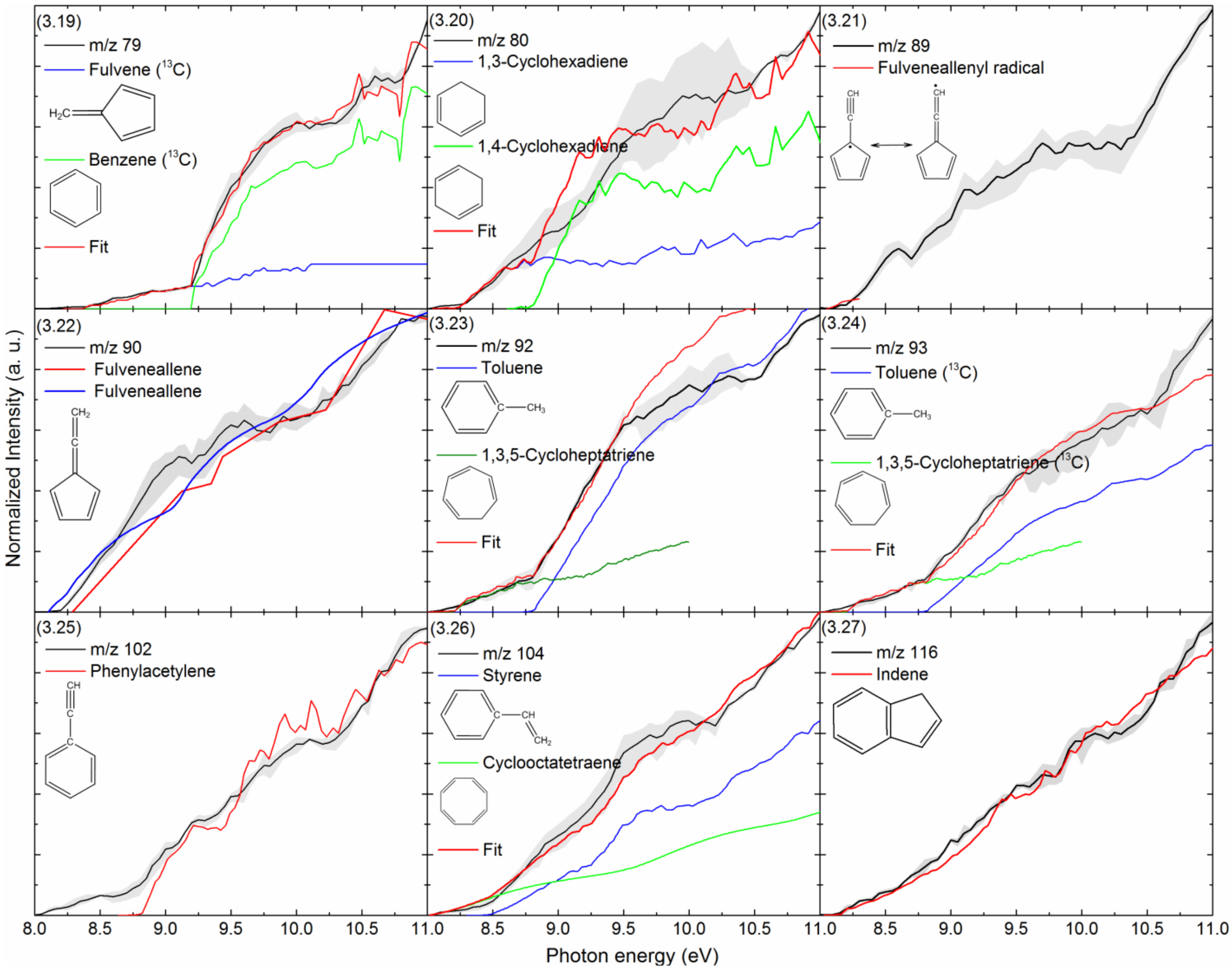

**Figure 3 (cont.).** Experimental photoionization efficiency curves (PIE) recorded from the decomposition of JP-10 at 1,600 K at mass-to-charges of 79, 80, 89, 90, 92, 93, 102, 104, and 116 (black line) along with the experimental errors (grey area), and the reference PIE curves (blue, green, or red). In case of multiple contributions to one PIE curve, the red line resembles the overall fit.

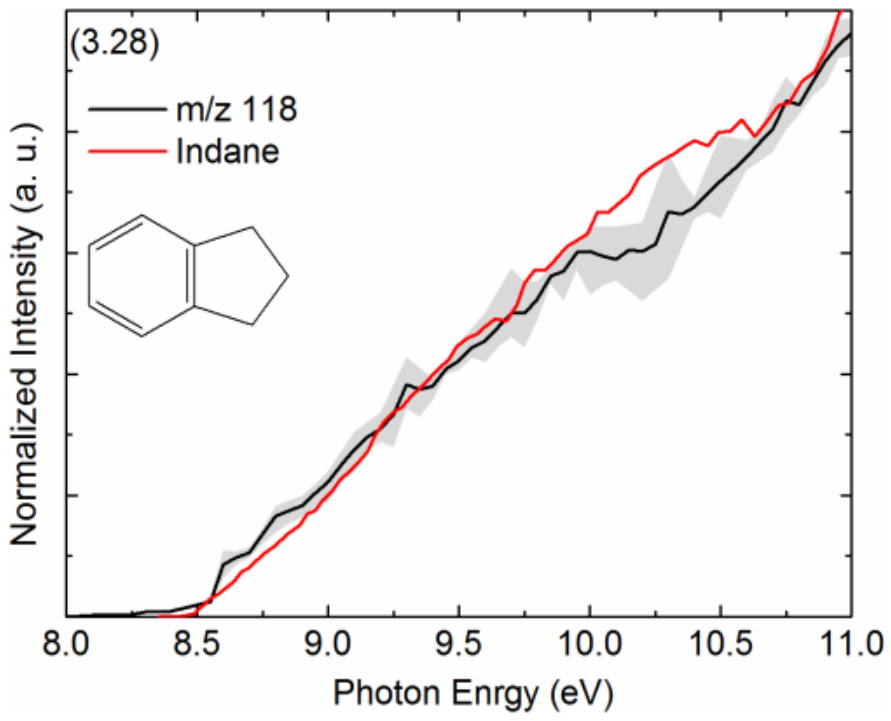

**Figure 3 (cont.).** Experimental photoionization efficiency curves (PIE) recorded from the decomposition of JP-10 at 1,600 K at mass-to-charges of 118 (black line) along with the experimental errors (grey area), and the reference PIE curve (red line).

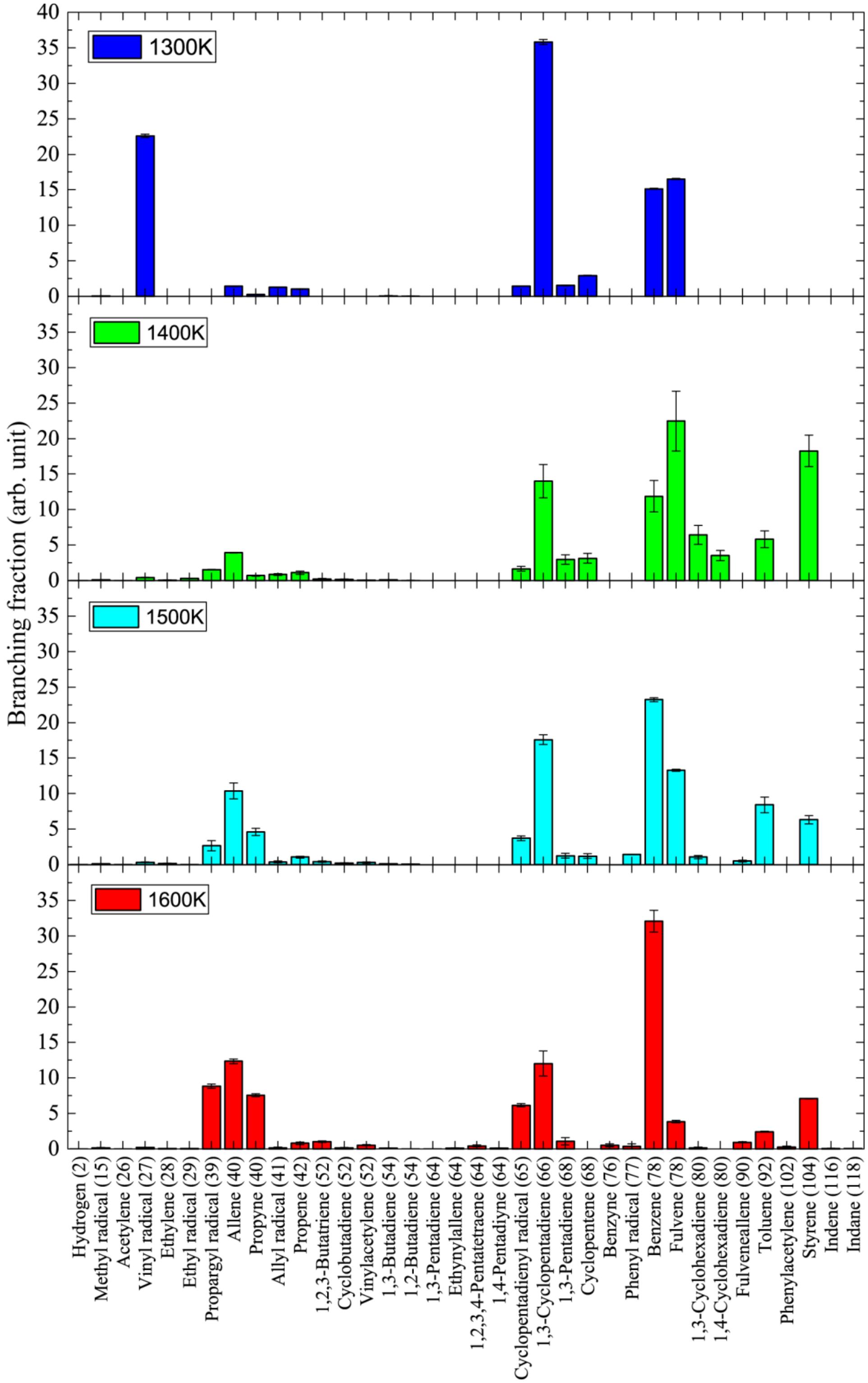

**Figure 4.** Overall branching ratios of the products obtained in the decomposition of JP-10 at temperatures of 1300, 1400, 1500, and 1600 K.

**class I**

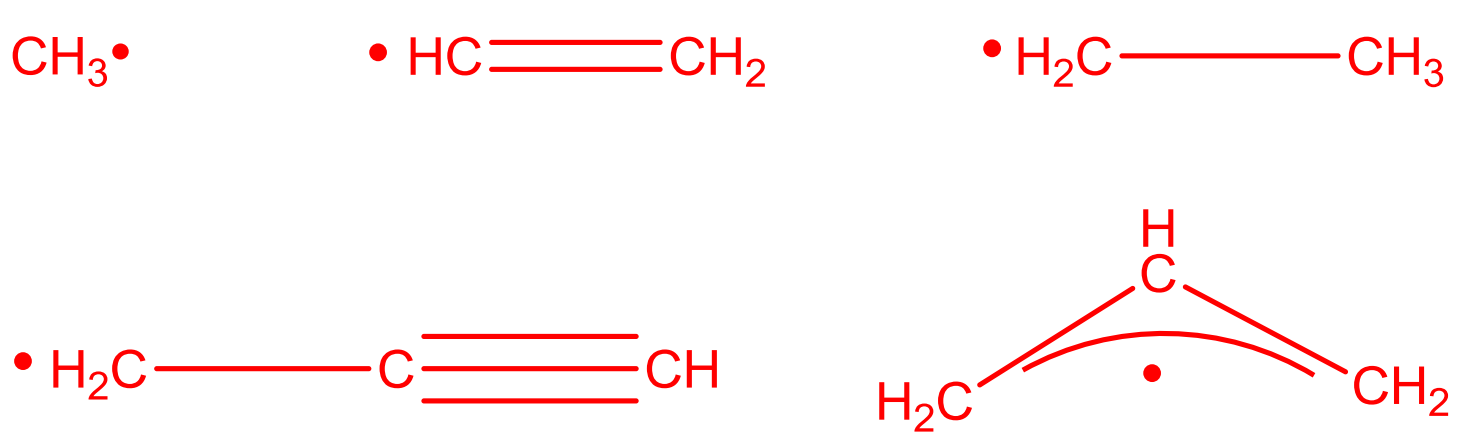

**class II**

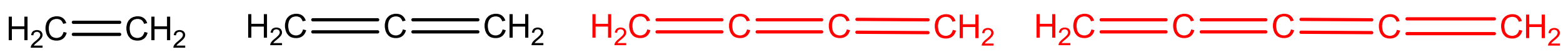

**class III**

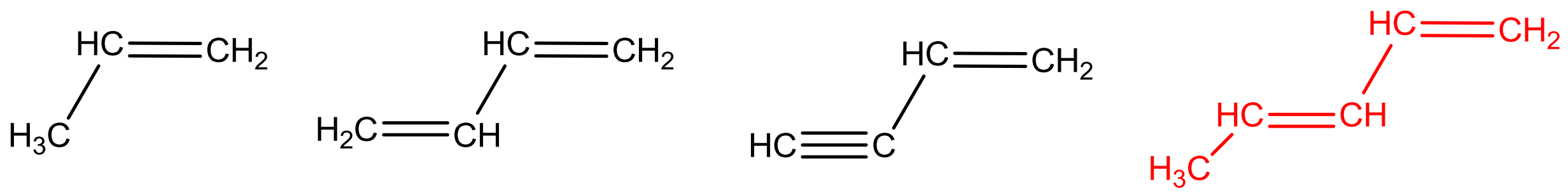

**class IV**

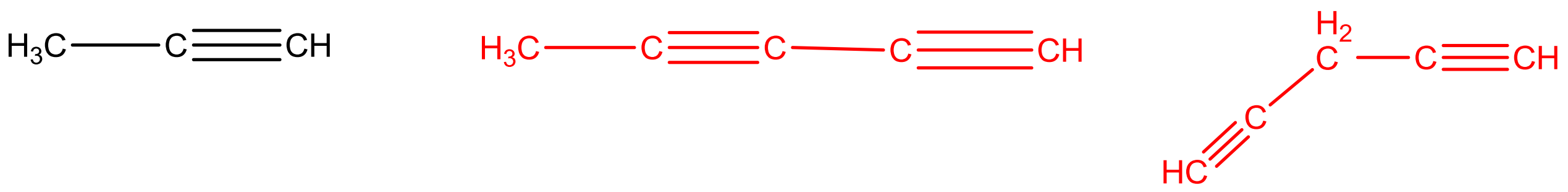

**class V**

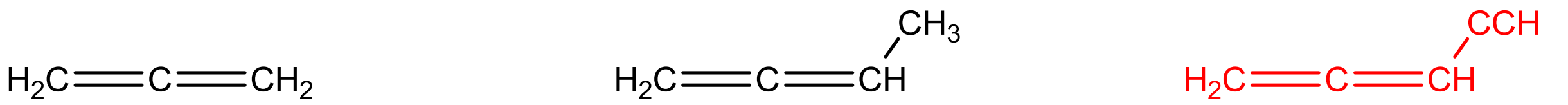

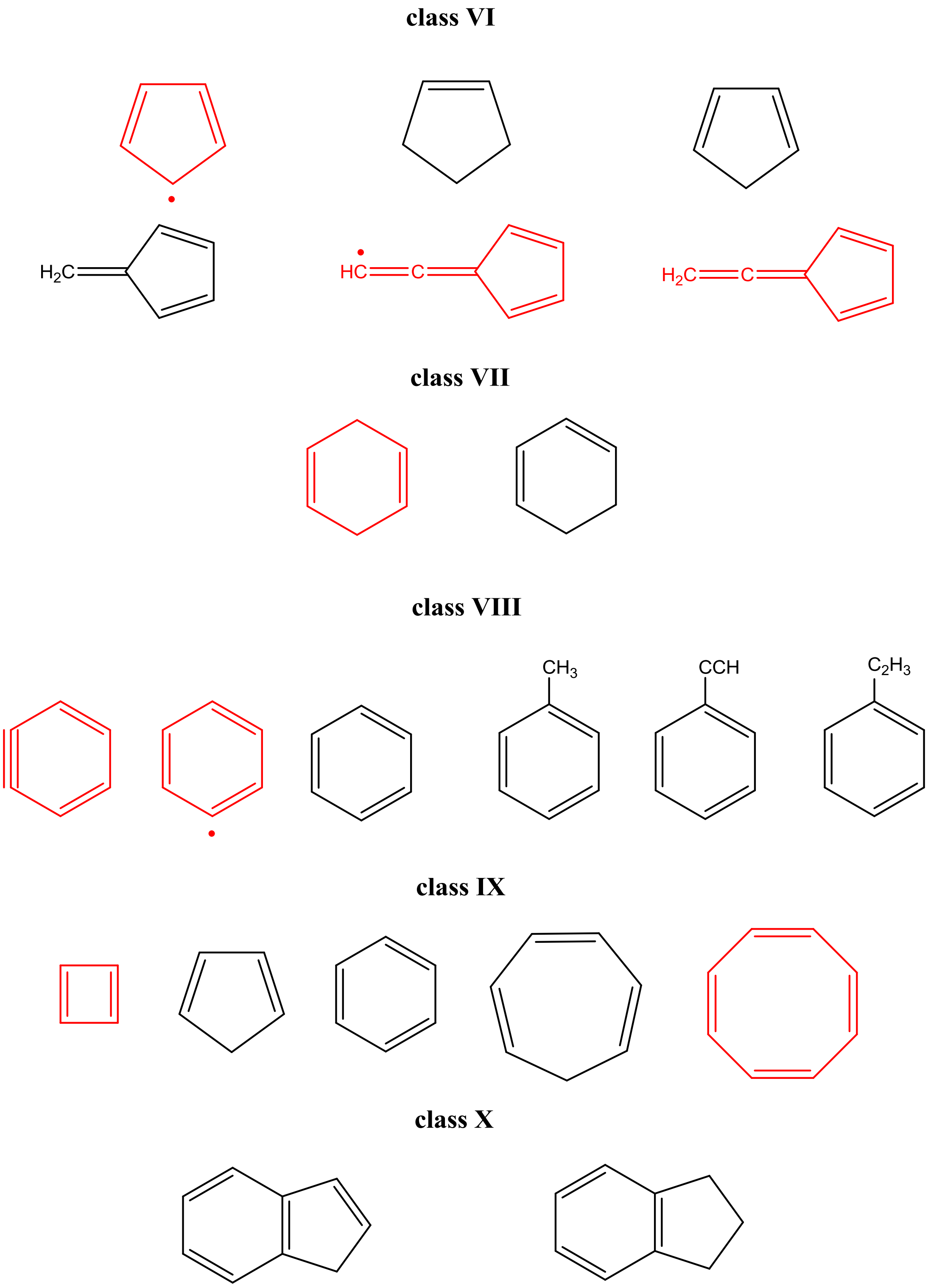

**Figure 5.** Ten product classes observed in the present studies on the decomposition of JP-10. Species colored in red were detected for the first time in the decomposition of JP-10.

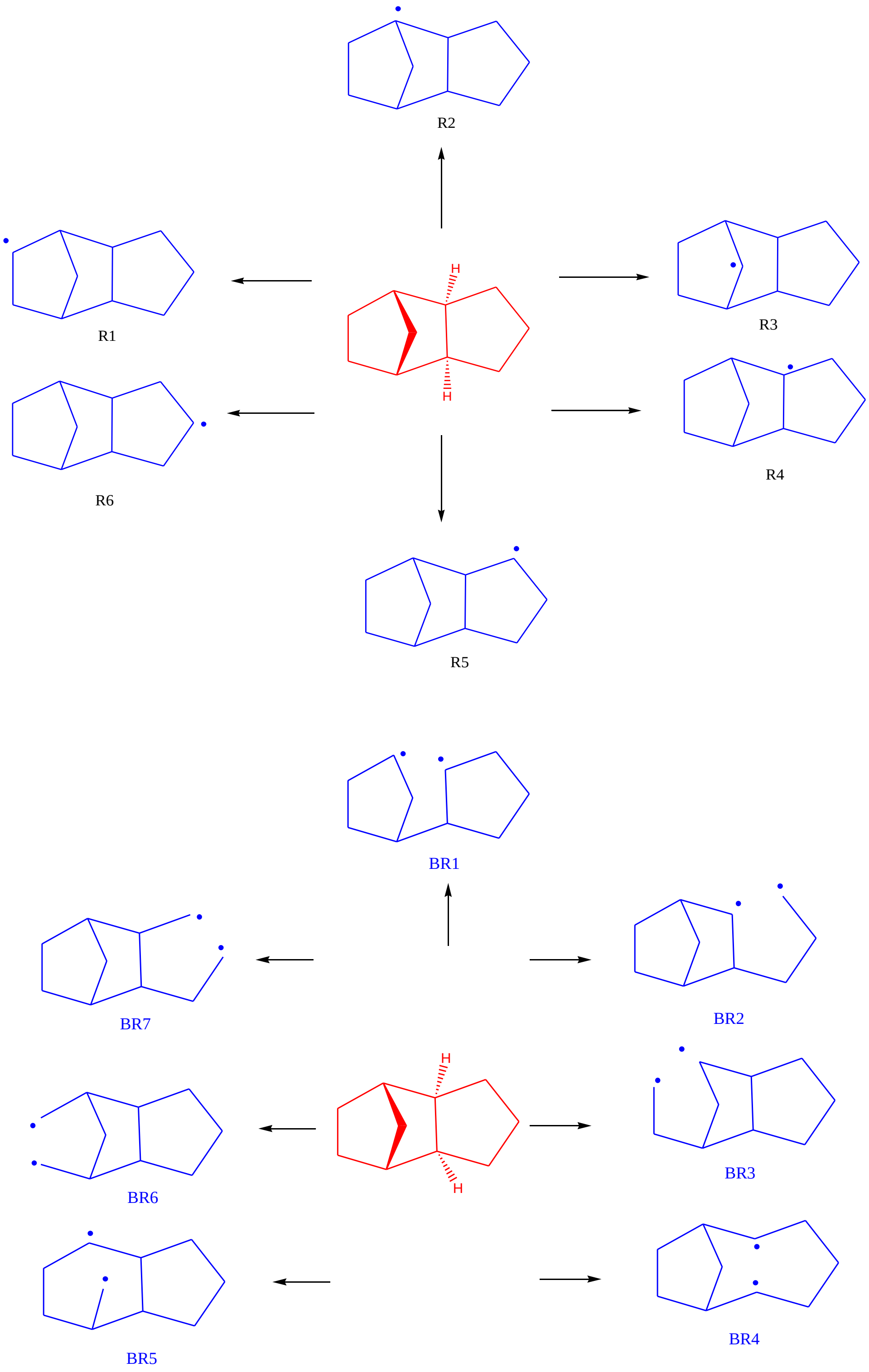

**Figure 6.** Top: Primary radicals R1-R6 (blue) formed via hydrogen abstraction from JP-10 (red). Bottom: Homolytic carbon-carbon bond in JP-10 (red) ruptures resulting in the formation of seven biradicals BR1 – BR7 (blue).

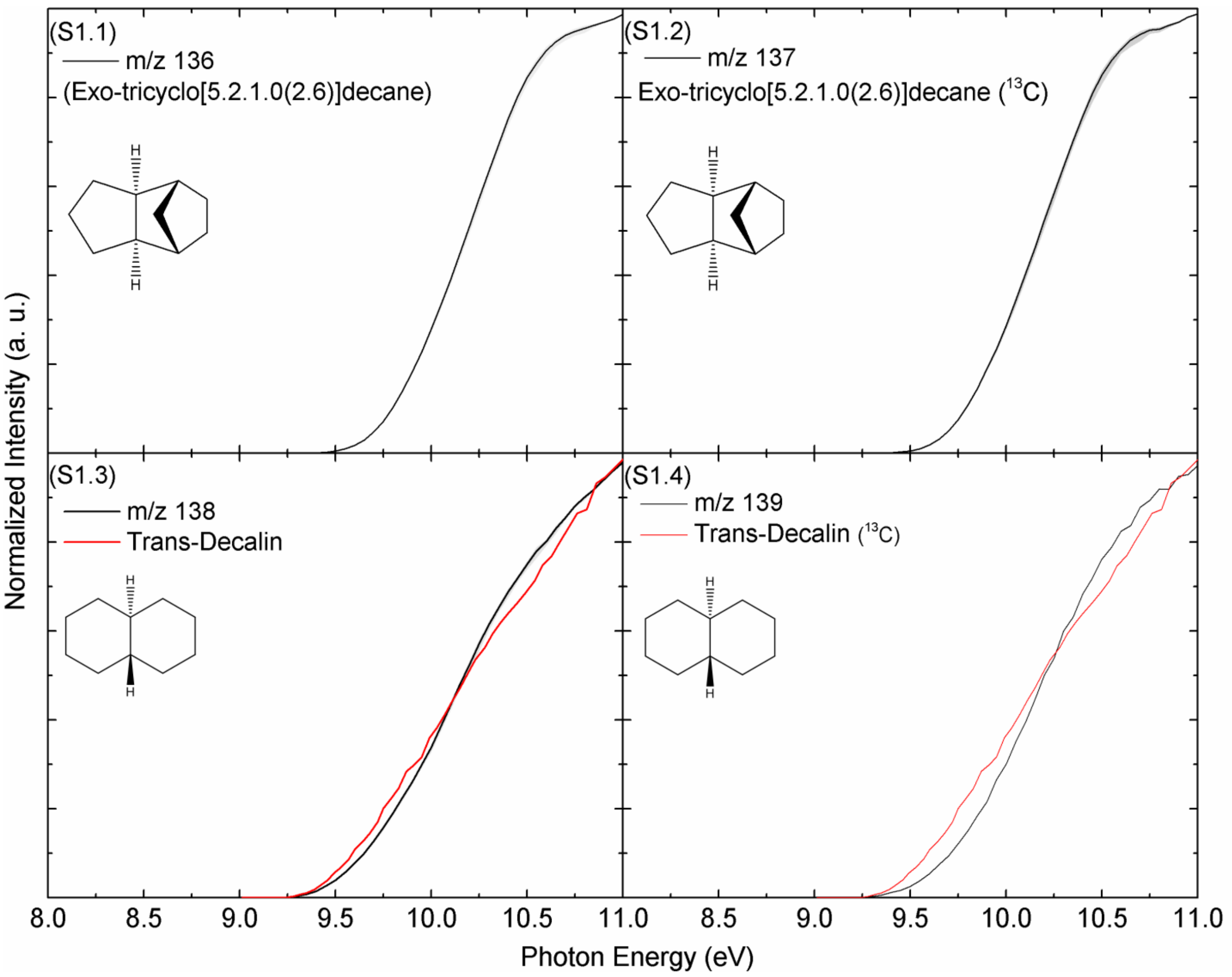

Figure S1. Experimental photoionization efficiency curves (PIE) recorded from the pyrolysis of JP-10 at 298 K at mass-to-charges of (m/z) 136, 137, 138 and 139 (black line) along with reference curves (red line).

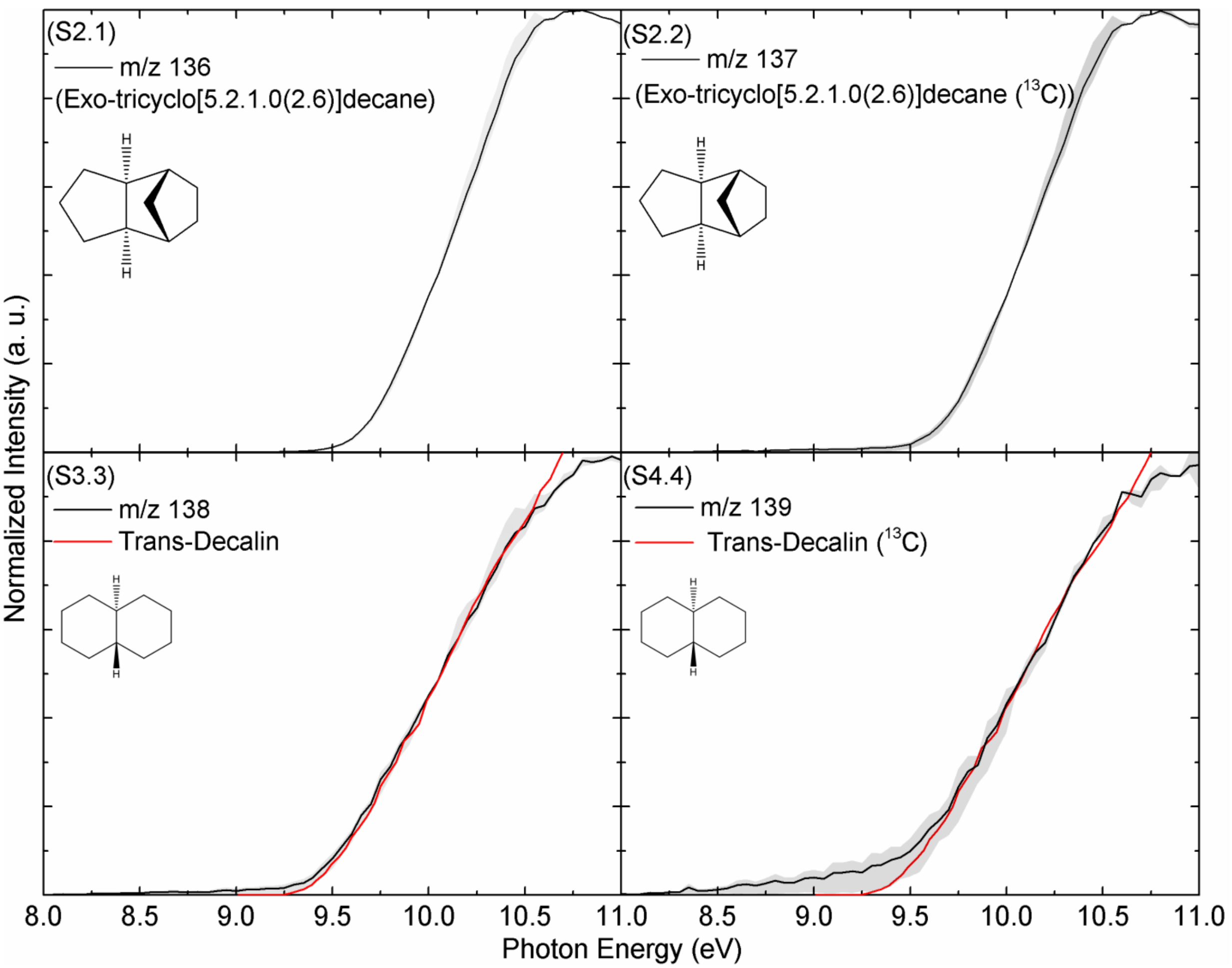

Figure S2. Experimental photoionization efficiency curves (PIE) recorded from the pyrolysis of JP-10 at 1,100 K at mass-to-charges of (m/z) 136, 137, 138 and 139 (black line) along with the experimental errors (grey area) and the reference curves (red line).

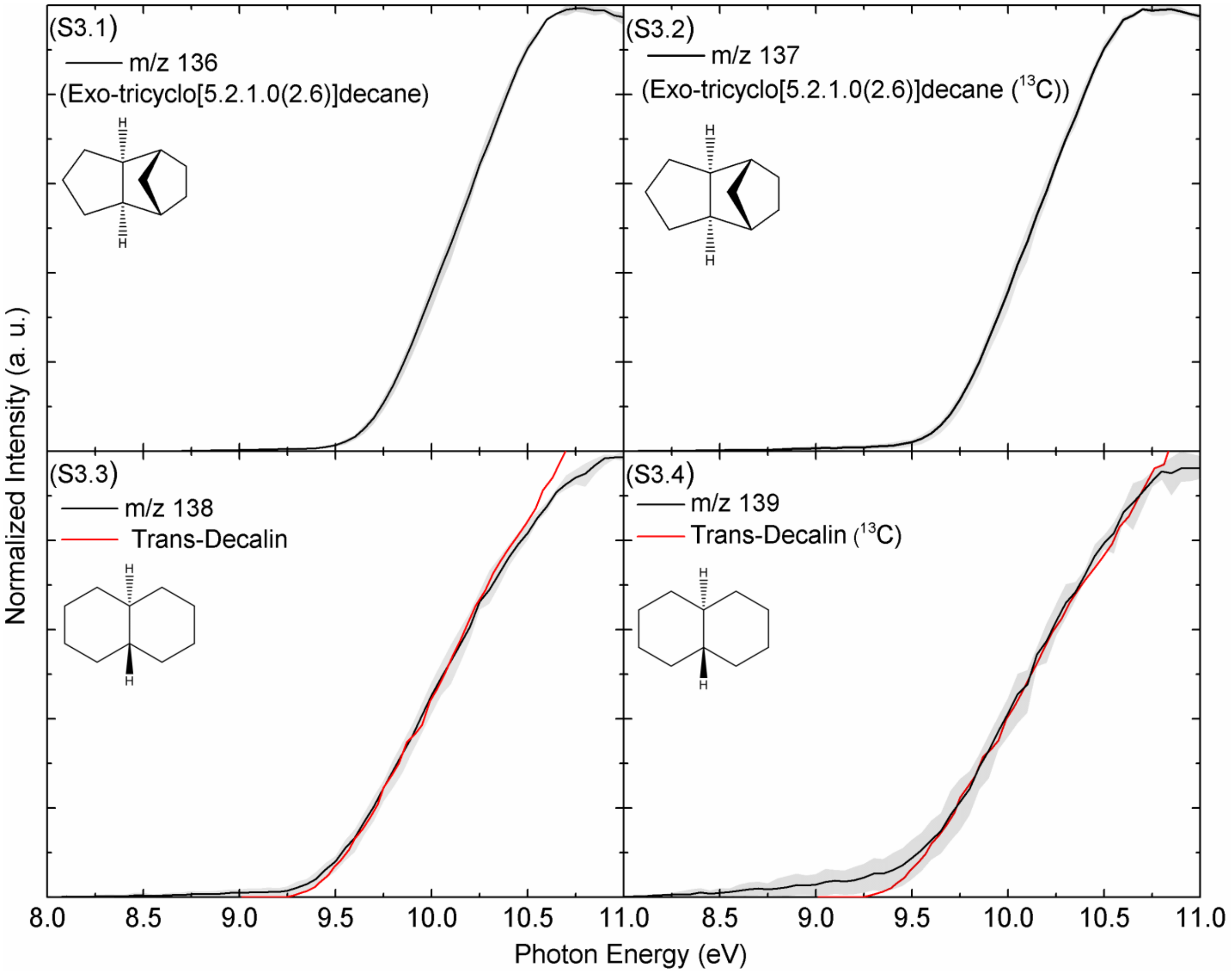

Figure S3. Experimental photoionization efficiency curves (PIE) recorded from the pyrolysis of JP-10 at 1,200 K at mass-to-charges of (m/z) 136, 137, 138 and 139 (black line) along with the experimental errors (grey area) and the reference curves (red line).

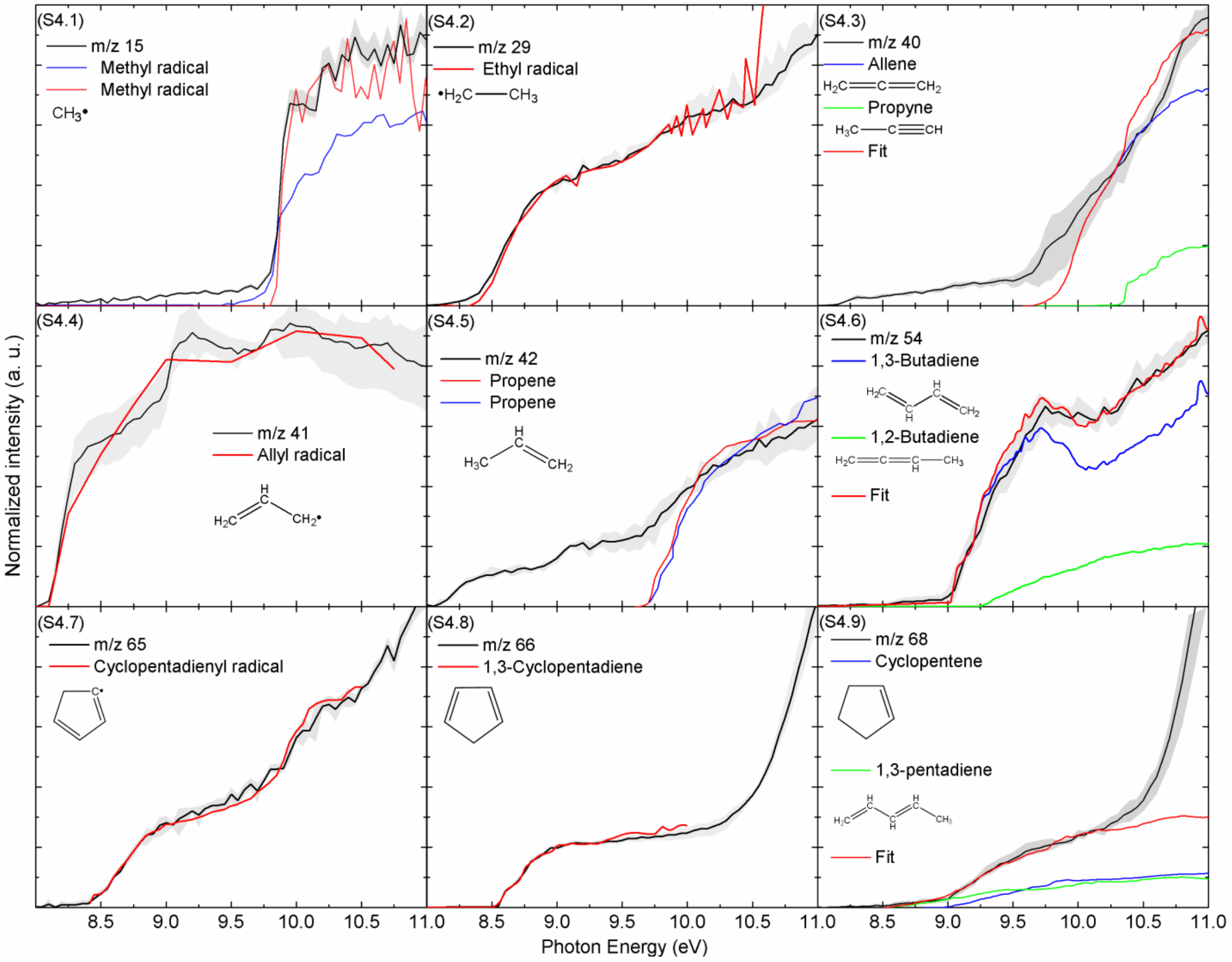

Figure S4. Experimental photoionization efficiency curves (PIE) recorded from the pyrolysis of JP-10 at 1,300 K at mass-to-charges of (m/z) 15, 29, 40, 41, 42, 54, 65, 66 and 68 (black line) along with the experimental errors (grey area), and reference PIE curves (blue, orange, green, or red). In case of multiple contributions to one PIE curve, the red line resembles the overall fit.

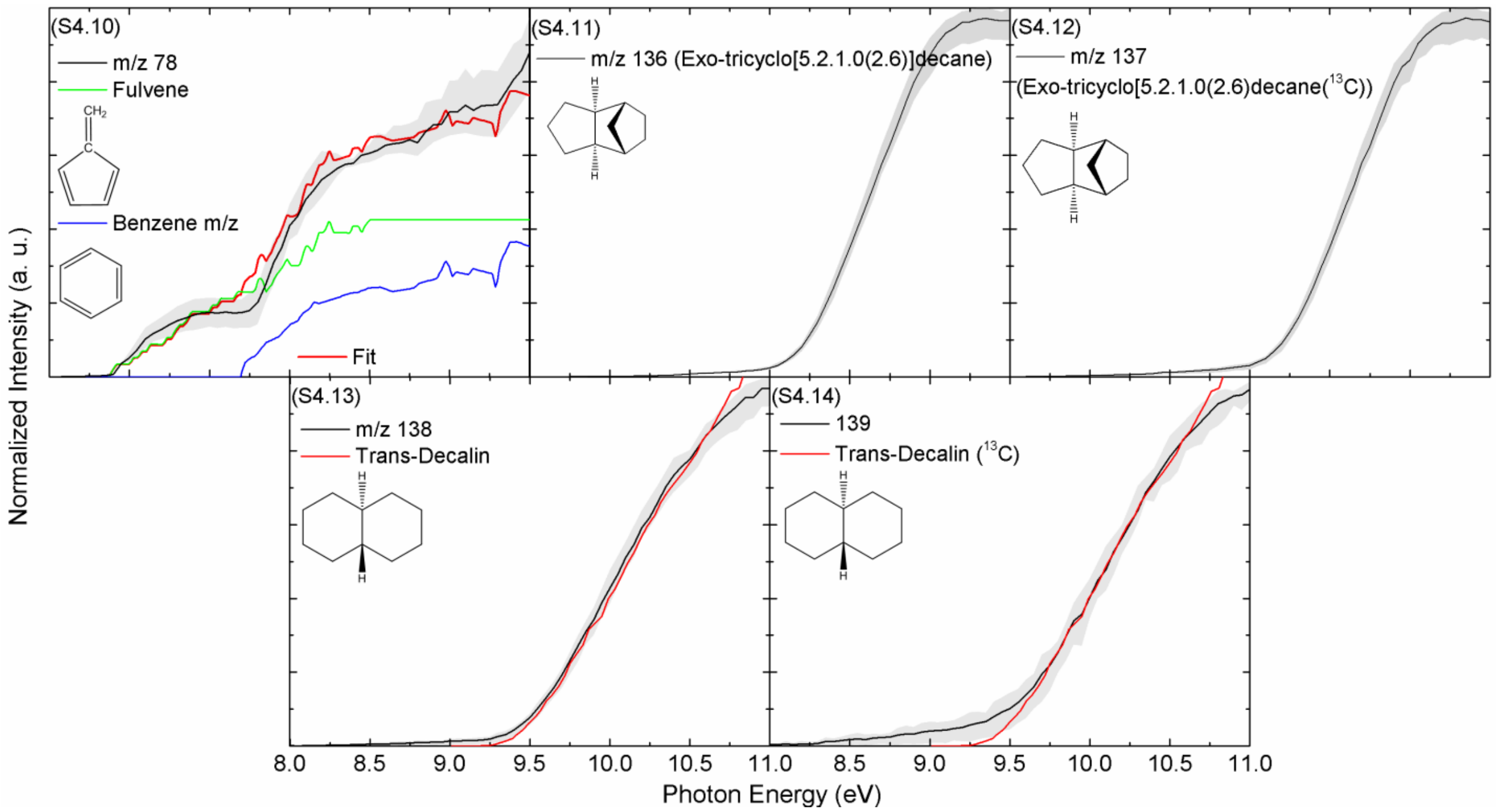

Figure S4 (cont.). Experimental photoionization efficiency curves (PIE) recorded from the pyrolysis of JP-10 at 1,300 K at mass-to-charges of (m/z) 78, 136, 137, 138, and 139 (black line) along with the experimental errors (grey area), and reference PIE curves (blue, green, or red). In case of multiple contributions to one PIE curve, the red line resembles the overall fit.

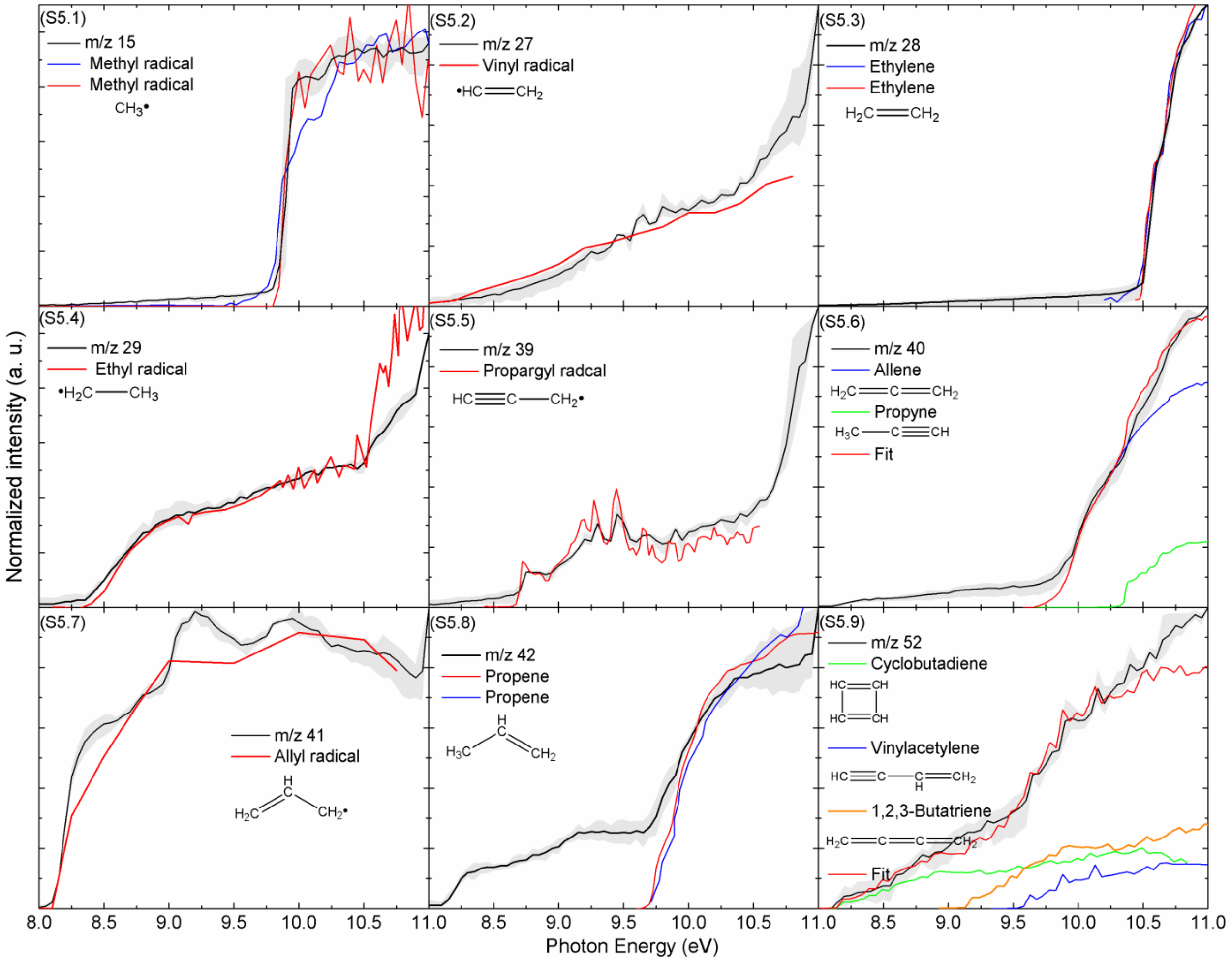

Figure S5. Experimental photoionization efficiency curves (PIE) recorded from the pyrolysis of JP-10 at 1,400 K at mass-to-charges of (m/z) 15, 27, 28, 29, 39, 40, 41, 42, and 52 (black line) along with the experimental errors (grey area), and reference PIE curves (blue, orange, green, or red). In case of multiple contributions to one PIE curve, the red line resembles the overall fit.

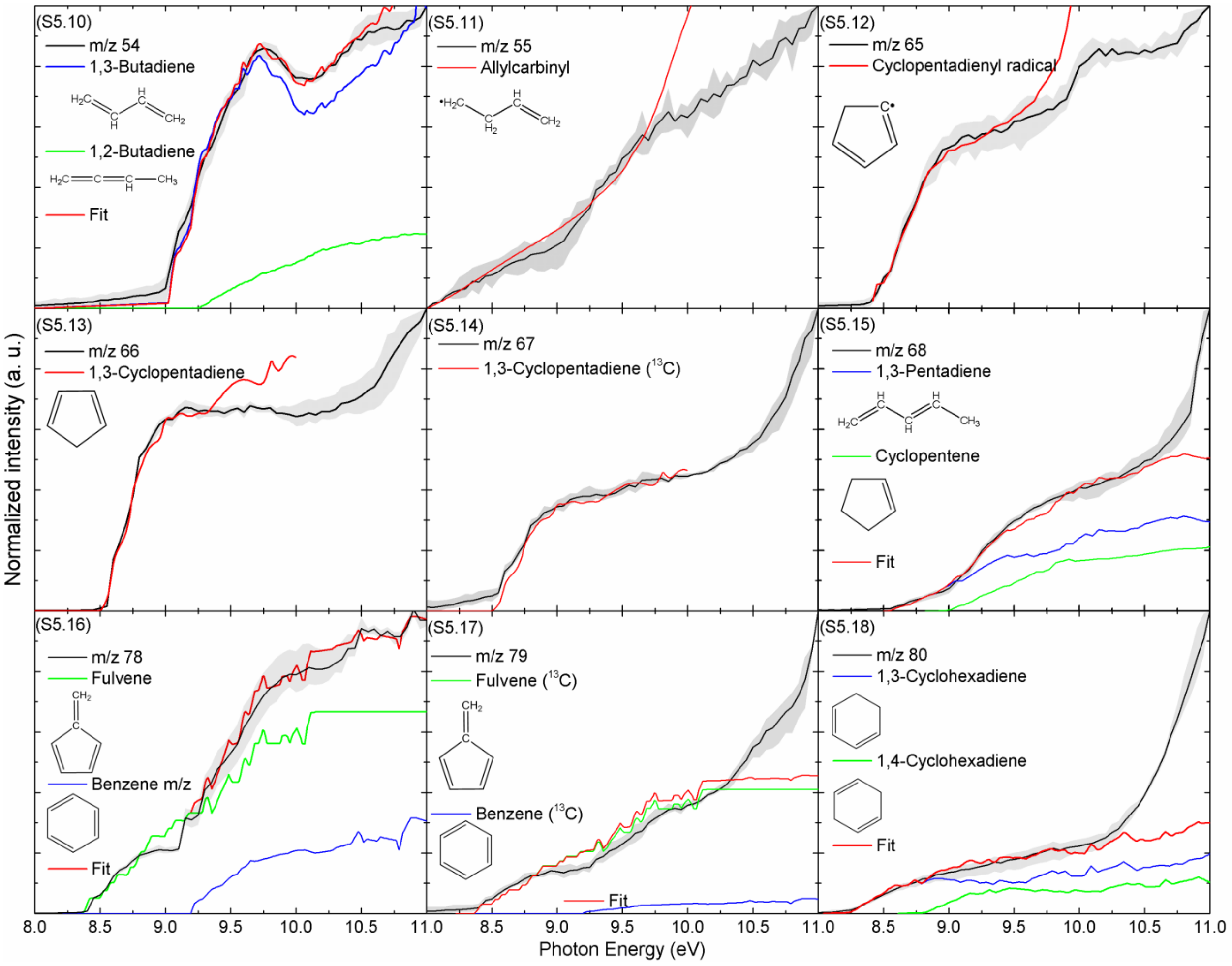

Figure S5 (cont.). Experimental photoionization efficiency curves (PIE) recorded from the pyrolysis of JP-10 at 1,400 K at mass-to-charges of (m/z) 54, 55, 65, 66, 67, 68, 78, 79 and 80 (black line) along with the experimental errors (grey area), and reference PIE curves (blue, green, or red). In case of multiple contributions to one PIE curve, the red line resembles the overall fit.

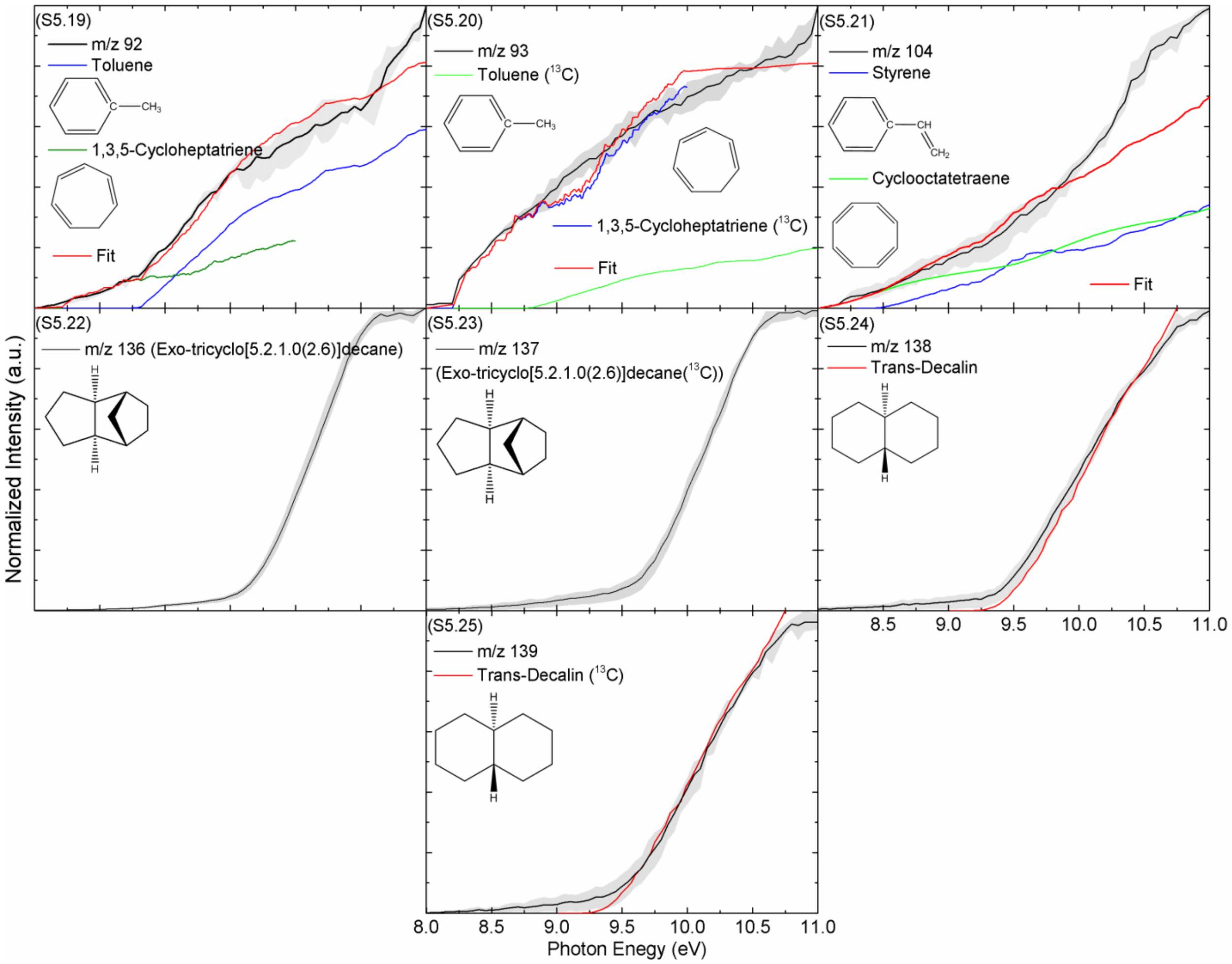

Figure S5 (cont.). Experimental photoionization efficiency curves (PIE) recorded from the pyrolysis of JP-10 at 1,400 K at mass-to-charges of (m/z) 92, 93, 104, 136, 137, 138 and 139 (black line) along with the experimental errors (grey area), and reference PIE curves (blue, green, or red). In case of multiple contributions to one PIE curve, the red line resembles the overall fit.

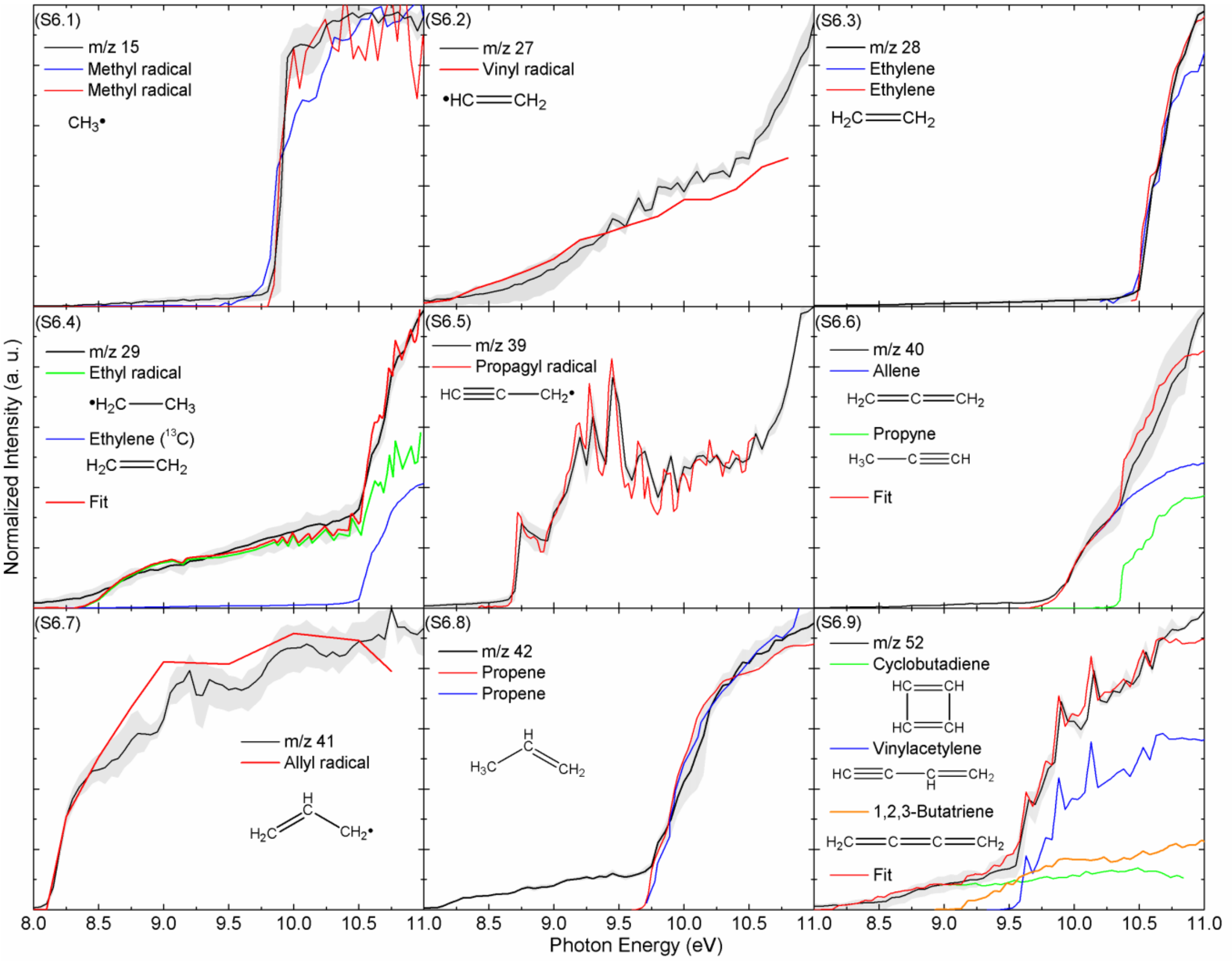

Figure S6. Experimental photoionization efficiency curves (PIE) recorded from the pyrolysis of JP-10 at 1,500 K at mass-to-charges of (m/z) 15, 27, 28, 29, 39, 40, 41, 42, and 52 (black line) along with the experimental errors (grey area), and reference PIE curves (blue, orange, green, or red). In case of multiple contributions to one PIE curve, the red line resembles the overall fit.

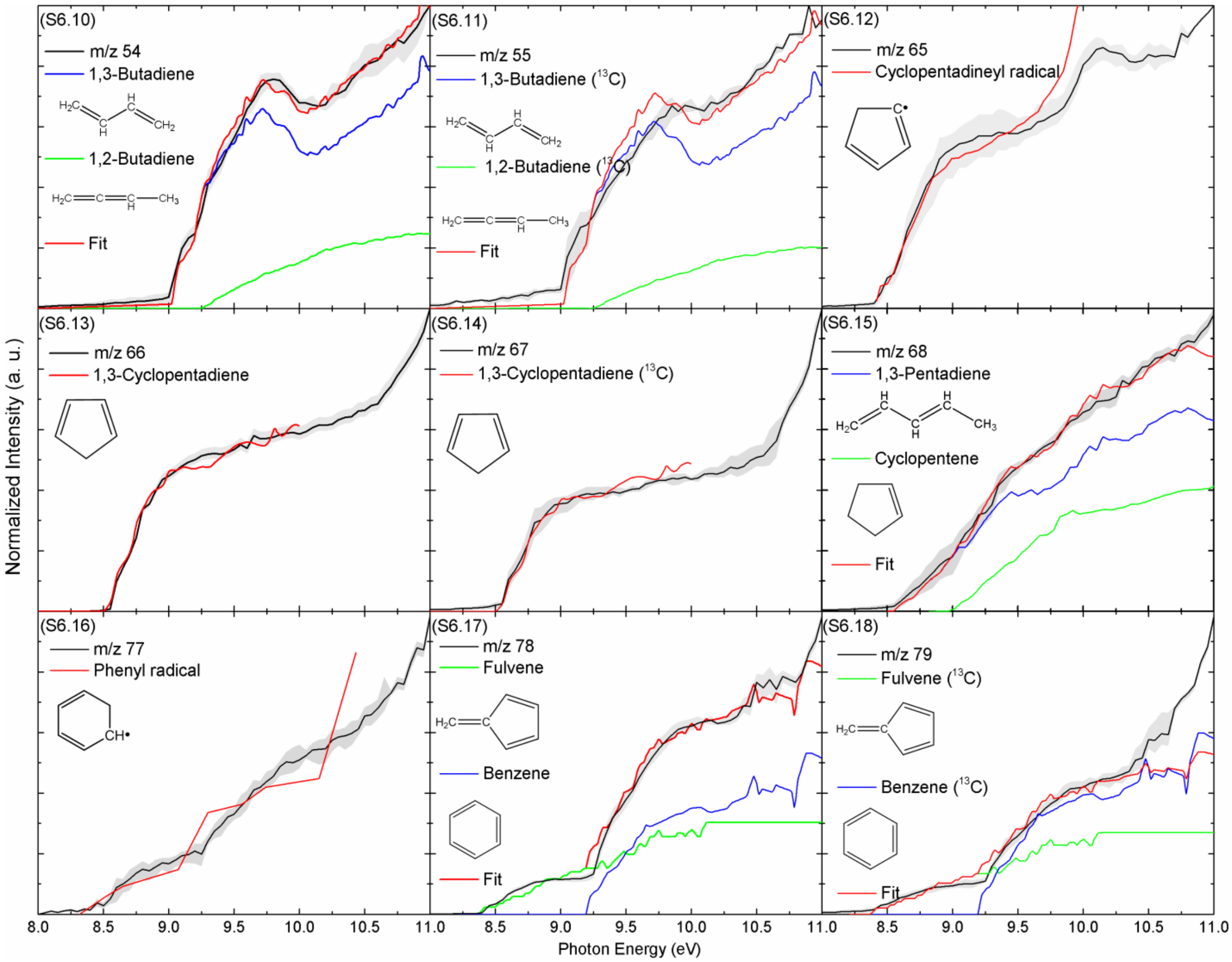

Figure S6 (cont.). Experimental photoionization efficiency curves (PIE) recorded from the pyrolysis of JP-10 at 1,500 K at mass-to-charges of (m/z) 54, 55, 65, 66, 67, 68, 77, 78 and 79 (black line) along with the experimental errors (grey area), and reference PIE curves (blue, green, or red). In case of multiple contributions to one PIE curve, the red line resembles the overall fit.

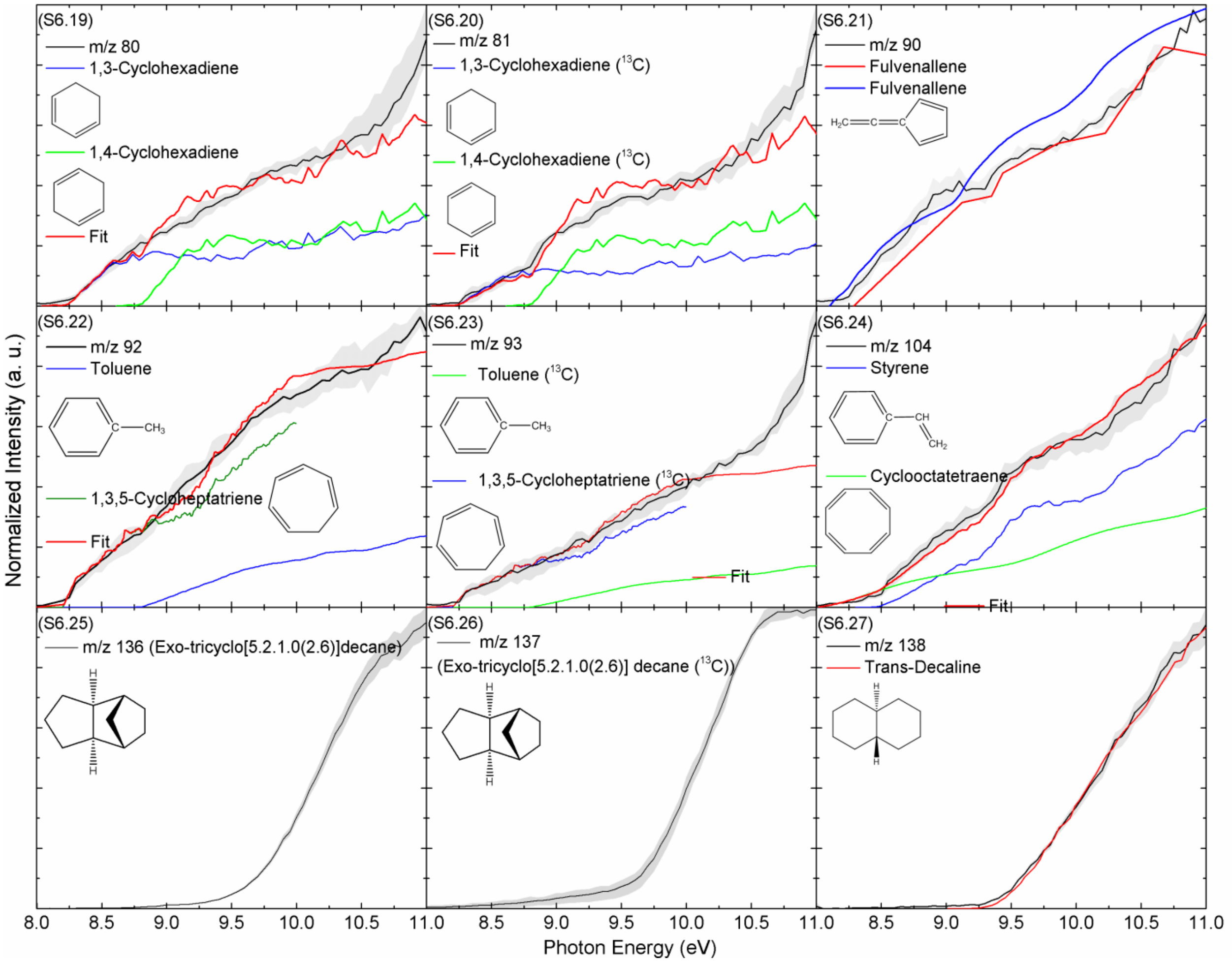

Figure S6 (cont.). Experimental photoionization efficiency curves (PIE) recorded from the pyrolysis of JP-10 at 1,500 K at mass-to-charges of (m/z) 80, 81, 90, 92, 93, 104, 136, 137, and 138 (black line) along with the experimental errors (grey area), and reference PIE curves (blue, green, or red). In case of multiple contributions to one PIE curve, the red line resembles the overall fit.

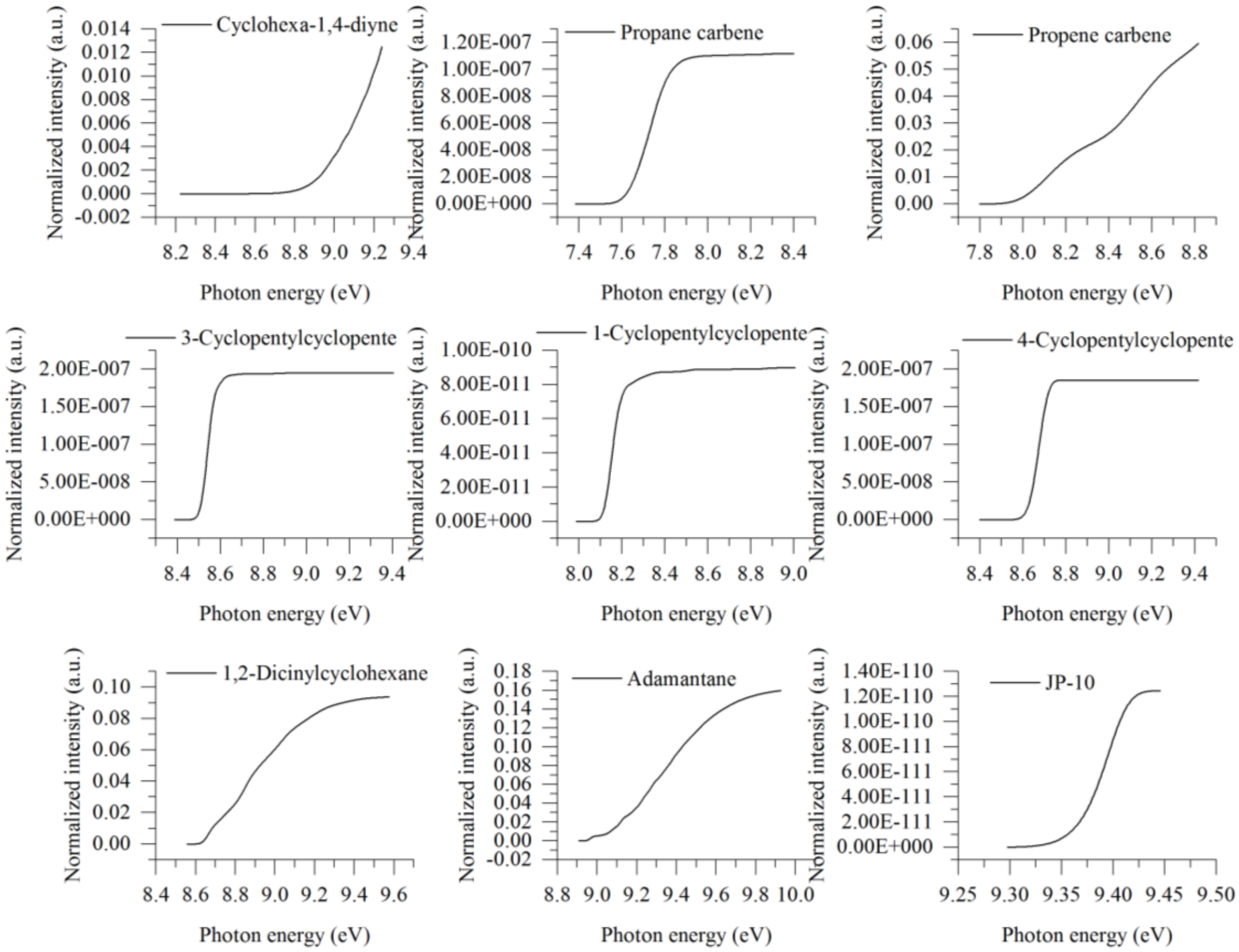

Figure S7. Computed photoionization efficiency curves at the CCSD (T)-F12/cc-pVTZ-f12//B3LYP/ 6-311G** level of theory.

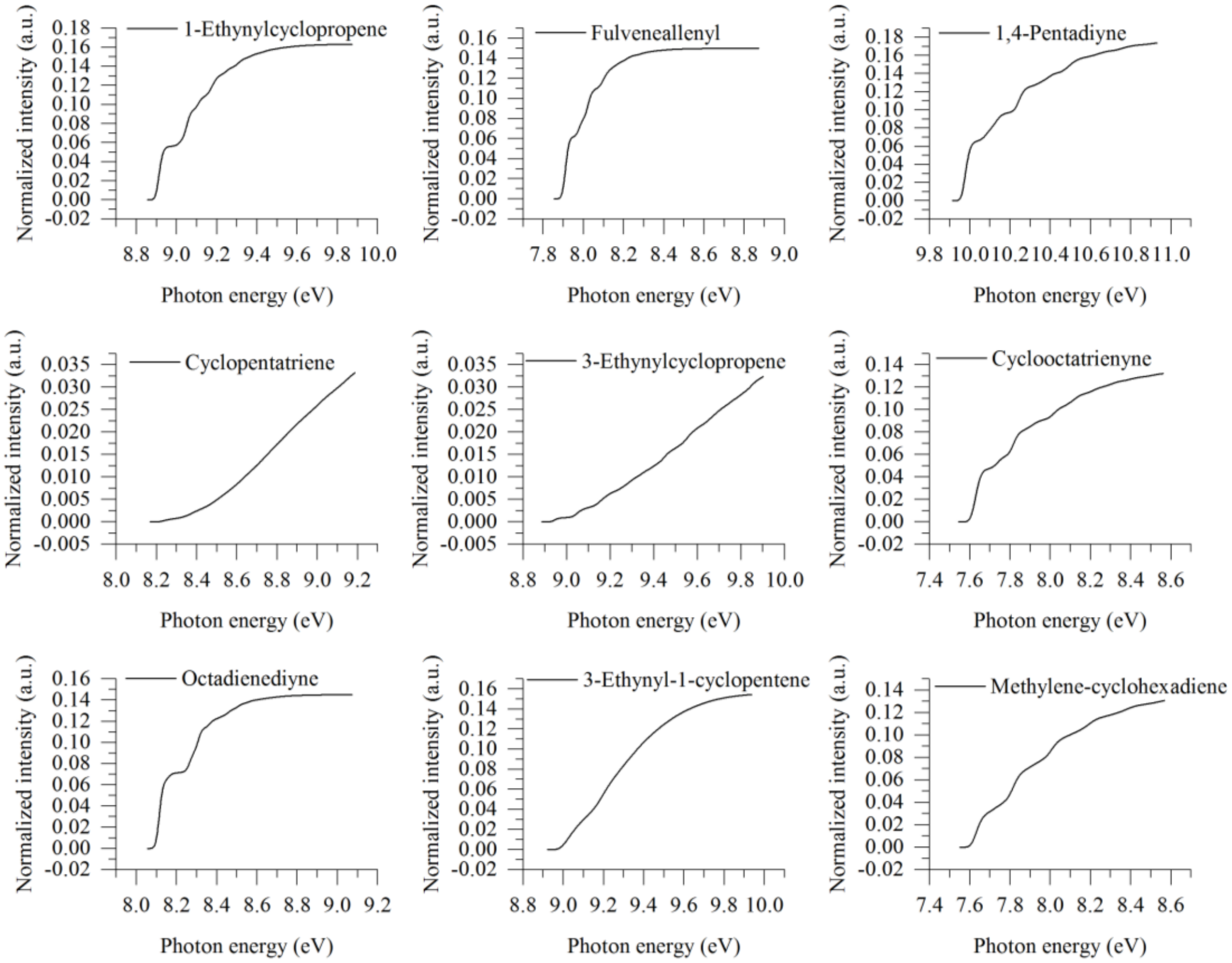

Figure S7 (cont.). Computed photoionization efficiency curves at the CCSD (T)-F12/cc-pVTZ-f12//B3LYP/ 6-311G** level of theory.

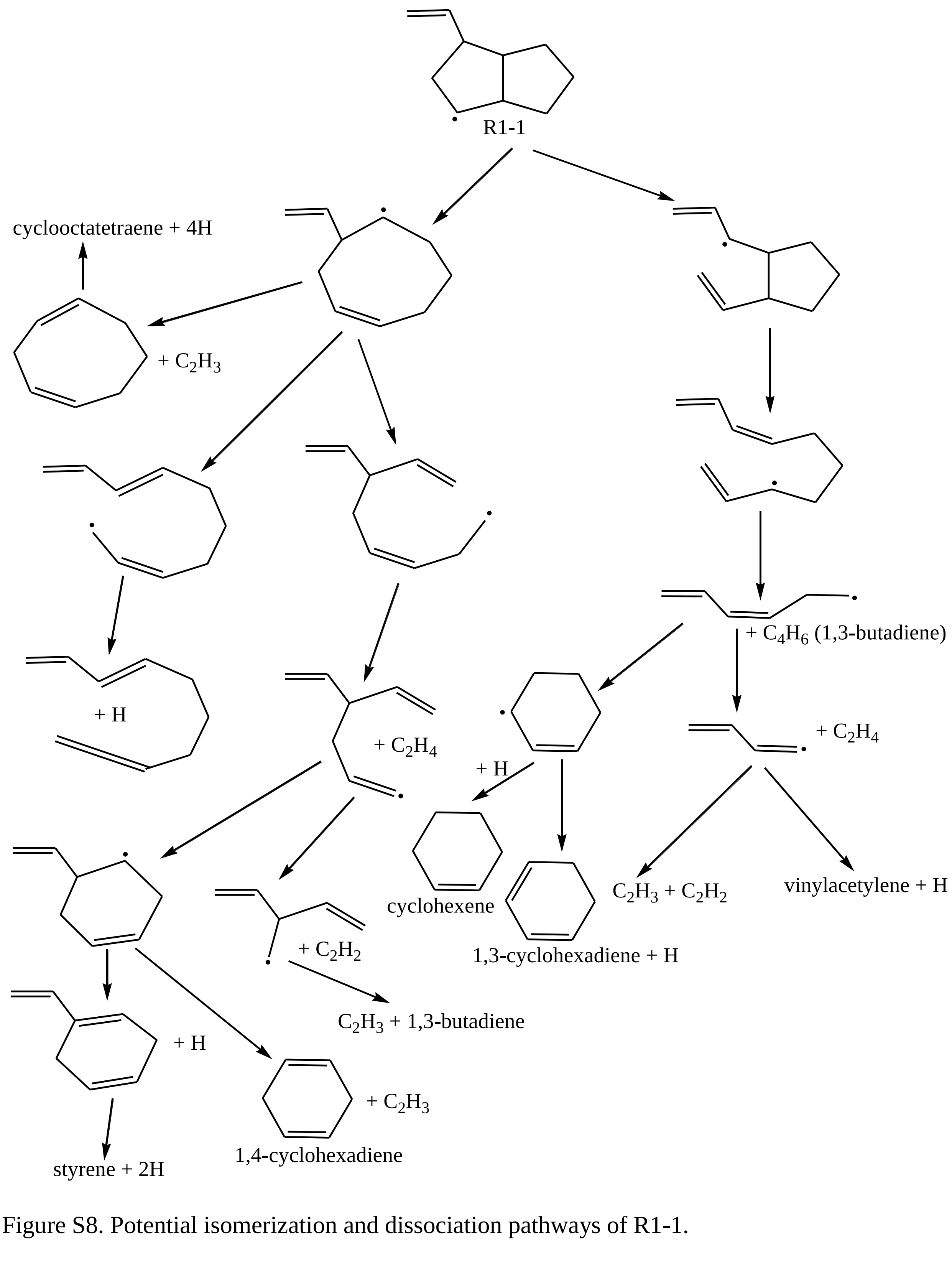

Figure S8. Potential isomerization and dissociation pathways of R1-1.

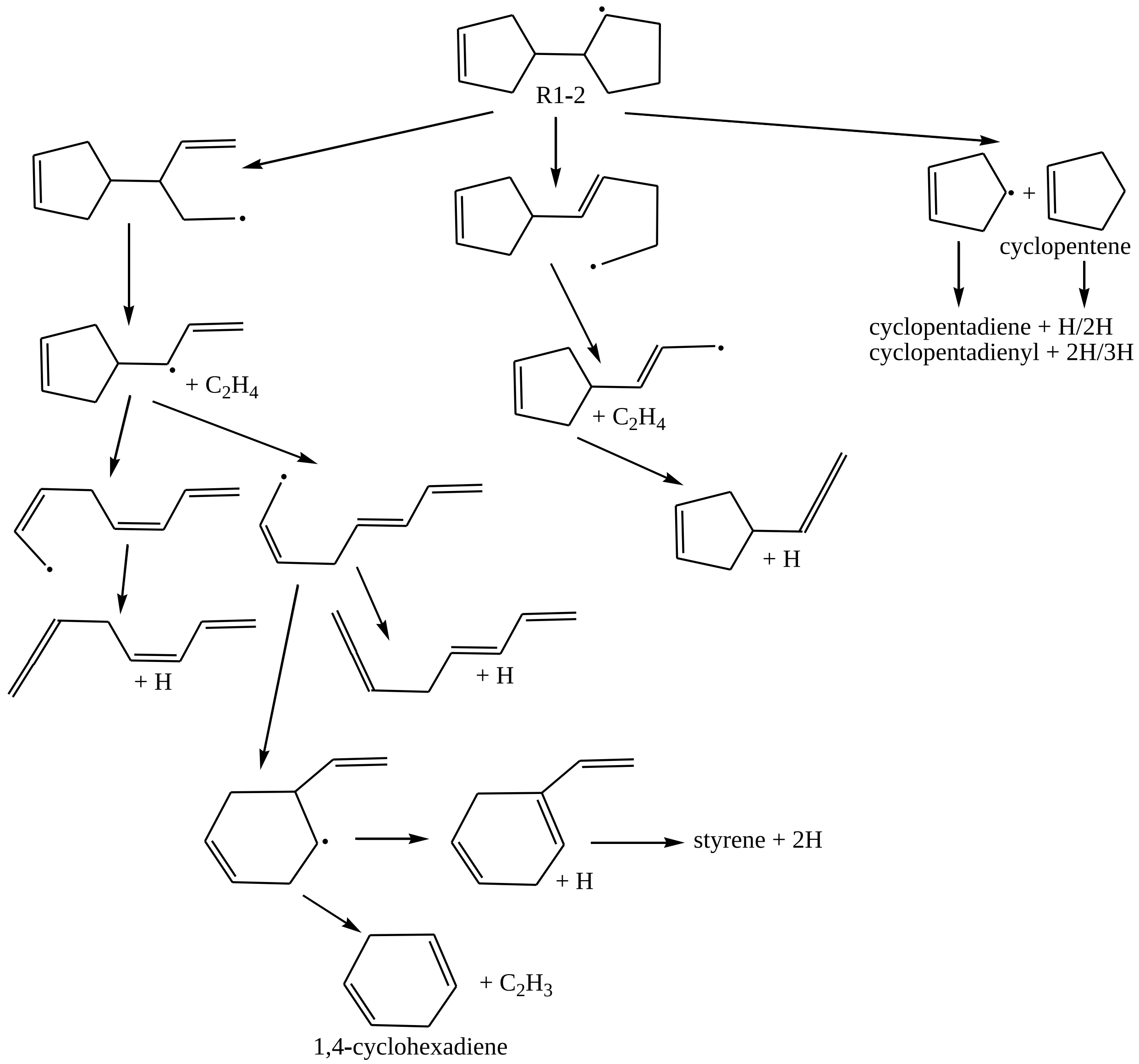

Figure S9. Potential isomerization and dissociation pathways of R1-2.

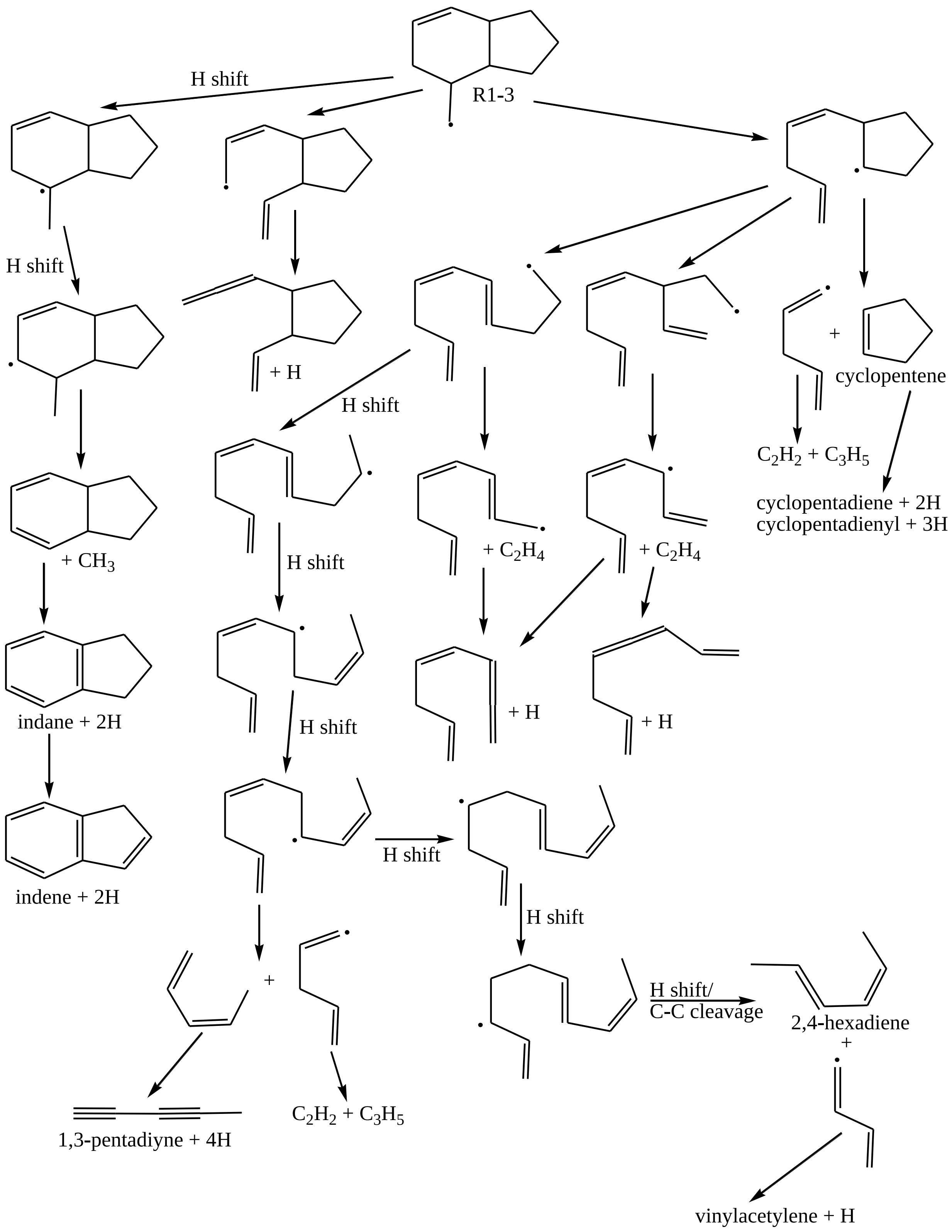

Figure S10. Potential isomerization and dissociation pathways of R1-3.

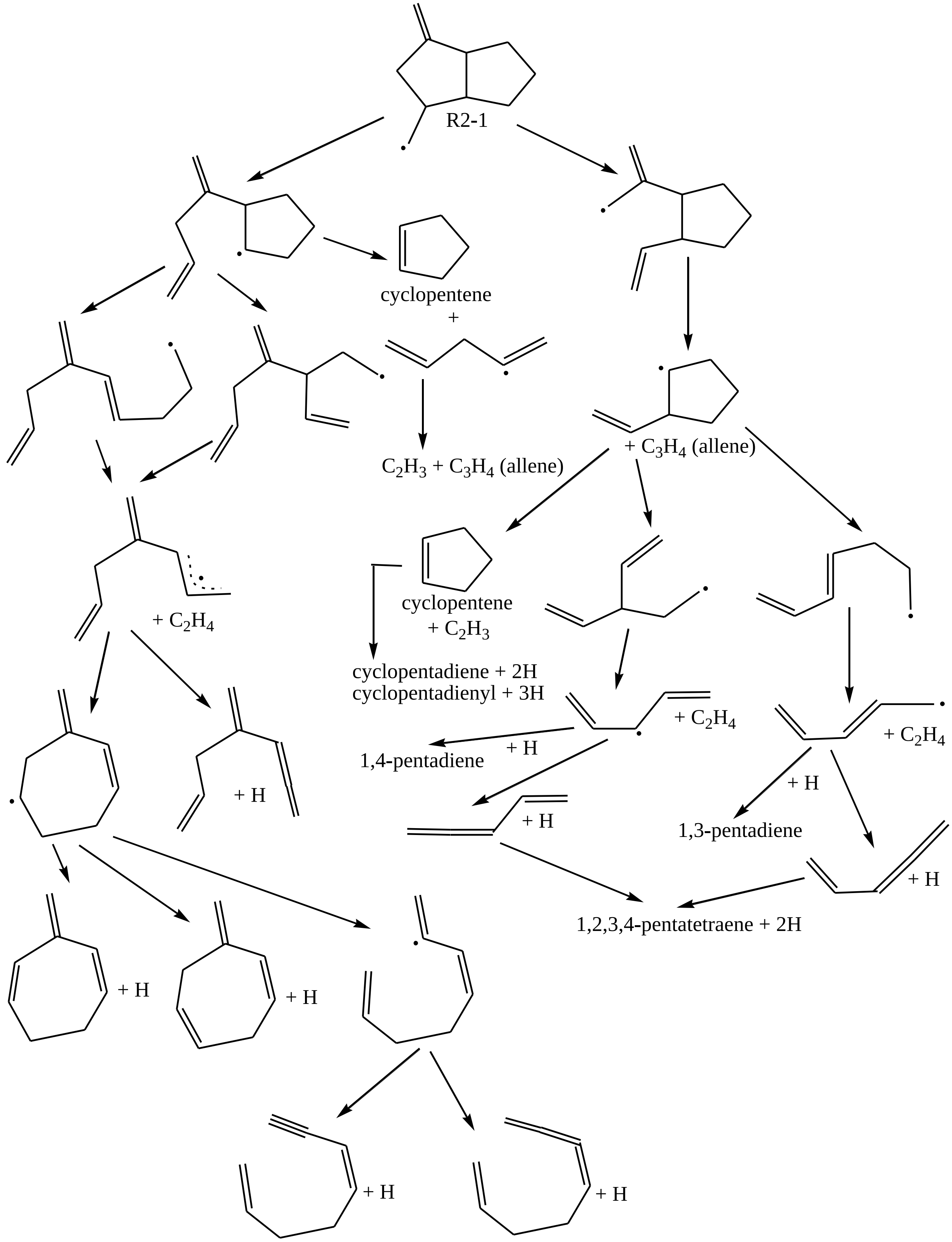

Figure S11. Potential isomerization and dissociation pathways of R2-1.

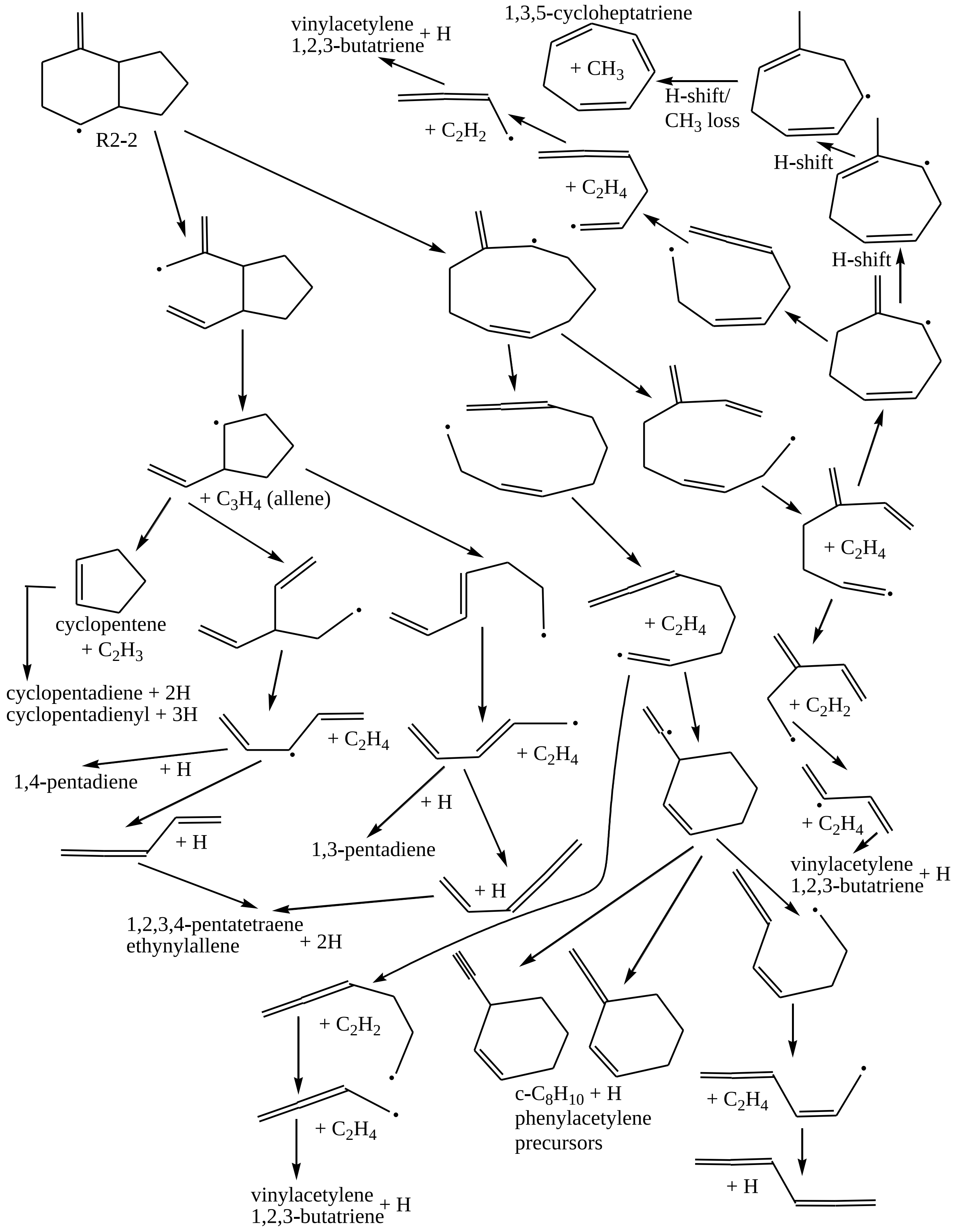

Figure S12. Potential isomerization and dissociation pathways of R2-2.

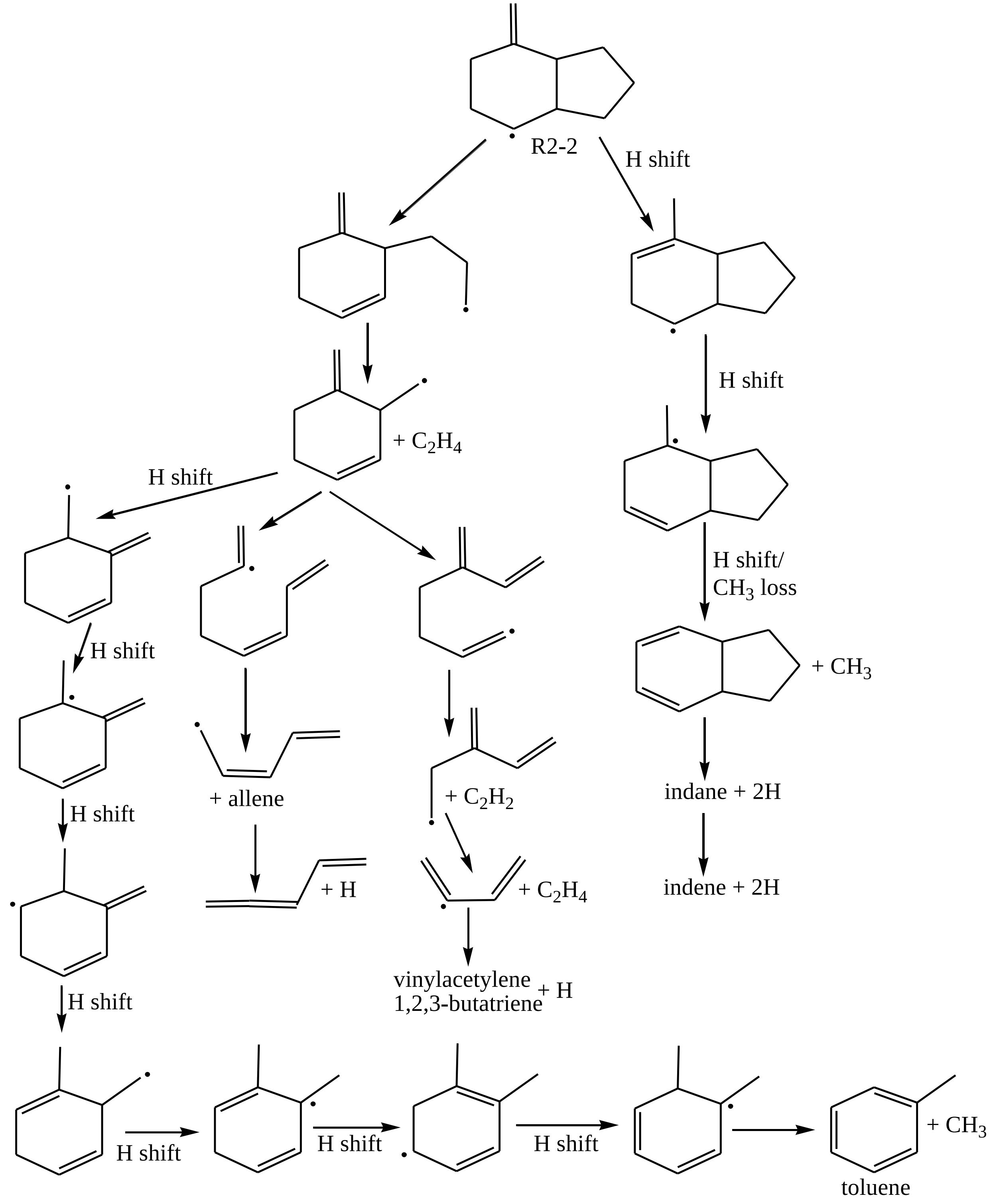

Figure S13. Potential isomerization and dissociation pathways of R2-2 (continued).

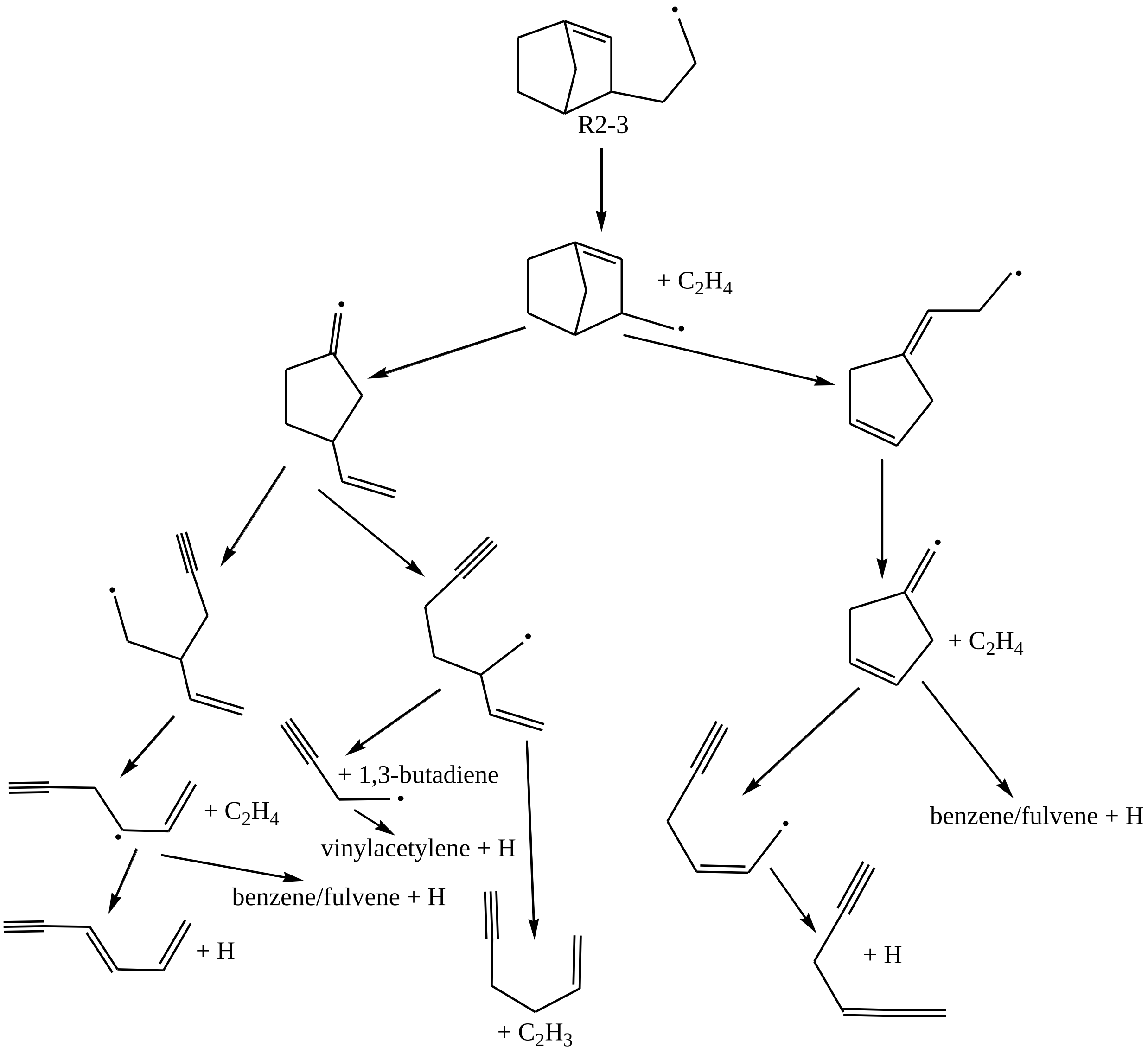

Figure S14. Potential isomerization and dissociation pathways of R2-3.

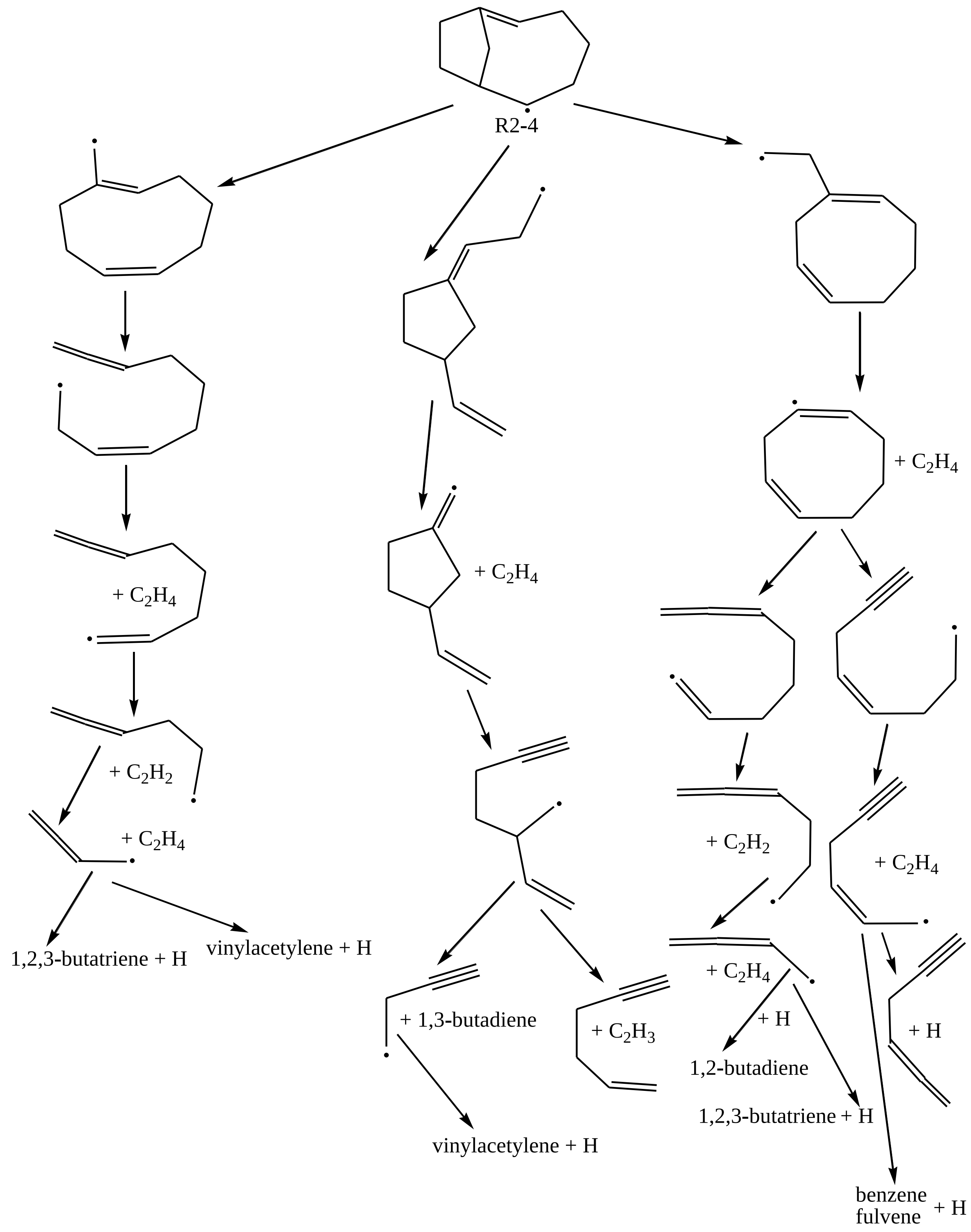

Figure S15. Potential isomerization and dissociation pathways of R2-4.

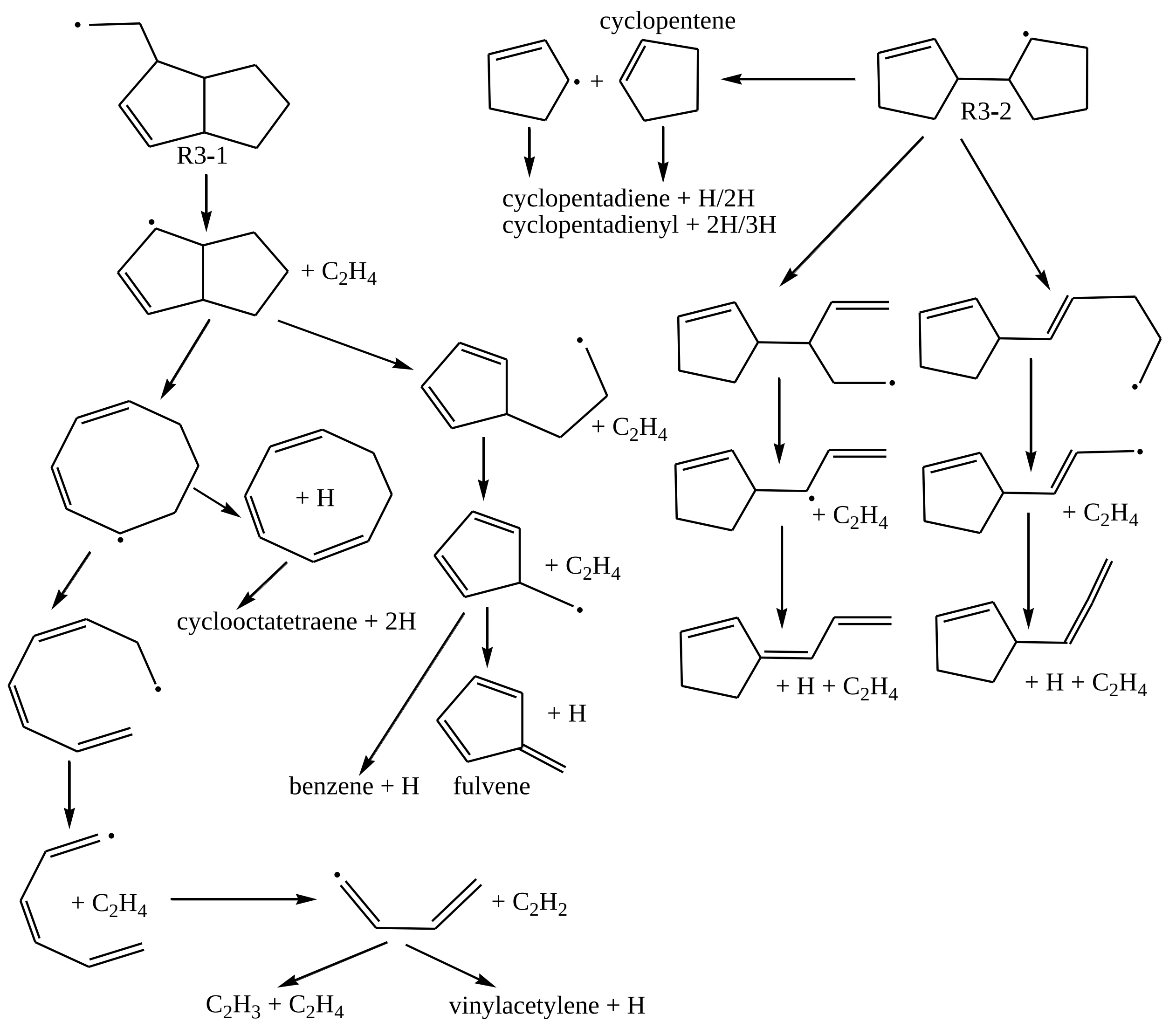

Figure S16. Potential isomerization and dissociation pathways of R3-1 and R3-2.

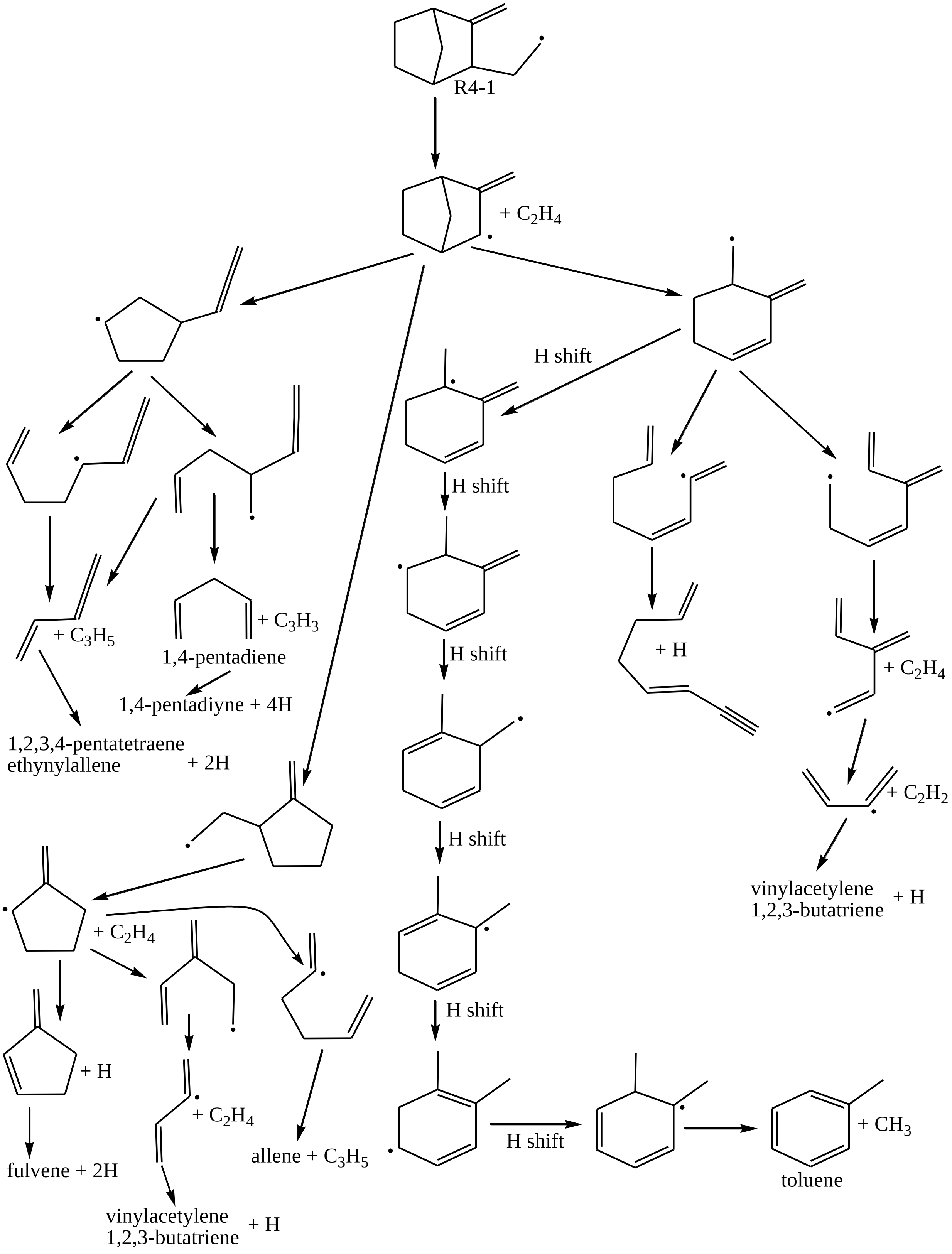

Figure S17. Potential isomerization and dissociation pathways of R4-1.

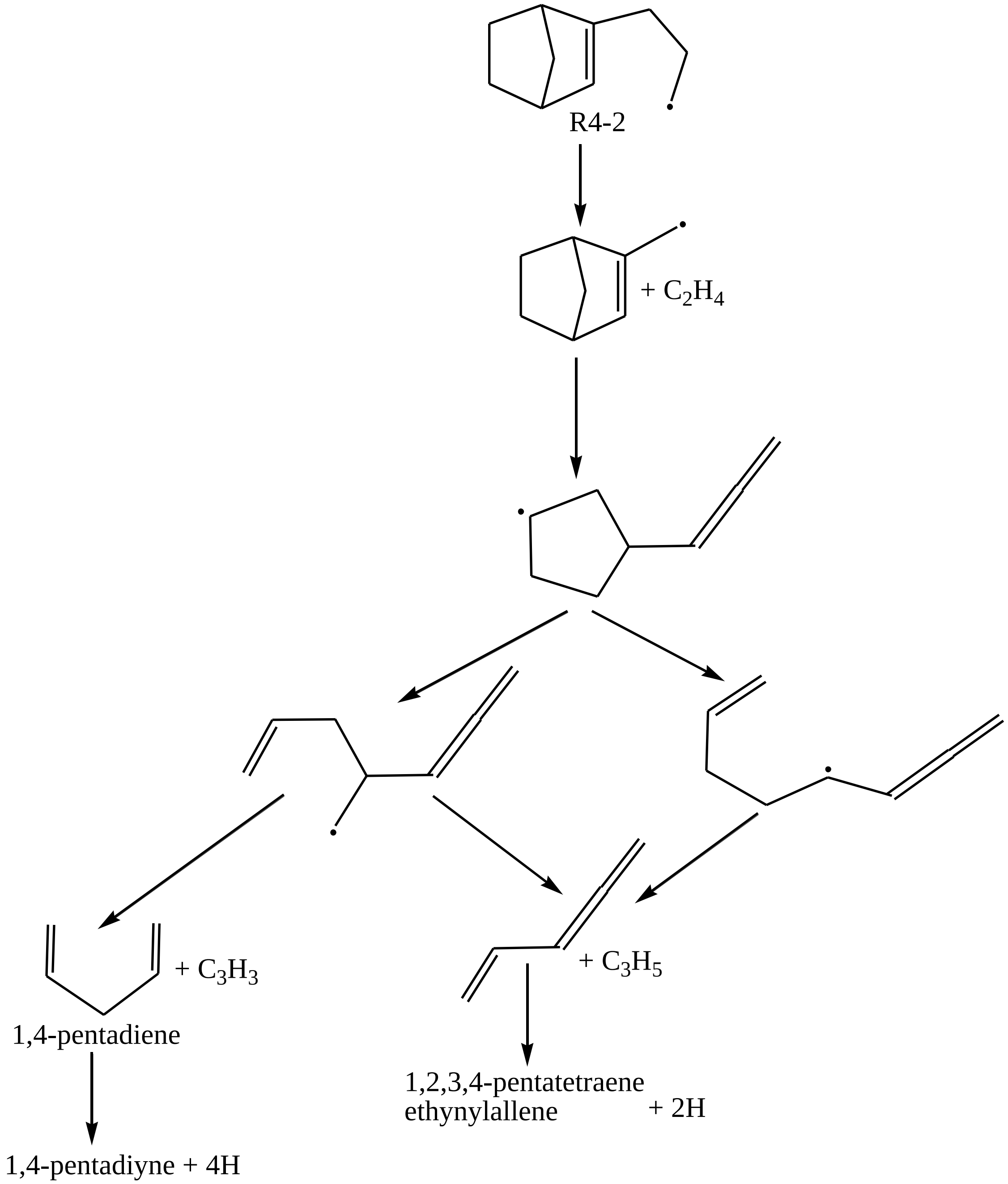

Figure S18. Potential isomerization and dissociation pathways of R4-2.

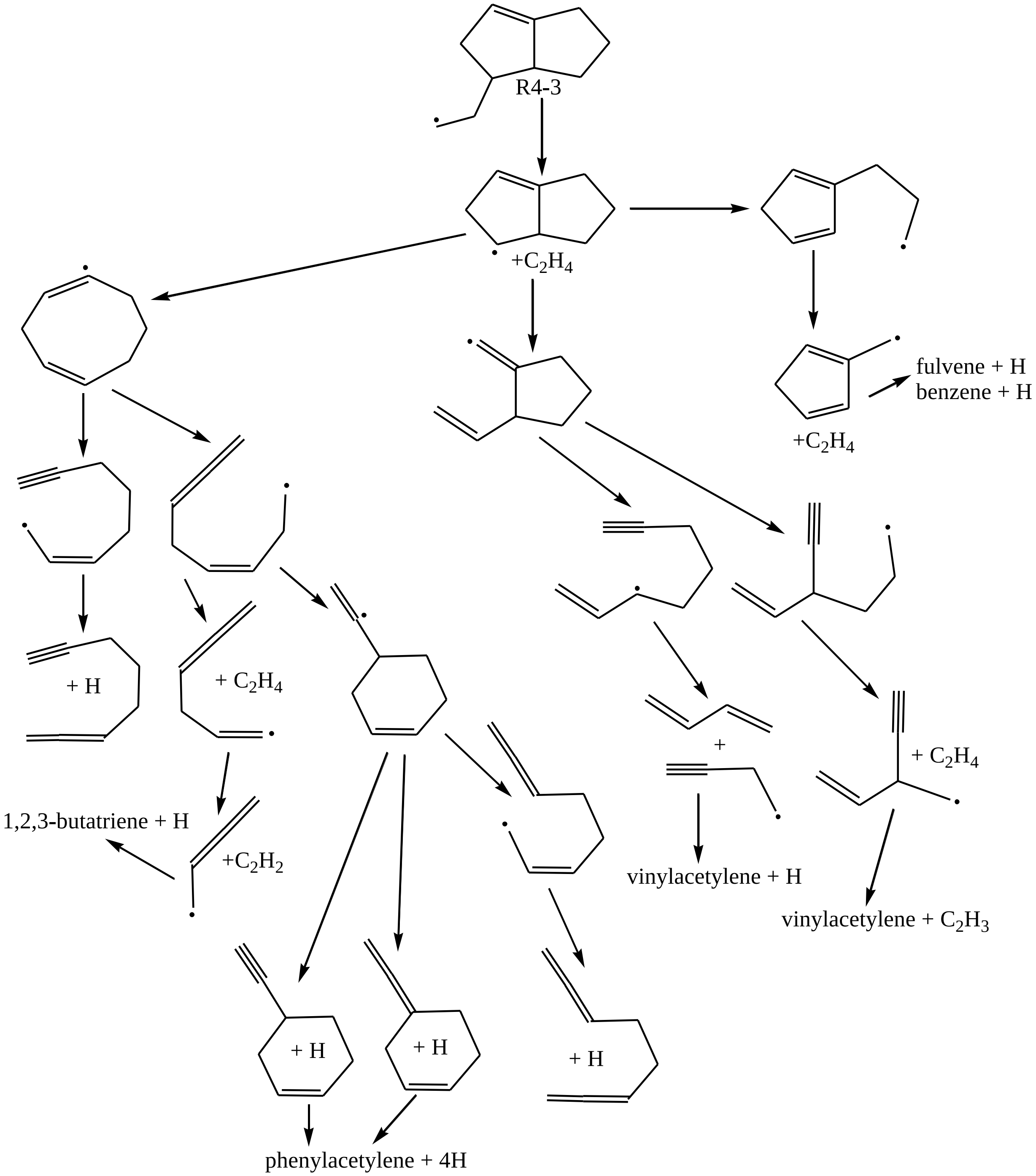

Figure S19. Potential isomerization and dissociation pathways of R4-3.

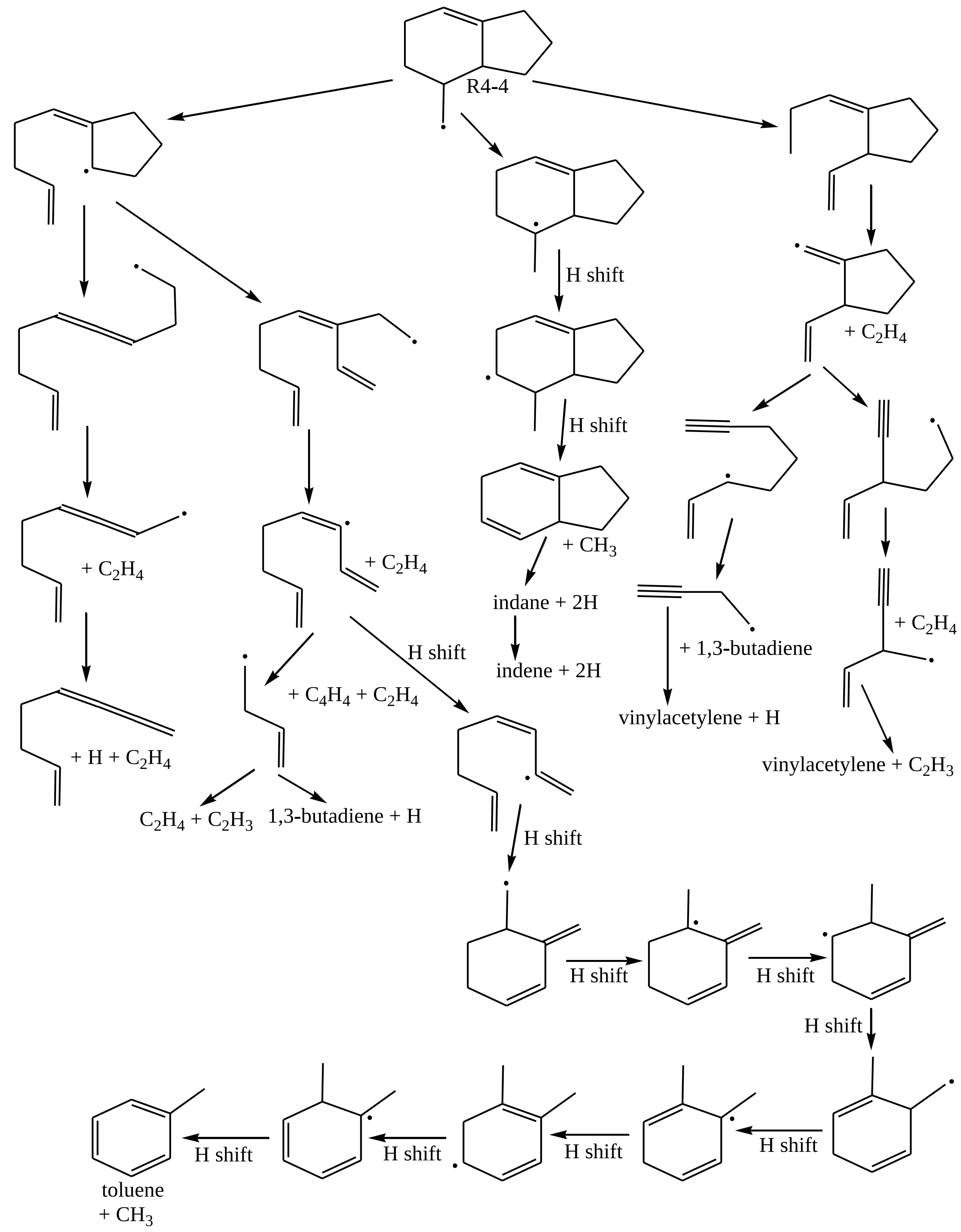

Figure S20. Potential isomerization and dissociation pathways of R4-4.

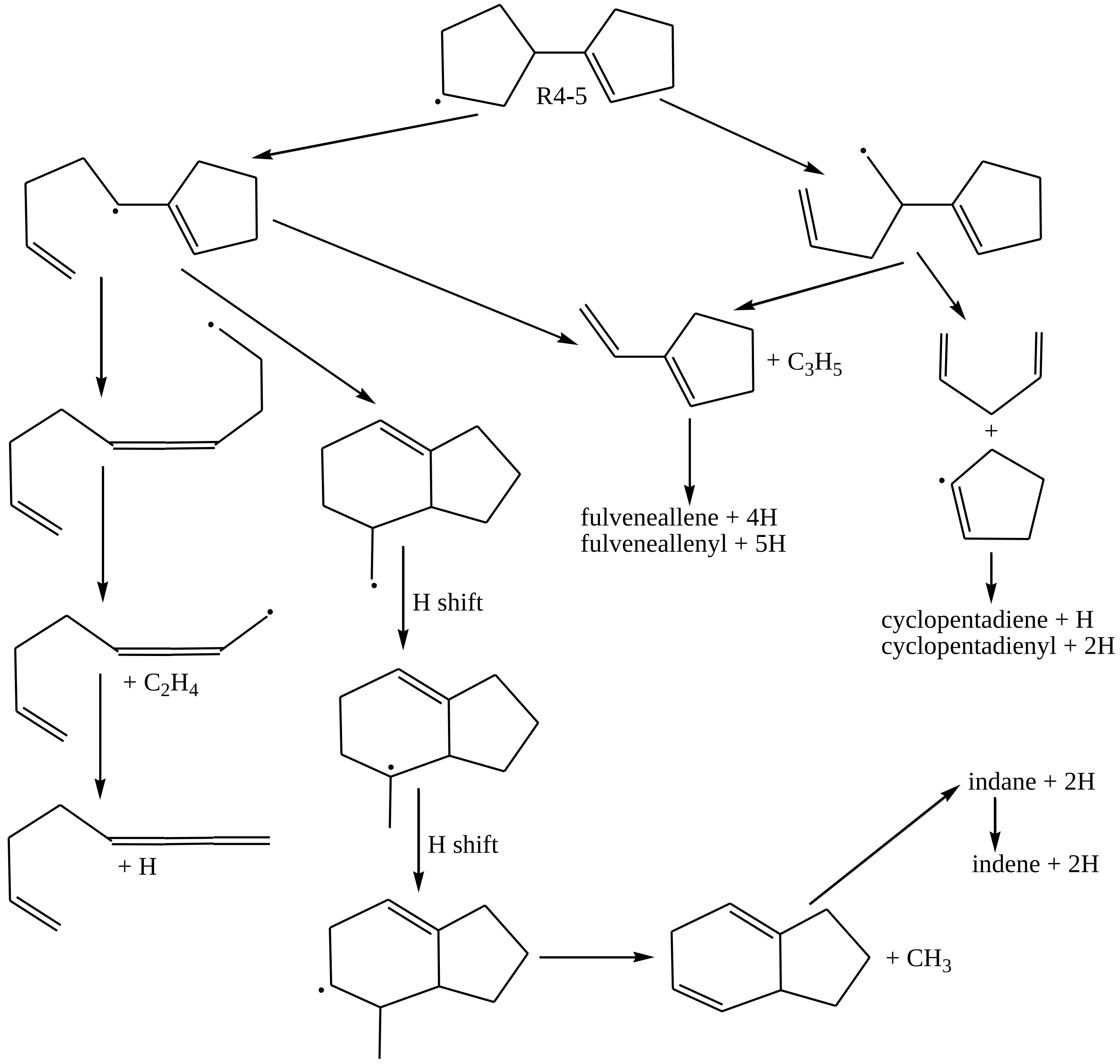

Figure S21. Potential isomerization and dissociation pathways of R4-5.

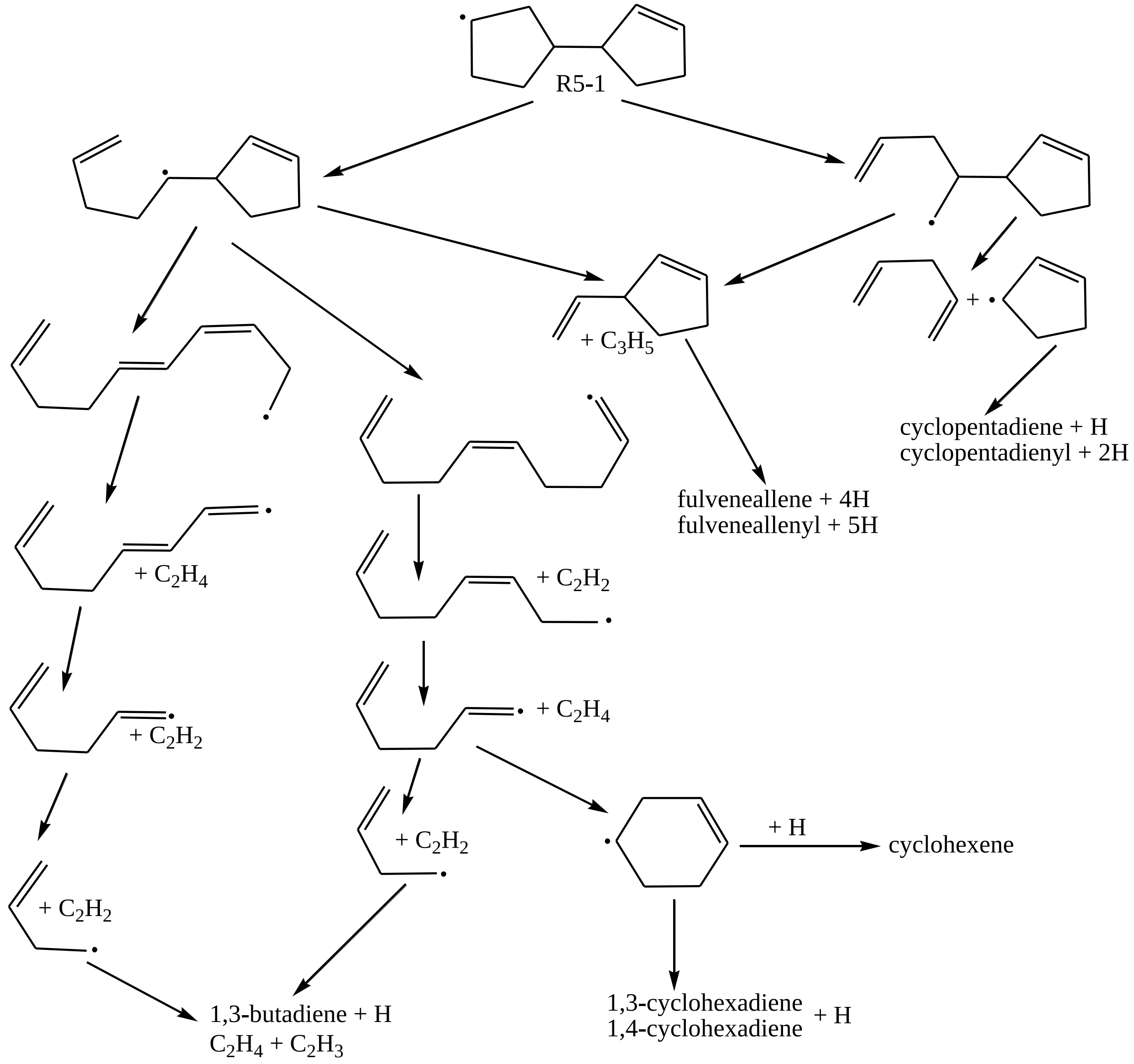

Figure S22. Potential isomerization and dissociation pathways of R5-1.

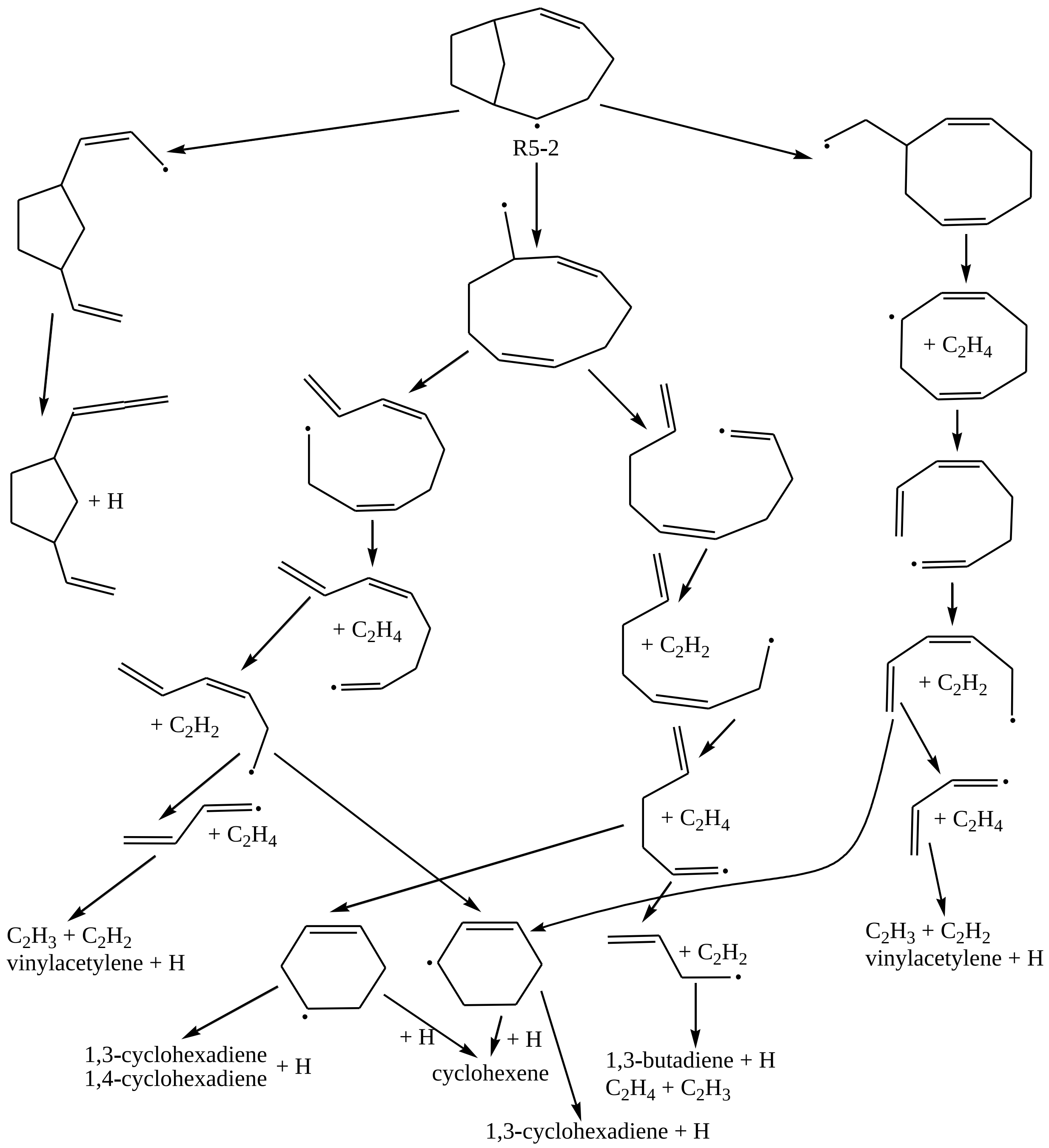

Figure S23. Potential isomerization and dissociation pathways of R5-2.

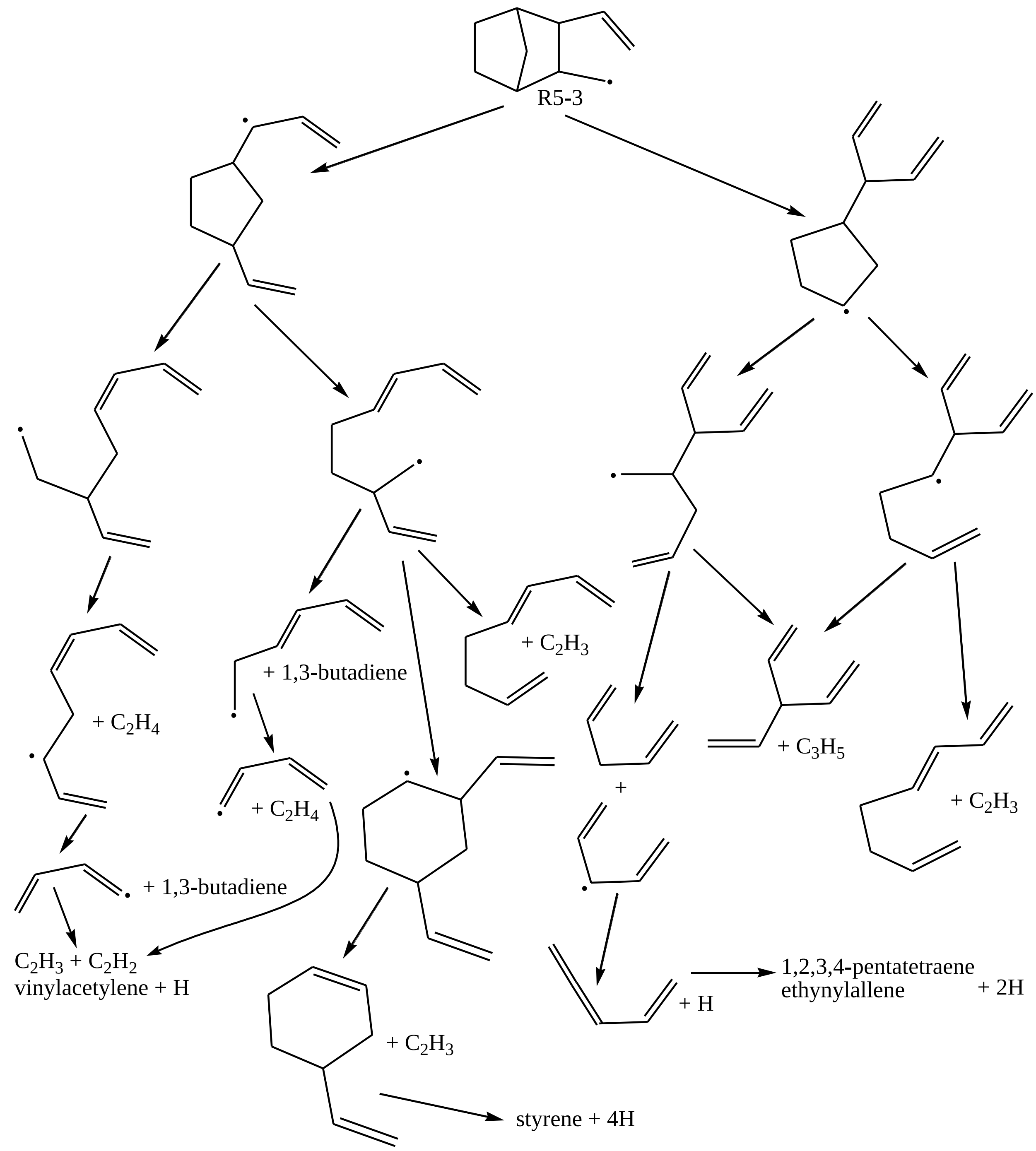

Figure S24. Potential isomerization and dissociation pathways of R5-3.

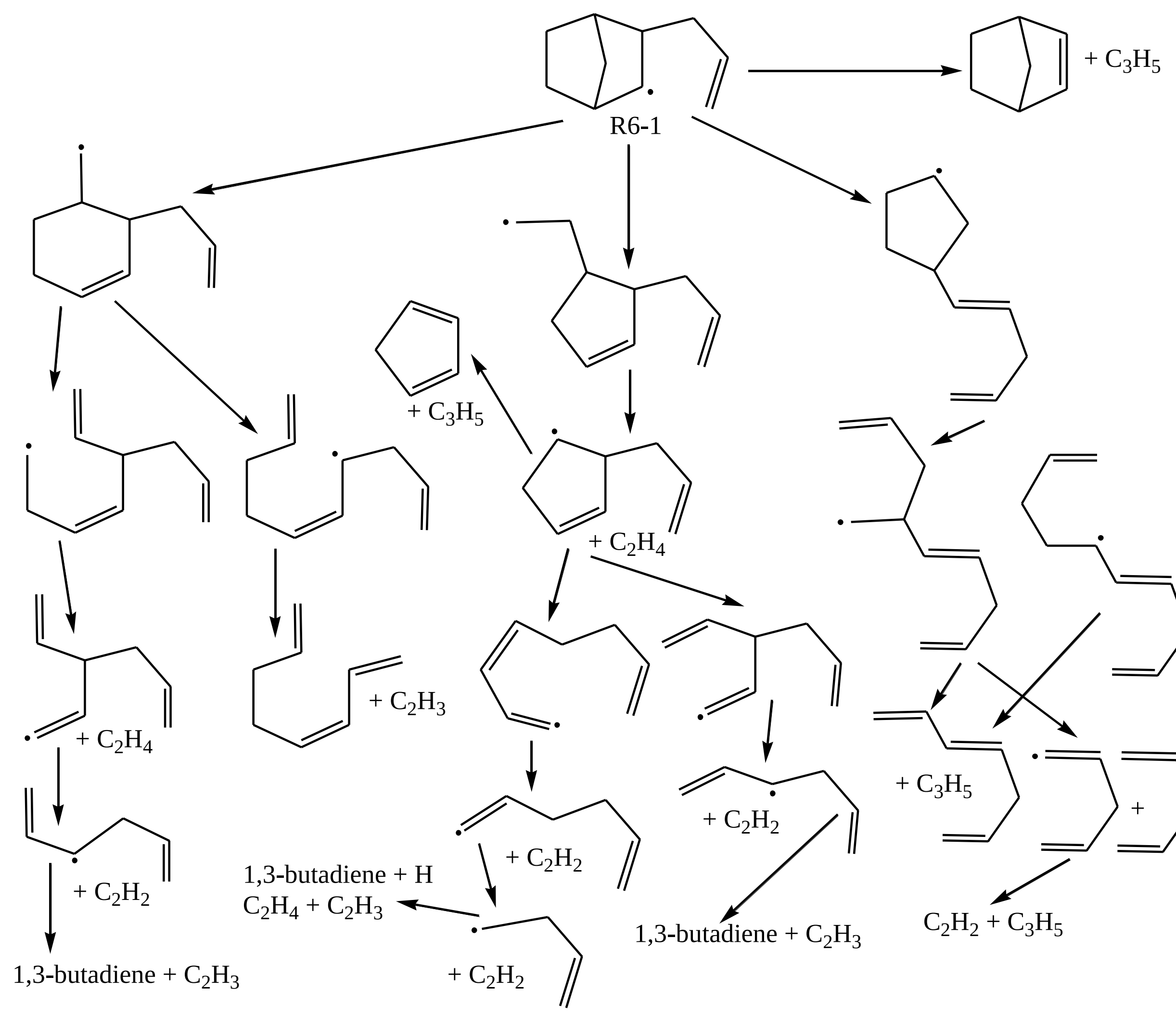

Figure S25. Potential isomerization and dissociation pathways of R6-1.

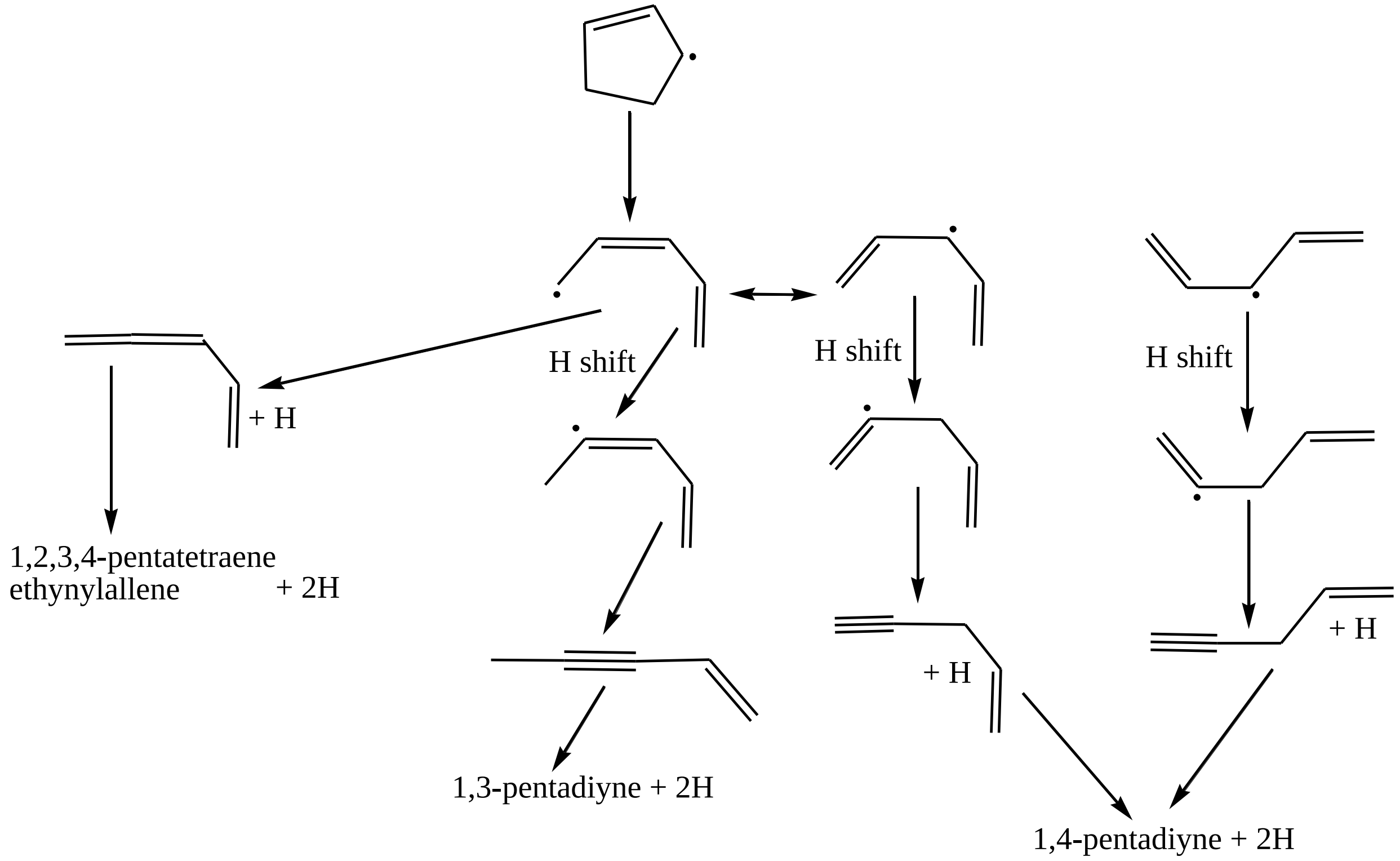

Figure S26. Potential mechanisms for the formation of acyclic $C_5H_4$ isomers.

Table S1. Calculated vertical and adiabatic ionization energies at the CCSD (T)-F12/cc-pVTZ-f12//B3LYP/6-311G** (eV) level of theory.

| Species | IE, vertical | IE, adiabatic |
|---|---|---|
| **$C_8H_6$** | | |
| (3E,5E)-3,5-Octadiene-1,7-diyne | 8.52 | 8.38 |
| Cyclooctatrienyne | 8.03 | 7.85 |
| **$C_7H_8$** | | |
| 1,6-Heptadiyne | 10.45 | 10.31 |
| 1-Ethynyl-1-cyclopentene | 8.93 | 8.69 |
| 3-Ethynylcyclopentene | 9.59 | 9.27 |
| 5-Vinyl-cyclopenta-1,3-diene | 8.71 | 8.42 |
| 6-Methylfulvene | 8.40 | 8.12 |
| Norbornadiene | 8.86 | 8.38 |
| Methylene-cyclohexadiene | 8.09 | 7.86 |
| **$C_7H_5$** | | |
| Fulveneallenyl | 8.23 | 8.17 |
| **$C_6H_4$** | | |
| (3E)-3-Hexene-1,5-diyne | 9.22 | 9.06 |
| Cyclohexa-1,4-diyne | 9.50 | 8.55 |
| 1-Hexene-3,5-diyne | 9.23 | 9.09 |
| **$C_5H_4$** | | |
| 1,4-Pentadiyne | 10.50 | 10.30 |
| 3-Ethynylcyclopropene | 9.76 | 9.24 |
| 1,2,4-Cyclopentatriene | 8.82 | 8.50 |
| 1-Ethynylcyclopropene | 9.47 | 9.21 |
| **$C_3H_6$** | | |
| Propane carbene (terminal) singlet | converges to propene | |
| Propane carbene (terminal) triplet | 8.70 | 6.48 (goes to propene cation) |
| Propane carbene (central) singlet | 8.40 | 7.69 |
| Propane carbene (central) triplet | 7.88 | 7.56 |
| **$C_3H_4$** | | |
| Propene carbene singlet | 8.58 | 8.12 |
| Propene carbene triplet | 8.41 | 8.36 |
| Cycloropane carbene singlet | 9.51 | 6.85 (goes to allene cation) |
| Cycloropane carbene triplet | 8.40 | 6.12 (goes to allene cation) |
| **$C_{10}H_{16}$[a)]** | | |
| 3-cyclopentylcyclopentene | 9.12 | 8.72 |
| 1-cyclopentylcyclopentene | 8.71 | 8.31 |
| 4-cyclopentylcyclopentene | 9.18 | 8.74 |
| bicyclopentylidene | 8.28 | 7.94 |
| 1,2-divinylcyclohexane | 9.37 | 8.9 |
| adamantane | 9.96 | 9.26 |
| JP-10 | 10.1 | 9.47 |

[a)]Calculated at the G3(MP2,CCSD)//B3LYP/6-311G** level of theory.

## Theoretical Calculations

The vertical and adiabatic ionization energies for various isomers of potential fragments of the JP-10 pyrolysis were calculated theoretically to assist in their assignment based on the experimental photoionization energy (PIE) curves. Geometries of neutral molecules and the cations were optimized using the hybrid density functional B3LYP method[1, 2] with the 6-311G** basis set and their vibrational frequencies were evaluated at the same B3LYP/6-311G** level of theory. Next, the vertical and adiabatic ionization energies were refined by single-point explicitly correlated coupled clusters CCSD(T)-F12 calculations[3] with Dunning's correlation-consistent cc-pVTZ-f12 basis set.[4] It is expected that the CCSD(T)-F12/cc-pVTZ-f12 energies provide a close approximation for CCSD(T) energies at the complete basis set limit. The CCSD(T)-F12 calculations were not feasible for $C_{10}H_{16}$ isomers and hence their energies were refined employing the G3(MP2,CCSD) modification[5, 6] of the G3 model chemistry approach.[7] The adiabatic ionization energies included zero-point vibrational energy corrections (ZPE) evaluated using B3LYP/6-311G**-computed vibrational frequencies. The CCSD(T)-F12 ionization energies are expected to be accurate within about 0.02 eV, where the anticipated accuracy of the G3(MP2,CCSD) results is ± 0.1 eV. Ionization Franck-Condon factors were calculated utilizing B3LYP/6-311G** vibrational frequencies and normal modes for the neutral and ionized species using the theoretical approach developed by Barone and coworkers.[8] All calculations were carried out employing the GAUSSIAN 09[9] and MOLPRO 2010[10] program packages.